\documentclass[Journal]{IEEEtran}
\usepackage{esvect}
\usepackage{enumitem}
\newlist{abbrv}{itemize}{1}
\setlist[abbrv,1]{label=,labelwidth=1in,align=parleft,itemsep=0.1\baselineskip,leftmargin=!}
\usepackage{colortbl}
\usepackage{boldline,multirow}

\usepackage{hhline}
\usepackage[withpage]{acronym}
\usepackage{graphicx}
\usepackage{amsmath}
\usepackage{MnSymbol}
\usepackage{dblfloatfix}
\usepackage{wasysym}
\usepackage{graphicx} 
\usepackage[table]{xcolor}
\usepackage{url}
\usepackage{lipsum}
\usepackage{textcomp}
\usepackage{tabularx}
\usepackage{lettrine}
\usepackage{cite}
 \usepackage{amsmath}
\usepackage{tabularx}
\usepackage[ampersand]{easylist}
\usepackage{multirow}
\usepackage{tikz}
\usepackage{pdflscape}
\usepackage{longtable}
\usepackage{rotating}
\usepackage{pifont}
\usepackage{ctable, fancybox}
\usepackage{bold-extra}
\usepackage{csquotes}
\usepackage{lineno,hyperref}
\usepackage{chronology}
\usepackage{array}
\usepackage{caption}
\usepackage{makecell}
\usepackage{boldline}
\usepackage{subfigure}
\usepackage{lineno} 
\definecolor{almond}{rgb}{0.94, 0.87, 0.8}

 \newcommand{\tick}{\ding{52}}
 \newcommand{\key}{\rotatebox{90}}
\begin{document}

\title{Opportunistic Routing  Metrics: A Timely One-Stop Tutorial Survey }

\author{{Mostafa~Abdollahi,~\IEEEmembership{Student~Member,~IEEE,}~{Farshad~Eshghi},~\IEEEmembership{Member,~IEEE,} Manoochehr~Kelarestaghi,~\IEEEmembership{Member,~IEEE,} Mozafar Bag-Mohammadi
	}
	\thanks{M. Abdollahi, F. Eshghi and M. kelarestaghi are with the Dept. of Electrical \& Computer Engineering, Faculty of Engineering, Kharazmi University, Tehran , Iran. e-mail: \{ std\_abdolahi68, farshade, kelarestaghi\}@khu.ac.ir}
	\thanks{M. Bag-Mohammadi was with the Wireless Network Labs, Engineering Faculty, Ilam University, Ilam, Iran. e-mail: \{ mozafar\}@ilam.ac.ir}
}

 \maketitle

\begin{abstract}
High-speed, low latency, and heterogeneity features of 5G, as the common denominator of many emerging and classic wireless applications, have put wireless technology back in the spotlight. Continuous connectivity requirement in low-power and wide-reach networks underlines the need for more efficient routing over scarce wireless resources, in multi-hp scenarios. In this regard, Opportunistic Routing (OR), which utilizes the broadcast nature of wireless media to provide transmission cooperation amongst a selected number of overhearing nodes, has become more promising than ever. Crucial to the overall network performance, which nodes to participate and where they stand on the transmission-priority hierarchy, are decided by user-defined OR metrics embedded in OR protocols. Therefore, the task of choosing or designing an appropriate OR metric is a critical one. The numerousness, proprietary notations, and the objective variousness of OR metrics can cause the interested researcher to lose insight and become overwhelmed,   making the metric selection or design effort-intensive. While there are not any comprehensive OR metrics surveys in the literature, those who partially address the subject are non-exhaustive and lacking in detail. Furthermore, they offer limited insight regarding related taxonomy and future research recommendations. In this paper, starting with a custom tutorial with a new look to OR and OR metrics, we devise a new framework for OR metric design. Introducing a new taxonomy enables us to take a structured, investigative, and comparative approach to OR metrics, supported by extensive simulations. Exhaustive coverage of OR metrics, formulated in a unified notation, is presented with sufficient details. Self-explanatory, easy-to-grasp, and visual-friendly quick references are provided, which can be used independently from the rest of the paper. Finally, a new insightful framework for future research directions is developed. This tutorial-survey has been organized to benefit both generalists and OR specialists equally, and to be used not only in its entirety but selectively as well. 
\end{abstract}

\begin{IEEEkeywords} Opportunistic routing; Metric; Multi-hop; Wireless network; Ad hoc; Wireless sensor network; Transmission cooperation; Candidate forwarder list.   \end{IEEEkeywords}

\section{ \large\bf Introduction} \label{sec1}
Routing, the process of choosing the (possibly) most cost-efficient path(s) between two endpoints in a multi-hop network, has long been of special interest to network researchers. The routing process is one of the main functionalities of the network layer. Different modes of transmission (uni/broad/multi/any-cast) combined with the consideration of various factors, such as traffic volume and type, number of hops, node density, mobility, energy requirements, and Quality of Service (QoS), have led to the creation of different routing algorithms/protocols. Early routing algorithms were introduced for wired networks. Thus, they were adapted to specific characteristics of those networks, such as low-error/high-bandwidth links and point-to-point (unicast per link) packet delivery mode.
\par
\begin{figure*}[t!]
  	\centering
  	\includegraphics[width=125mm]{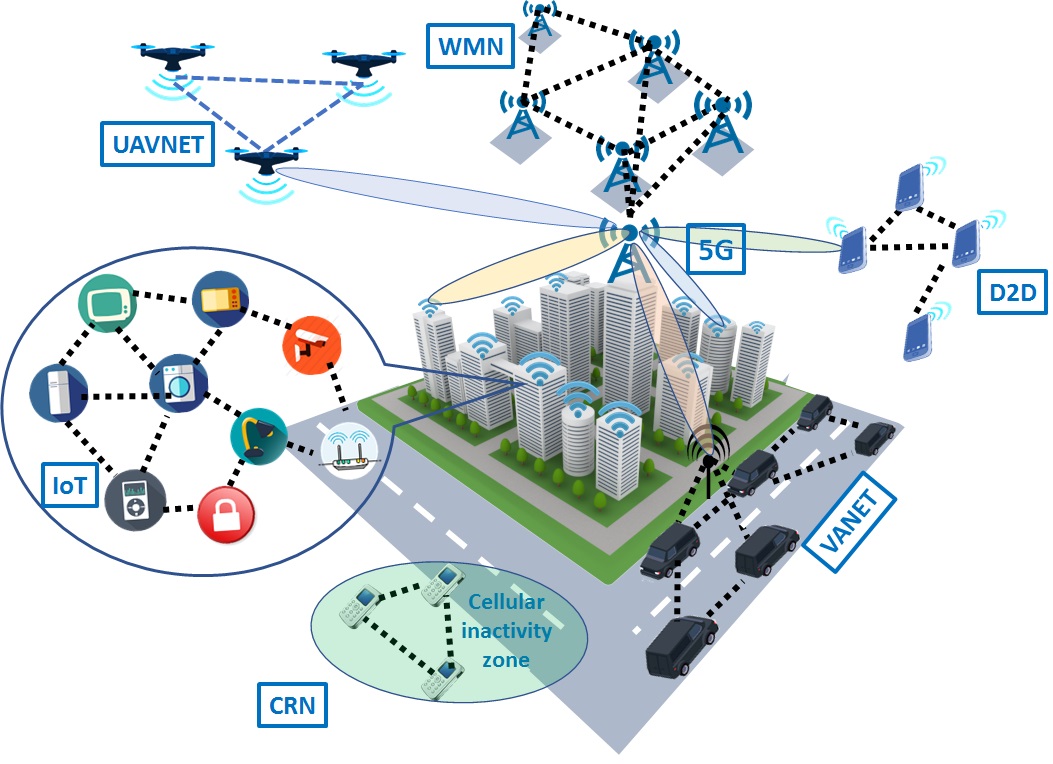}
  	\caption{Classic and new emerging multi-hop wireless networks.}
  	\label{fig:pic2018}
  \end{figure*}

 In recent years, a variety of Wireless Networks (WNs) have been deployed extensively due to their ease of access, large areas of applicability, low set-up cost, and fast implementation. On the other hand, the prevalence of multi-hop-in-nature wireless infrastructures in Wireless Mesh Networks (WMNs), the delivery of information at a limited number of sink nodes in Wireless Sensor Networks (WSNs), the ubiquitous connectivity requirement in the context of the Smart City \cite{P_2018_27}, the topology variation and geographical spread of Vehicular Ad hoc Networks (VANETs) \cite{P_2018_28,P_2019_3_032301}, the emerging Vehicular Energy Networks (VENs) \cite{P_2019_2_101347}, and Unmanned Aerial Vehicles Networks (UAVNETs, also called Flying Ad hoc Networks (FANETs))  \cite{P_2019_1_279}\cite{P_2019_1_278}\cite{P_2019_4_010053}, the enormity of low-power devices in IoT \cite{P_2018_29}\cite{P_2019_2_279}, the D2D communications \cite{P_2018_300}\cite{P_2019_3_280407} support in 5G \cite{P_2017_12}, and the ad hoc networks of secondary users in cognitive wireless networks \cite{P_2018_32} all raise serious demands for more promising multi-hop routing protocols in wireless networks (Fig.~\ref{fig:pic2018}).\\ The WNs are characterized by high-error/low-bandwidth links, time-varying, and multi-path communication channels, a broadcast medium (which implies that all nodes within radio range hear each other), a tight power budget, and possibly mobile nodes. These characteristics require a custom-designed dynamic routing algorithm/protocol. Other than the higher dynamics due to mobility and time-varying link qualities, traditional wireless routing, i.e., the single-path routing, does not differ substantially from wired networks routing. In single-path routing, the route information generated through the route discovery process is included in the packet header. The route information determines how a packet traverses a single path, hop-by-hop, towards the final destination.\par
\begin{figure}[h!]
  	\centering
  	\includegraphics[width=0.9\linewidth]{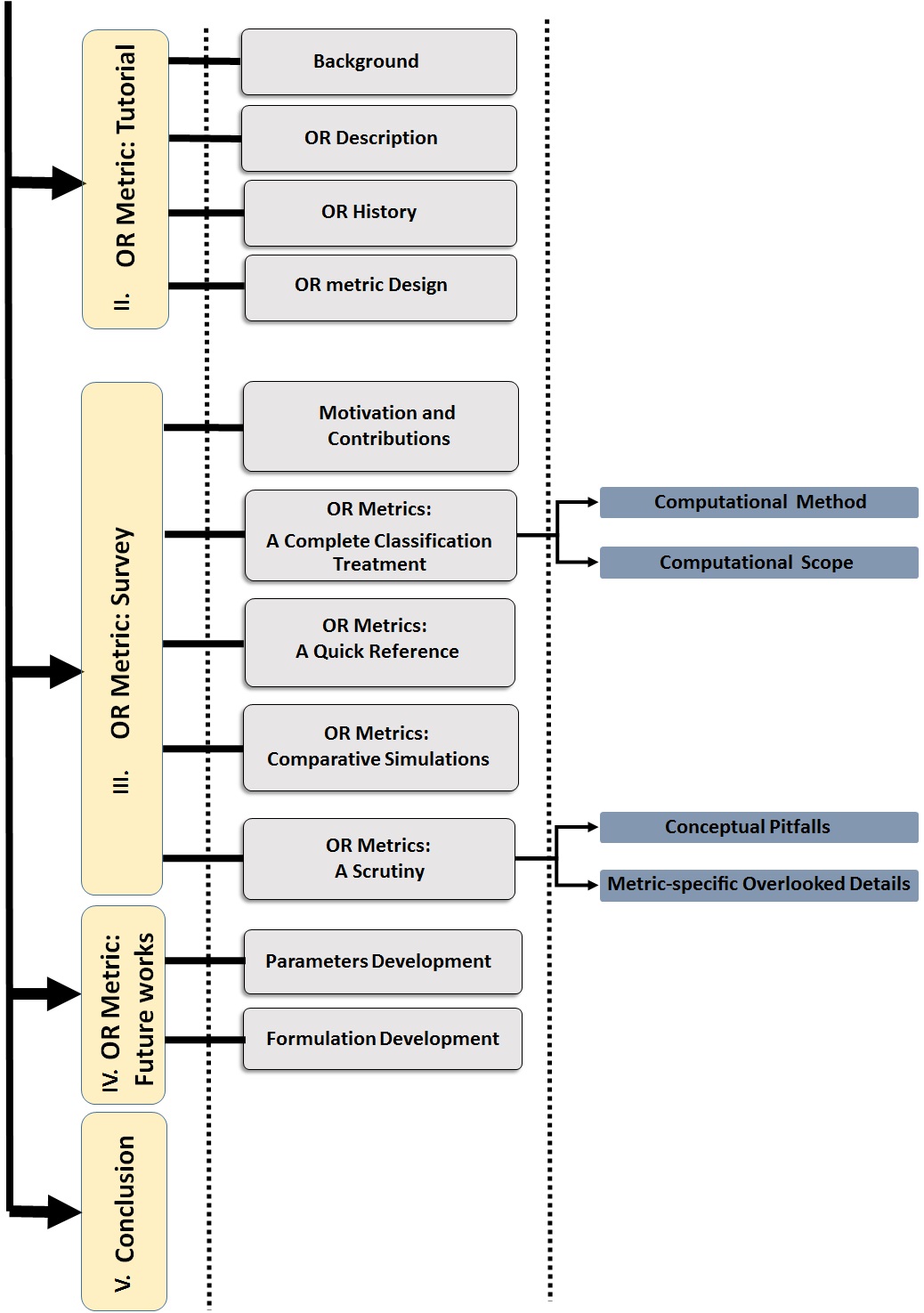}
  	\caption{Organization of the paper.}\label{fig:chart}
  \end{figure}

The rest of this paper is organized according to Fig.~\ref{fig:chart}. To make this work more useful, even for generalists, we start with a self-explanatory tutorial, including background on diversity in wireless, a description of  OR, the OR history, and finally, the OR metric design process. The second section, a survey on OR metrics, starts with the motivation for and the contributions of this work. The section continues with a classification of OR metrics presented in section \ref{sec3}. After that, a complete collection of OR metrics is classified and described using the introduced unified notation in Table \ref{table:tab11}.  In section \ref{sec33}, two easy-to-use quick-reference tables and one timeline illustration are presented.  A simulation comparison between main representatives of OR metric classes, and a discussion of implicit and difficult-to-notice network performance dependencies appear in section \ref{sec4}. Some fundamental and critically important questions regarding OR metric computation, in general, and some specific OR metrics are raised in section \ref{Sec:Scrutiny}. Ideas for future research opportunities are proposed in section \ref{sec5}. The concluding remarks complete the discussion.

  \section{\large\bf OR Metric: A Tutorial}
   \begin{figure}[h!]
   	\centering
   	\includegraphics[width=0.5\textwidth]{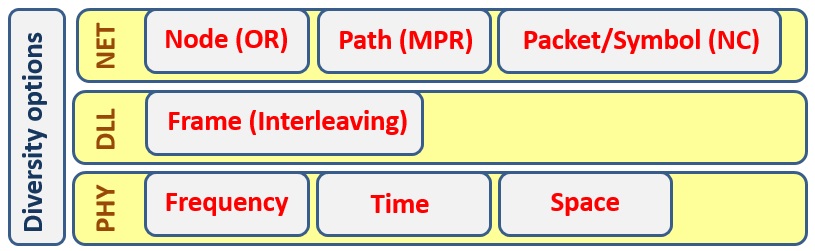}
   	\caption{Diversity options in wireless transmission.}
   	\label{fig:Diversity}
   \end{figure} 
   
   \begin{table*}[!htbp]
   	\centering
   	\caption{List of acronyms.}
   	\label{table:Abbrevations}
   		\resizebox{1.0 \textwidth}{!}{
   	\begin{tabular}{p{1.3cm} p{7cm} p{1.7cm} p{7.2cm}}
   		\hline
   		ACK & Acknowledgement & APR & Alternate-Path Routing \\
	 
	 BS & Base Station &	CFS & Candidate Forwarding Set \\ 
	 CPS  & Cluster Parent Set & CR-SIoT & Cognitive Radio Social Internet of Things\\ 
	 CR & Cognitive Radio & 	CRN & Cognitive Radio Networks \\
	 CRAHN & Cognitive Radio Ad-Hoc Networks & D2D & Device-to-Device \\
   		DIFS & DCF Interframe Space & DL & Deep Learning \\
   		DLL & Data Link Layer & DS-CDMA & Direct Sequence-Code Division Multiple Access \\
   		
   		EOF & Energy Objective Function & EWMA & Exponential Weighted Moving Average\\
   		
   		FANET & Flying Ad hoc Networks & IoT & Internet of Things\\
   		IoV & Internet of Vehicles & LIVE & Link Validity Estimation \\
   		
   		LQI & Link Quality Indicator  &  MAC & Medium Access Control\\
   		MANET & Mobile Ad hoc Network & MDF & Multi-user Diversity Forwarding\\
   		MIMO & Multiple-Input and Multiple-Output & ML & Machine Learning\\
   				
   			MPR & Multi-Path Routing & NB-IoT & Narrowband Internet of Things \\
   			NC & Network Coding & NET & Network Layer\\	
   			OFDM & Orthogonal Frequency Division Multiplexing & OR & Opportunistic Routing \\

   			PDP & Packet Delivery Probability & PHY & Physical Layer\\	
   			
   			QoS & Quality of Service & RFC & Request For Comment\\
   			RTT & Round-Trip Time & RSSI & Radio Signal Strength Indicator \\  
   			SDF & Selection Diversity Forwarding & SIFS & Short Inter Frame Space\\
   			
   				STC & Space-Time Coding & TCP & Transmission Control Protocol \\
   				UAWSN & Acoustic Wireless Sensor Network & UOWN& Underwater Optical Wireless Network \\
   				
   				UWSN & Underwater Wireless Sensor Networks & VEN & Vehicular Energy Networks\\
   				
   				VANET & Vehicular ad hoc Networks & WANET & Wireless Ad hoc Networks\\
   				WN & Wireless Networks & WMN & Wireless Mesh Networks\\
   				WSN & Wireless Sensor Networks& 5G & Fifth Generation Cellular System  \\ 
   		\hline
   	\end{tabular}
}
   \end{table*}
 \subsection{\bf Background} \par

 The unreliability of wireless links necessitates considering extra provisions, particularly in multi-hop applications. Employing diversity can mitigate the unreliability by benefiting from the broadcasting nature of the wireless medium. In the wireless literature, there are two different, though somewhat close, points of view regarding diversity.
 The first point of view regards diversity as having a variety of available wireless transmission means and choosing the best one. Some related techniques are Multi-user Diversity Forwarding (MDF) and Selection Diversity Forwarding (SDF), which choose from the available multiple transmitter-receiver pairs and downstream forwarders, respectively \cite{P_2019_2_5113}. The second point of view regards diversity as the collaboration or cooperation between various transmission means. In the context of OR, we stick to the latter approach in what follows.\\ To provide a structured diversity discussion, we opt to present it in the framework of the TCP/IP network protocol stack. As Fig.~\ref{fig:Diversity} suggests, the diversity in wireless transmission can be introduced in the three bottom layers of the network protocols stack. It is essential to note that diversity implementation is generally a cross-layer task. However, we place each diversity option at the layer which performs its core part. At the lowest layer, PHY, the diversity is over the transmission channel and based on the fact that individual channels experience independent fading phenomena \cite{P_2019_1_23}. The independent channels are generated by 1) spreading the signal over a broader frequency band (frequency diversity), e.g., Direct Sequence-Code Division Multiple Access (DS-CDMA) \cite{P_2019_2_51029}, or carrying it on multiple frequency carriers, e.g., Orthogonal Frequency Division Multiplexing (OFDM) \cite{P_2019_2_51035}, 2) spreading the data over time (time diversity), and 3) providing multiple physical transmission paths per link (space diversity), e.g., Multiple-Input and Multiple-Output (MIMO) \cite{P_2019_2_51037} and Space-Time Coding (STC) \cite{P_2019_2_51040}. \\ 
 One layer up, at the DLL level,  a data frame can be spread over multiple transmitted data frames through a technique traditionally called frame interleaving \cite{P_2019_2_51105}.\\ 
 At the network level, i.e., NET, the content of a subject packet is transmitted by multiple carriers such as paths, packets, and nodes. The Multi-Path Routing (MPR), as opposed to the Alternate-Path Routing (APR) \cite{P_2019_4_191937} in wired networks, sends a packet through multiple paths simultaneously (path diversity) to compensate for the unreliability of individual paths \cite{P_2019_1_26}. The Network Coding (NC) provides diversity by transmitting carefully designed mathematical combinations of multiple original packets instead. The mathematical combination can be implemented at the packet level \cite{P_2019_1_24}\cite{ P_2019_1_25}, and the symbol level \cite{p_9_7}. Finally, the OR \cite{P6} takes advantage of the broadcasting nature of the wireless medium and makes all the overhearing nodes incrementally contribute to the forwarding process (node diversity).\\
 It is worth mentioning that diversity options at the same layer or different layers can be combined to provide better resilience against the unreliability of wireless medium, for instance, the use of joint OR and intra-flow network coding at the NET layer in \cite{P9} and \cite{P_2019_1_24}.

 \par
 \subsection{\bf OR Description}
Opportunistic routing is a highly capable scheme benefiting from the broadcast nature of the wireless medium, especially in, but not limited to, static and semi-static wireless environments. As well as its original application in WMNs and WSNs, OR has very recently found its place in the aforementioned emerging wireless applications, including the Smart City \cite{P_2018_25}, VANET\cite{P_2018_20,P_2018_33,P32}, IoT \cite{P_2017_7,P_2018_13,P_2018_21}, D2D communications \cite{P_2018_22}\cite{P_2018_26}, and cognitive wireless networks \cite{P_2018_23}\cite{P_2018_24}.
\par  OR forces a subset of nodes overhearing a batch of in-transit packets, to cooperatively participate in the forwarding process. In fact, OR relaxes the notion of the next hop in traditional wireless routing and replaces it with an ambitious scheme that opportunistically exploits all the potential forwarders (which are closer to the destination than the sender). In other words, OR provides the possibility of packet delivery over different paths. Therefore, the merit of OR lies in extending the forwarding role to the members of the forwarder list. In contrast to the single-path (multiple) routing, which specifies a single (multiple) path(s) for forwarding a packet, OR creates an ordered set of nodes wherein a node takes a forwarding action when its predecessors on the list fail to transmit.
\par 
In the following, the basics of OR are discussed for a better understanding. The main steps in a conventional OR protocol are:
\begin{itemize}
	\item {In-advance calculation and dissemination of OR metric of all nodes for each source-destination pair,}
	\item {candidate forwarder set (CFS)  formation,}
	\item {ordering of the CFS according to their OR metric values,}
	\item {forwarder list selection, }
	\item {members of the forwarder list becoming aware of their selection and location on the list, }
	\item {and transmission scheduling.
		}
\end{itemize}

 \begin{figure}[h!]
 	\centering
 	\includegraphics[width=0.5\textwidth]{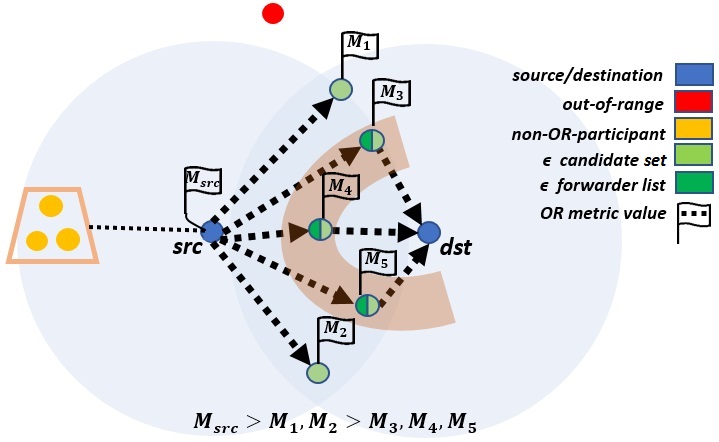}
 	\caption{A typical two-hop apart source-destination OR scenario.}
 	
 	\label{fig:OR2}
 \end{figure}
 \begin{figure*}[t!]
 	\centering
 	\includegraphics[width=1.0\textwidth]{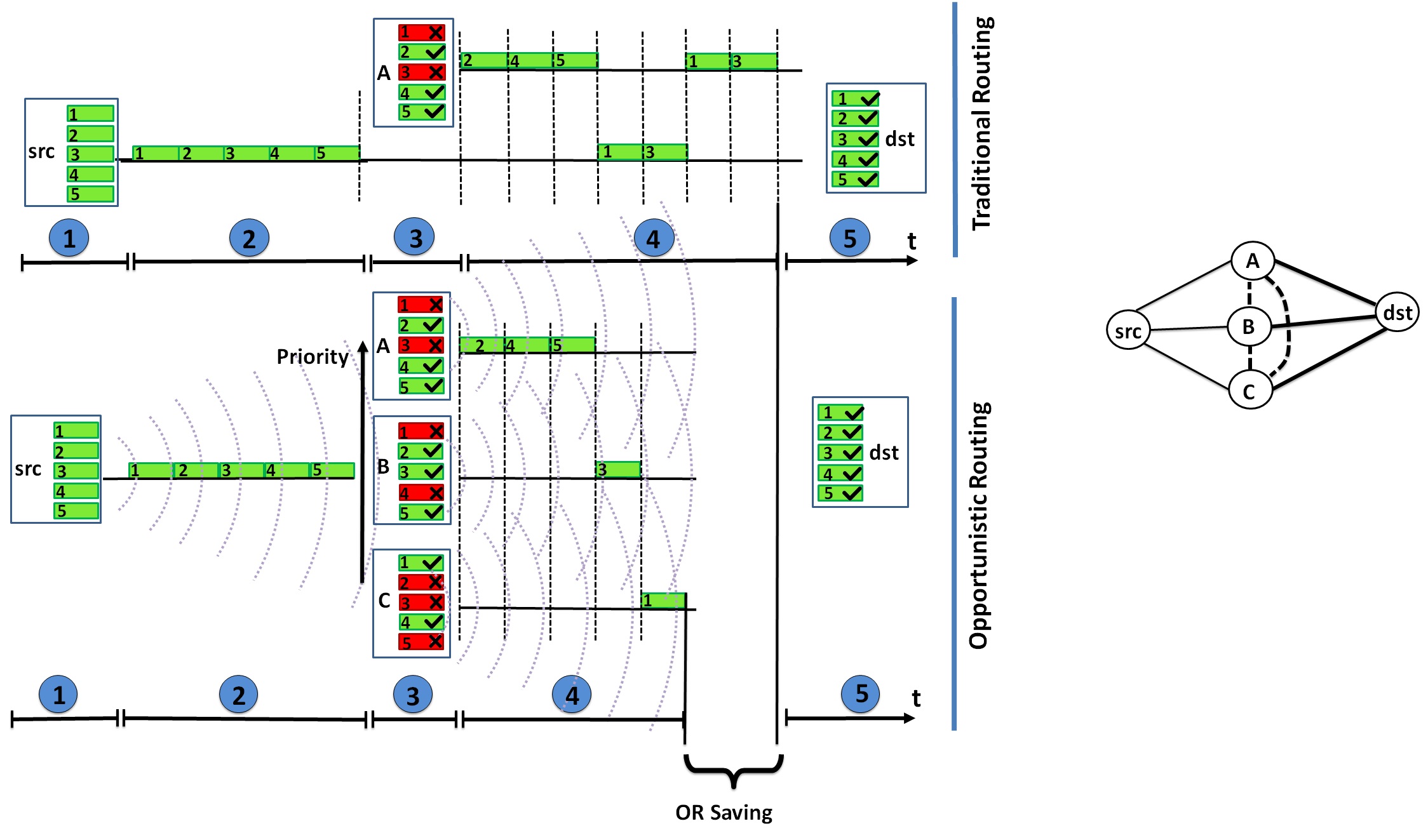}
 	\caption{An example of OR vs. traditional routing.}
 	\label{fig:OR1}
 \end{figure*} 
Figure~\ref{fig:OR2} illustrates an OR scenario associated with a two-hop-apart source-destination pair. Based on the inherent overhearing capability resulting from the broadcasting nature of the wireless medium, all the nodes which can hear a specific node's transmission can be put on a list called the CFS. However, while a node's OR metric generally represents its forwarding merit for a specific source-destination transmission, the eligibility of the node to be placed on the CFS, i.e., its incremental contribution capability, might depend on the status of other potential contributors as well. Orange-color nodes in Fig.~\ref{fig:OR2} demonstrates a related situation wherein they are prevented from consideration for forwarding, as they cannot offer more contribution than what is already provided by $src$. Regarding the incremental contribution, a somewhat similar situation arises in the case of nodes located on correlated wireless paths. By assuming in-advance knowledge of all OR metric values, the members of the CFS (i.e., the light-green color nodes in Fig.~\ref{fig:OR2}) are ordered according to their OR metric values. A subset of the ordered CFS is selected, according to a criterion specific to the employed OR protocol, as the forwarder list (i.e., dark-green color nodes in Fig.~\ref{fig:OR2}). The members of this list are the ones who can incrementally contribute to forwarding the heard transmissions. The nodes on the forwarder list with lower metric values have higher priorities regarding the transmission cooperation. Since the transmissions from the members of the forwarder list are scheduled based on their priorities, not only should they become aware of their forwarder list selection, they must know their priorities as well. It is fair to say that different OR protocols might differ in OR metric calculation, forwarder list selection scheme, and/or transmission scheduling.
\newline Figure~\ref{fig:OR1} is a simplified illustration of the transmission cooperation in OR compared with the traditional routing in a wireless network. As the sample topology in the figure shows, the source and the destination nodes ($src$ and $dst$) are not in direct reach of each other. The potential forwarding nodes $A$, $B$, and $C$, which can hear each other, all have error-free and erroneous links to $dst$ and $src$ (shown by thick and light solid lines), respectively. Moreover, node $C$ has the worst link to $src$ amongst the three, while the other two have equal link qualities. We assume that $src$ intends to deliver five packets to $dst$. 
\\The upper part of Fig.~\ref{fig:OR1} shows the $src-dst$'s packet delivery and channel business using the traditional best-path routing via node $A$. For the sake of simplicity, it is assumed that erroneously-received packets at node $A$ on the first attempt (i.e., the packets 1 and 3) are received successfully on the second attempt.\\
The lower part of Fig.~\ref{fig:OR1} illustrates the delivery of packets using OR. Based on the perfect-second-hop links assumption, nodes $A$, $B$, and $C$ have the same priorities in the context of OR. Thus, just due to implementation, we assign
the highest and the lowest priorities to nodes A and C, respectively. The first time step in the figure shows that the packets queuing at $src$ ready for transmission. The second time step shows the broadcast transmission of the packets by $src$. In the third time step, the different reception patterns at nodes $A$, $B$, and $C$, due to different link qualities, is illustrated. The fourth time step shows the scheduling concept and cooperative forwarding. The cooperative forwarding in OR reduces the channel contention in traditional MACs. 
 It should be noticed that the time-aggregate transmissions by nodes $A$, $B$, and $C$, on the second hop, is the same as the transmissions on the first hop. For simplicity, it is implicitly assumed that the whole batch is collectively received by the intermediate nodes on the first attempt. Finally, the fifth time step demonstrates successful reception of data packets at $dst$.\\
The shorter fourth time step can observe the achievable OR saving, and in turn, shorter total channel business compared to the longer fourth time step in the traditional routing.
   
\subsection{\bf OR History} 
The exploitative manner of OR has proven to be successful in increasing the overall throughput seen by a single flow \cite{P6_2}\cite{P9}. There have been earlier non-OR, yet somewhat similar, attempts of limited and prioritized rebroadcasting according to some criteria (e.g., \cite{P_2020_06_251106} introduces adaptive broadcast for reliable delivery of emergency warning packets in inter-vehicle communication as opposed to simple flooding). However, OR was first introduced through the EXOR protocol \cite{P6_2}\cite{P6}. Typically, OR uses the notion of the data batch instead of a single packet as the primary data unit to facilitate cooperation between forwarding nodes. Since this consideration intrinsically increases the data delivery delay, OR is a viable forwarding method for bandwidth-intensive multimedia applications with elastic delay requirements. A useful numerical example that shows the difference between OR and traditional routing protocols appears in appendix~\ref{Appendix}.
\par The coordination between forwarding nodes, commonly known as scheduling, is a critical and challenging problem which was addressed in pioneer OR proposals \cite{P6_2,P8,P_8_1,P_60_6,P_60_4,P_41_20,P37,P17,p_9_1,p_9_2,P36,P_9_2_2,P30,P_43_41,P_42_21,P_43_44,P_50_4,P_50_5,P_50_6,P7} where the goal was to prevent duplicate packet transmissions. Alternatively, the seminal work of the MAC-independent opportunistic routing (MORE)\cite{P9} protocol suggests the use of NC combined with the concept of distributively implemented transmission credits to eliminate the need for coordination between forwarding nodes. Network coding, best known for its ability to reduce the number of packet retransmissions, can substantially increase network throughput \cite{p_9_2_1}. MORE and its derivatives (e.g., \cite{p_9_666}) combine individual packets of the same batch using random network coding (i.e., intra-flow network coding). 
 COPE \cite{P4} is another successful method of integrating OR with inter-flow network coding. COPE reduces the number of transmissions by combining data packets from different OR flows. Several consequent OR proposals \cite{P_60_2,p_9_3,p_9_4,p_9_5,p_9_6,P_43_45,P_43_46,p_9_7,p_9_8,p_9_9} mixed OR with NC in different ways following the footsteps of MORE and COPE.
 \begin{figure*}[t!]
 	\centering
 	\includegraphics[width= 0.8 \textwidth]{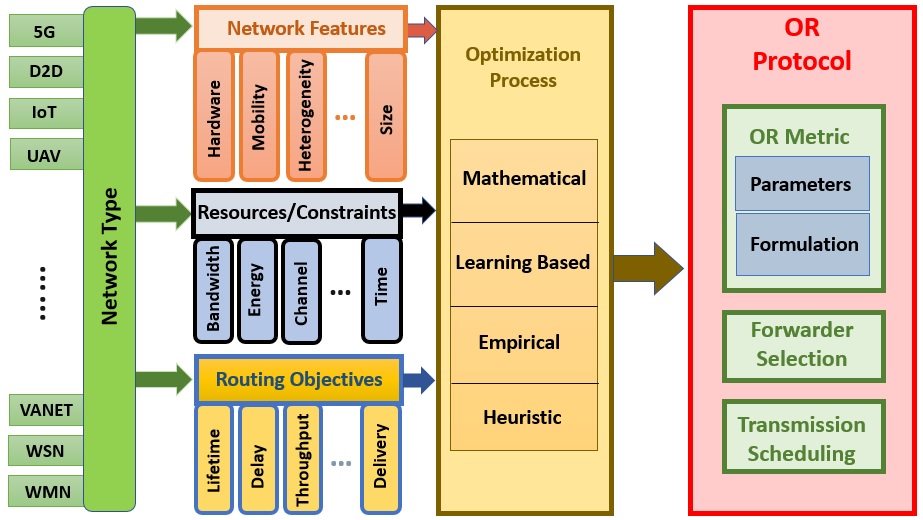}
 	\caption{OR metric design process.}
 	\label{fig:process}
 \end{figure*} 
 \subsection{\bf OR Metric Design  }
 \par The varying characteristics of a wireless channel along with time, distance, environmental conditions, and interference/noise levels make the quality of links between a node and its neighbors unpredictable. Various routing metrics have been introduced to measure and fairly compare various path costs (as a precursor for applying any routing algorithm) \cite{P_40_1}. These metrics can accommodate several parameters, such as mobility, energy consumption, and QoS. As was pointed out earlier, in the context of OR, a metric is typically used to select and prioritize forwarding nodes from a set of candidate nodes and to set their level of cooperation. The performance of an OR protocol strongly depends on the selected forwarding nodes \cite{p_9_10}\cite{p_9_11}. Therefore, the choice of a good and representative routing metric is of crucial importance to the overall network performance.\\
 As Fig.~\ref{fig:process} suggests, at its highest generality, a wireless routing problem starts with a given particular network type, for instance, a WSN. The specification of the network type carries the possible related sets of imposed constraints, available resources, and network features. The network features are usually of characteristic or capability nature, such as trust, mobility, or hardware, which are mostly specific to the network nodes and the wireless transmission channel. The constraints and resources are parameters usually of consumptive nature, such as bandwidth, time, or energy.  Based on the constraints and the resources, a routing optimization problem is formed by selecting one or more performance measures (e.g., network lifetime), as the routing objectives, to be optimized. The solving process of this routing optimization problem results in a particular OR protocol, as well as a new OR metric if needed. The parameters present in the formulation of the resulting OR metric are usually the same as or derived from the items which appear in the features, resources/constraints, and objectives. The formulation of the OR metric is also very much dependent on the adopted optimization method, such as mathematical, learning-based, etc. 
 The DSTT metric is a good representation of such a design process. Presented in the context of an Underwater Acoustic Wireless Sensor Network (UAWSN), the network features the particular water transmission channel and the immobile sensor nodes with limited available energy as the primary constraint. DSTT and its related OR protocol try to heuristically optimize a combination of three OR objectives, the network's lifetime, the Packet Delivery Probability (PDP), and the geographical advancement per hop.\\
 Regarding the OR metric design process, the following considerations  should be taken into account:
 \begin{itemize}
 	\item Not necessarily every OR metric goes through all the design steps of Fig.~\ref{fig:process}.
 	\item Starting with a particular network type, the challenge of designing a good OR metric consists choosing the set of right representative parameters (from the related network features, resources/constraints, and routing objectives), and establishing an appropriate relation (i.e., the OR metric's formulation) between them using an optimization process. The resulting OR metric's formulation is very much dependent on the choice of the optimization process.      	  
 	\item In particular platforms, for instance, CRNs, the OR metric design process might require extra considerations. The stochastic intermittent appearance and mobility (if applicable) of the primary and secondary users, make the availability of network resources such as time and frequency (i.e., spectrum) very fluid. In this situation, the OR metric design, and certainly its employing OR protocol, should address some extra challenges such as the spectrum availability, the interruption time, and the deafness problem (listening on the wrong channel) \cite{P_2019_2_71950}. 
 \end{itemize} 

\begin{table*}[htb]
	\centering
	\caption{ Related survey papers}
	\label{table:tabsurveys}
 
 	\begin{tabular}{| p{3cm}| p{0.5cm} |p{7cm} | p{5cm}| }
		\Xhline{1.5pt}
		{\bf ~~~~~~~~Reference} & \bf Year& {\bf  ~~~~~~~~~~~~~~~~~~~~~~~Contributions}& \bf ~~~~~~~~~~~~~~~~Limitations\\ \Xhline{1.8pt}		
		\multirow{6}{*}{\parbox{2.5cm}{\centering A Survey on Routing Metrics \cite{P_40_1}}}& \multirow{6}{*}{\parbox{1cm}{2007}}&
		\begin{itemize}
			\item The first survey on wireless routing.
			\item Investigates the metrics using the following aspects, the influence factors, the mathematical properties, the design goal, the implementation characteristics, and the evaluation manner.
			\item States the taxonomy for routing metrics based on their mathematical properties. 
			\item Provides a standard summary for each metric.
		\end{itemize} & \begin{itemize}
		\item Covers routing metrics rather than OR metrics.
		\item 	The taxonomy gives minimal insight.
		\item Lack of a comparative study.
		\item	No future direction recommendations.
	\end{itemize} \\ \hline
	\multirow{6}{*}{\parbox{2.5cm}{\centering	Survey on Opportunistic Routing in Multihop \cite{P_2020_06_272108}}} & \multirow{6}{*}{\parbox{1cm}{ 2011}} & \begin{itemize}
		\item Reviews OR protocols. 
		\item Provides an OR-protocols categorization based on the hop-count nature of the underlying metric. 
			
	\end{itemize} & \begin{itemize}
	\item Primarily on OR protocols. 
	\item Considers only a few OR metrics already available back then; now is very outdated.
	
\end{itemize} \\ \hline

\multirow{6}{*}{\parbox{2.5cm}{\centering	Routing Metrics of Cognitive Radio Networks: A Survey \cite{P_2019_2_71950}}} & \multirow{6}{*}{\parbox{1cm}{ 2014}} & \begin{itemize}
	\item Discusses metric design challenges adequately; emphasizes on a cross-layer approach.
	\item Provides the single- /multi–path metric categorization.
	\item  Simulation comparison between three metric subclasses.
	\item Useful future research directions including hybrid metrics.
\end{itemize} & \begin{itemize}
\item Limited to CRNs.
\item Lack of formulation details on OR metrics.
\item An insightful taxonomy at the time of publication; now seems insufficient.
\end{itemize} \\ \hline

\multirow{12}{*}{\parbox{2.5cm}{\centering Opportunistic Routing in Wireless Networks: Models, Algorithms, and Classifications \cite{P25}}} & \multirow{12}{*}{\parbox{1cm}{2015}}& \begin{itemize}
	\item Following a detailed discussion of OR, divides OR issues under three topics, OR metric, candidate (forwarder) selection, and candidate coordination (transmission scheduling).
	\item	Provides a taxonomy for each of these issues: local/end-to-end OR metric, control-based/ data-based candidate coordination, topology-based/geographical-based candidate selection.
	\item Some types of OR protocols are discussed.
	\item Recommends some future directions on each OR issue, as well as on some, then emerging, scenarios involving mobility, multicasting, security, etc. 
\end{itemize} & \begin{itemize}
\item Few OR metrics are mentioned. 
\item	More about OR protocols and candidate coordination/selection algorithms rather than OR metrics.
\item	Incomplete OR metric categorization with limited insight.
\item	Lack of comparative discussions/ simulations.
\item	Insufficient and unstructured OR metric future research directions.
\end{itemize} \\ \hline
\multirow{12}{*}{\parbox{2.5cm}{\centering A Survey on Opportunistic Routing in Wireless Communication Networks \cite{P_43_41}}}& \multirow{12}{*}{\parbox{1cm}{2015}} & \begin{itemize}
	\item Describes in detail the OR building blocks, CFS formation, OR metric calculation, forwarder list selection, and transmission scheduling, as well as their underlying techniques.
	\item	Categorizes OR approaches under five classes: geographic, link-state aware, probabilistic, optimization-based, and cross-layer.
	\item Two easy-to-grasp quick references for OR protocols taxonomy and features.
	\item	Provides some OR future research directions.
\end{itemize}& \begin{itemize}
\item	limited material about OR metrics with insufficient details.
\item The OR protocols taxonomy, though relevant, gives minimal perspective.
\item	The quick reference of OR protocols' taxonomy includes protocols that are not OR at all.
\item The future work recommendations comprise a mix of metric, protocol, and network type topics with no clear classification.
\end{itemize} \\ \hline
\Xhline{1.8pt}
\end{tabular}

\end{table*}

\section{\large\bf OR Metrics: A Survey}

\subsection{\bf Motivation}
It is generally understood that OR is an efficient diversity technique to combat the unreliability of the wireless medium, particularly in multi-hop scenarios. Looking at the related literature, newly emerged OR metrics \cite{P_2019_5_181545, P_2019_2_192147, P_2019_3_280030,P_2019_2,P_2019_12_051127, P_2019_1, P_2018_43, P_2018_41, P_2020_06_230000}, recently-devised opportunistic-based routing protocols \cite{P_2019_3_051447, P_2019_3_051453, P_2019_3_051456, P_2019_3_051506,P_2018_41,P_2019_3_290326, P_2020_07_031859, P_2020_07_031845, P_2020_07_031851, P_2019_10_041116,P_2019_10_151334, P_2020_03_280106, P_2020_07_031830}, extensive use of OR in new environments \cite{P_2019_2_101347}, networks \cite{P_2018_20}\cite{ P_2019_1_279} and applications \cite{P_2019_2_192147}, and also the publication of recent OR surveys \cite{P_2019_2_71937} \cite{P_2019_1_24} all point to the fact that OR and, certainly, OR metrics are still a very alive topic and of huge interest to wireless researchers, justify the timeliness of this survey. \\To justify the need for doing this survey, we proceed by summarizing the most relevant existing surveys in Table~\ref{table:tabsurveys}. Table~\ref{table:tabsurveys} briefly mentions the main contributions and limitations of each past work. \cite{P_40_1} provides a formerly complete treatment of routing metrics rather than OR metrics. However, several of its provided traditional routing metrics have been employed in the context of OR later. \cite{P_2020_06_272108} is a brief survey primarily on OR protocols. It considers
a few OR metrics already introduced back then and classifies the OR protocols according to the hop-count nature of their underlying metrics. The third survey \cite{P_2019_2_71950} gives limited coverage of OR metrics, due to its focus on a particular network type, the CRNs, with insufficient details. The most recent related surveys \cite{P25} \cite{P_43_41} discuss the whole concept of OR and, of course, the OR metrics partially. Consequently, their contributions to OR metrics are lacking in detail and are non-exhaustive. Furthermore, they offer limited insight regarding related taxonomy and future research recommendations. There are other OR-related surveys in the literature (not mentioned in Table~\ref{table:tabsurveys}) with little to no OR-metrics discussions. \cite{P_2019_2_71937} and \cite{P_2019_1_24} explain a few OR metrics as part of their surveys on the joint OR and intra/inter-flow network coding in WMNs. \cite{P_2019_2_71944} is a survey on just the mobility impacts on the OR algorithms, with no reference to OR metrics.
It is worth mentioning an interesting survey on opportunistic routing, which should be distinguished from the conventional OR we are dealing with herein. The survey \cite{P_2020_06_282301} proposes a framework for analyzing the routing algorithms in complex dynamic networks featuring a stochastic nature (e.g., time-variant random topology) due to the random behavior (e.g., mobility) of its nodes. However, the opportunism therein regards taking advantage of the casual, and possibly intermittent, occurrence of the communication opportunities for message advancing, as opposed to the opportunistic collaborative message advancing in OR.
 
\par Interested OR researchers studying the related literature encounter a large number of OR metrics with proprietary notations and presentations (formulations). Cookbook-style OR surveys do not provide a clear insight into the matter and do not satisfactorily fulfill research demands. Considering the limitations, deficiencies, non-exhaustiveness, and outdatedness of the existing OR-metric review studies in the literature, we strongly feel the need for a survey such as the one presented herein.
\subsection{\bf Contributions}
 In order to make this survey self-contained, a tutorial tailored to the structured approach of the survey is required. As such, the tutorial has been organized so as within which: a background on OR from a newly presented perspective, transmission diversity options across different layers of the communication protocol suite is provided; OR is described and compared with traditional routing in terms of performance improvement; a framework for OR metric design process is developed. Given the constantly-changing field of wireless, by adopting a structured, investigative, and comparative approach to OR metrics, we aim at keeping this tutorial survey relevant for as much longer as possible.
\par The main contributions of this tutorial survey are:\\

\begin{tabular}{!{\color{black}\vrule width 1.5pt}  >{\columncolor[gray]{.90}}
		p{0.2cm} | p{7.5cm}}  \hhline{-}\hhline{-} 
	
	\multirow{5}{*}{\centering\key {Tutorial}}   & \vspace*{-0.2cm}\begin{itemize}[leftmargin=0.4cm]
		\item A new representation of OR as a transmission diversity technique at the NET layer.
		\item A new framework for the OR metric design process. 
	\end{itemize} \\ \hhline{-} \hhline{-} \hhline{-}  \hhline{-}

	\multirow{26}{*}{\key{Survey}}   &  \vspace*{-0.2cm}\begin{itemize}[leftmargin=0.4cm]
		\item A new taxonomy of OR metrics based on computation method and scope.
		
		\item Exhaustive coverage of OR metrics in the literature, and providing sufficient details for each metric, rather than merely listing them, which obviates interested OR researchers of frequent and unnecessary referring to the related original papers.
		
		\item  Effort-intensive reformulation of OR metrics into a single unified-notation form (otherwise in differing proprietary notations) for better and easier understanding, without which any comparison between them is almost impossible. 
		
		\item Drawing the evolution of metrics introduced by almost the same authors, and the interrelation between similar metrics introduced by different authors.
		
		\item Self-explanatory, easy-to-grasp, and visual-friendly quick references which can be used independently from the rest of the paper. 
		\item Extensive simulations to compare the representatives of different OR metric categories in terms of network performance.   
		\item A critical look to the fundamental points missed in the classic approach to OR metrics and the investigation of overlooked details in the definition of specific OR metrics. 
	\end{itemize} \\ \hhline{-} \hhline{-} \hhline{-}

	\multirow{6}{*}{\key{Future Works}} & \vspace*{-0.1cm}\begin{itemize}[leftmargin=0.4cm]
		\item Comprehensive future research directions in line with the OR metric design process framework outlined earlier, which makes singling out research opportunities much more straightforward.
	\end{itemize} \\ \hhline{-}  
\end{tabular}
\\ \\ \newline 
 The inclusion of the tutorial part makes the paper self-contained, and beneficial to generalists as well as OR specialists. Moreover, the paper has been carefully organized and equipped with self-explanatory quick references, so that it can be referred to, not only in its entirety but selectively also.
 
\subsection{\bf OR Metrics: A Complete Classification Treatment}\label{sec3}

\begin{figure*}[b!]
	\centering
	\includegraphics[width=0.6\textwidth]{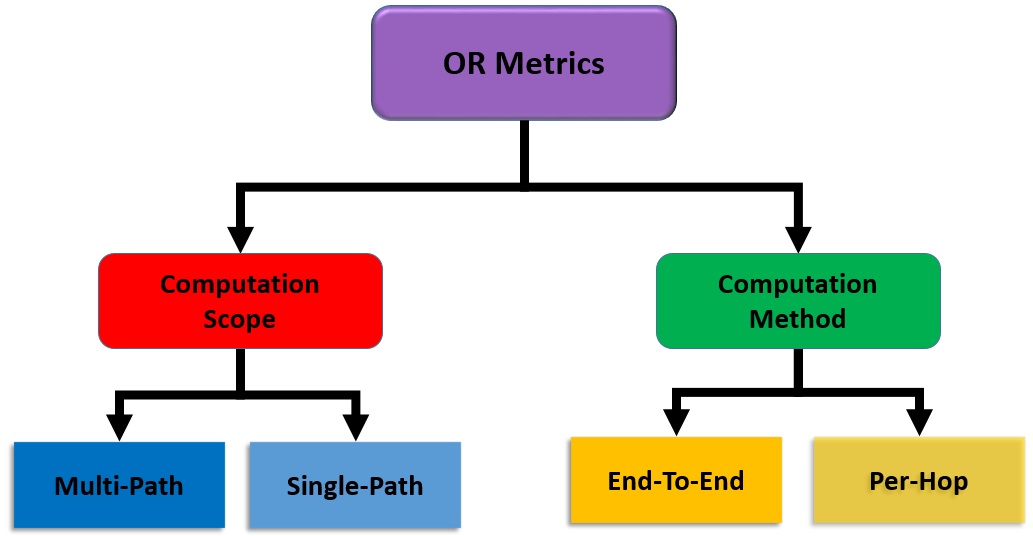}
	\caption{Taxonomy of OR metrics. }\label{fig:fig22_2}
\end{figure*}

\begin{figure*}[!ht]
 	\centering
 	\includegraphics[width=0.8 \textwidth]{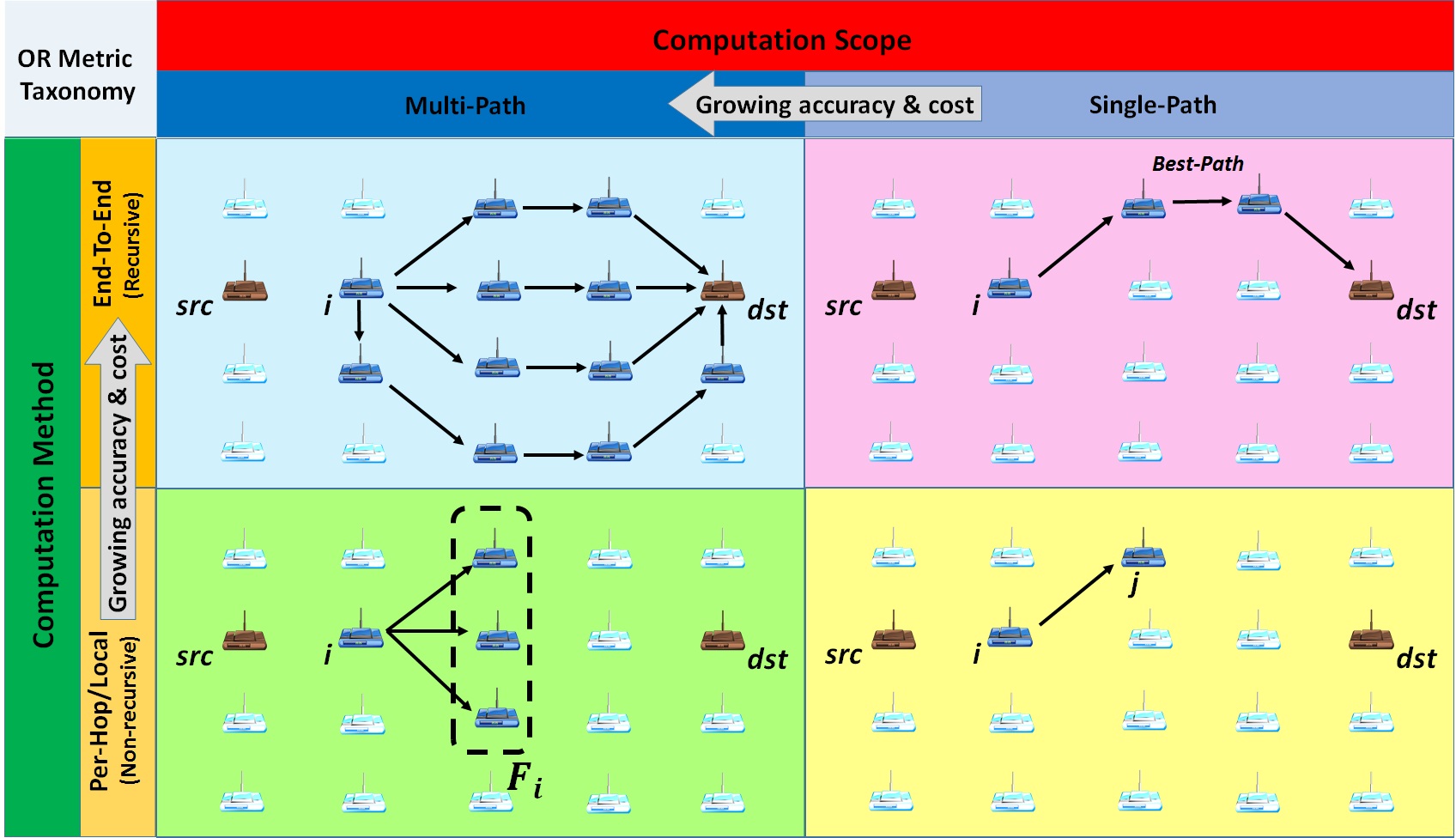}
 	\caption{Visualization of OR metrics classifications according to their computation methods and scopes.}\label{fig:FigMSpath.jpg}
 \end{figure*}

In this section, to provide perspective for interested researchers, we categorize OR metrics based on their two fundamental attributes, the computation scope, and the computation method. Then, in each category, the metrics are presented in chronological order, whereby it is interesting to follow how some later metrics evolved from earlier ones.
\\{\it\bf Classification Perspective:}
Routing metrics are used to prefer one routing solution to another. In traditional wired networks, \cite{P_2019_4_091555} divides routing metrics into two classes as local and global constraint metrics. The former includes the metrics defined over individual links (and possibly their immediate ingress/egress nodes). The latter includes the metrics defined over distinct paths calculated by combining each path's constituent local constraint metrics using some link combination operators  [6]. This classification falls short in providing a perspective in wireless OR networks (our main focus herein) since it does not consider the broadcast nature of wireless networks. In Figure \ref{fig:fig22_2}, we present a taxonomy of OR metrics in terms of their computation methods and computation scopes. \\
The computation scope of an OR metric determines the span of the network involved in calculating its value at a subject node. In a per-hop (local) metric, the span extends to those intermediate nodes (wirelessly reachable immediate neighbors), which can potentially help in relaying a flow towards a  specific final destination. In other words, the information used to calculate the metric concerns only the joining link(s) between a node and a subset of its immediate neighbor(s) which can contribute, more than the node itself, to the delivery at the destination (see the lower part of Fig.~(\ref{fig:FigMSpath.jpg})). However, in an end-to-end OR metric, the metric's computation scope spans until the final destination (see the upper part of Fig.~(\ref{fig:FigMSpath.jpg})). End-to-end OR metrics are calculated from the final destination back to the subject node in a recursive fashion. It is well understood that, compared to their per-hop counterparts, the end-to-end metrics feature higher accuracy, higher computational and information dissemination costs, and higher susceptibility to network changes. Regarding the latter, it should be noted that any topology changes beyond the next hop, excluding the destination (which undoubtedly changes the whole routing problem), will impact the related end-to-end, and not the per-hop, OR metric.    
 
 \par 
 
\begin{figure*}[!ht]
	\centering
	\includegraphics[width=0.8\textwidth]{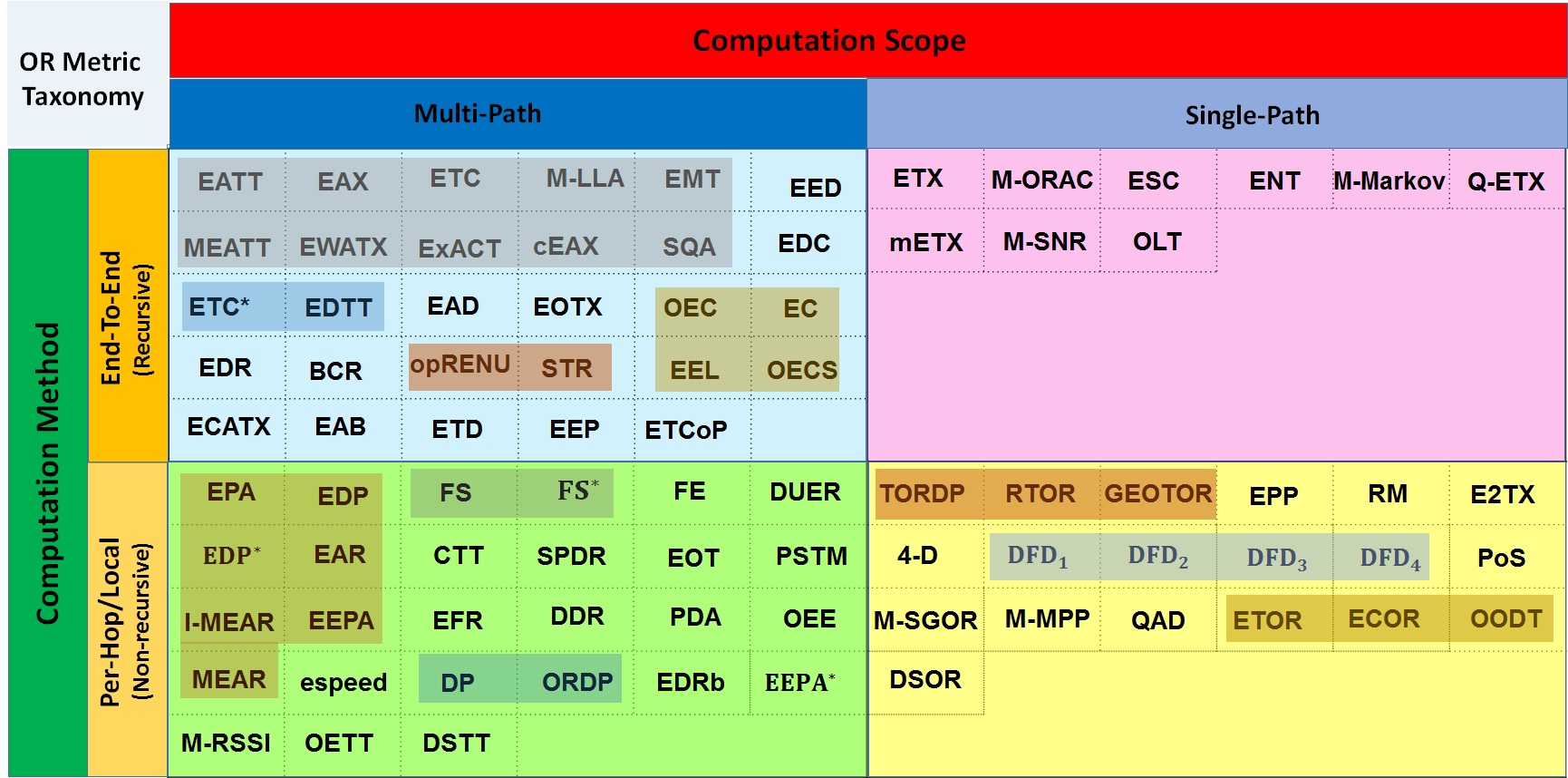}
	\caption{Categorization of OR metrics according to their computation methods and scopes, and also bundled based on high similarity or common root. }\label{fig:fig5}
\end{figure*} 

The computation method of an OR metric determines how many paths contribute to the metric's calculation. In single-path metrics, the value of the metric determines the merit of the best path (as in a unicast transmission) from a subject node to the destination (see the upper-right part of Fig.~(\ref{fig:FigMSpath.jpg})). However, a multi-path metric considers the contribution of many loop-free paths \cite{P17}  by taking advantage of the broadcast nature of wireless transmission (see the left part of Fig.~(\ref{fig:FigMSpath.jpg})). In fact, multi-path OR metrics are opportunistic as is OR itself.\\
While locality and globality concepts in routing protocols \cite{P_2019_5_261405} and in routing metrics are consistent, it is essential to distinguish between the concept of single-/multi-path in routing protocols \cite{P_2019_1_26} and in routing metrics. The single-/multi-path concept in routing protocols regards implementing the transmission policy through a single (multiple) path(s) which has (have) been previously ranked by their embedded routing metrics. The embedded routing metrics can be single-/multi-path regardless of the transmission policy choice of the corresponding routing protocol. 

 \par
Figure \ref{fig:fig5} illustrates how the complete set of OR metrics, to be explained shortly, is divided among the four categories above. Grouping together of OR metrics with high similarities or common roots facilitates metric searching in the literature.\par
One of the confusing facts about the OR metrics is the numerousness of the notations used, making any comparison between them almost impossible. For the sake of presentation consistency, we introduce a unified notation, herein, and reface, as much as possible, the original formulations of the OR metrics accordingly. The unified notation makes extracting similarity/non-similarity patterns between different OR metrics an easier task.\\ Before proceeding with the OR metric explanations, the definition of some basic terms in the unified notation is presented. \\

\begin{figure}[t!]
	\centering
	\includegraphics[width=0.5\textwidth]{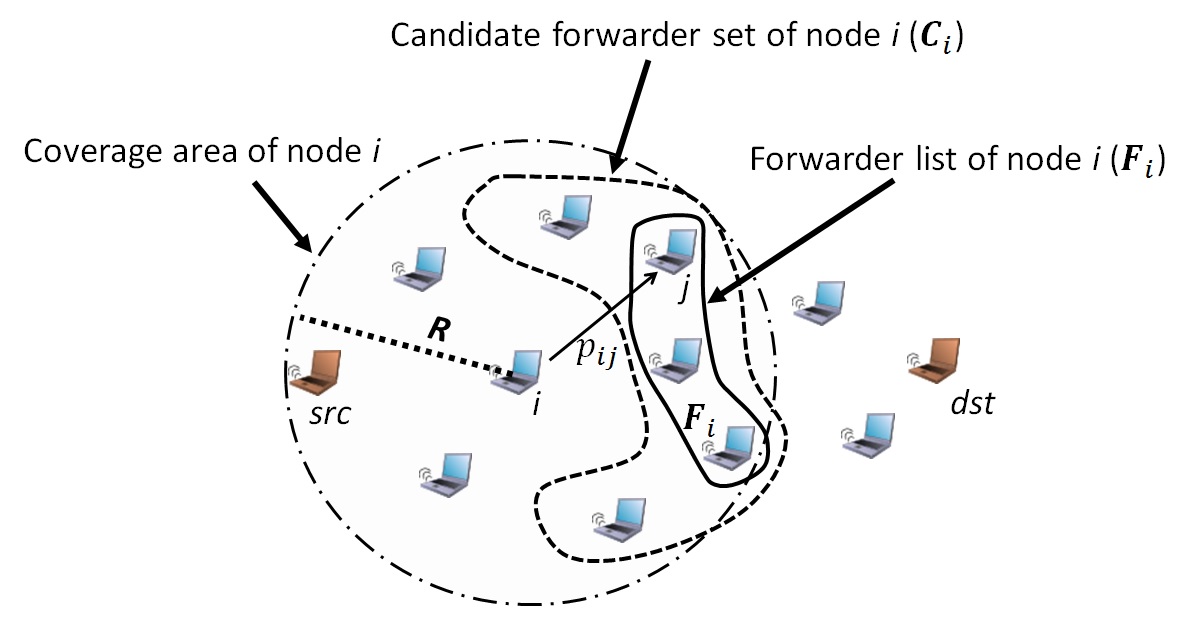}
	\caption{Sender $src$ communicates with destination $dst$. Node $i$ is an active forwarding node on behalf of $src$. $\boldsymbol C_i$ is a list of next-hop candidate nodes, i.e., candidate forwarder set, for node $i$. Node $i$ selects a sublist of these nodes $\boldsymbol F_i$ as the next-hop forwarding nodes, called forwarder list.}\label{fig:fig41}
\end{figure}

$\boldsymbol{ Definitions:}$ 
A simple common topology in OR is illustrated in Fig.~\ref{fig:fig41}. Let $src$, $dst$ and $\boldsymbol{C}_i$ be the source node, destination node and candidate forwarder set for an intermediate node $i$. The nodes in $\boldsymbol{C}_i$ can potentially help node $i$ in delivering data packets from $src$ to $dst$ (since they are wirelessly reachable by $i$ and have better relevant metric values than that of $src$). We assume that the members of $\boldsymbol{C}_i$ are sorted according to their respective priorities defined by their corresponding OR metrics (in which lower index shows a higher priority). The transmission by a higher-priority node prevents lower-priority nodes from transmitting. In OR, a subset of $\boldsymbol{C}_i$ is selected as the set of participating members to realize an opportunistic route between $src$ and $dst$ at level $i$. We denote this subset by $\boldsymbol{F}_i$, i.e., node $i$'s forwarder list. The goal of each OR protocol is to find $\boldsymbol{F}_i$ at each routing level $i$ to maximize the overall throughput of the selected opportunistic routes.\par 
In the definition of OR metrics, three important delivery probabilities appear frequently, which will be explained shortly.\par 
The link quality from node $i$ to node $j$ is commonly measured by the packet delivery probability, $p_{ij}$. For example, $p_{ij}=0.7$ means that on the average, 7 of 10 data packets transmitted by node $i$ are received by node $j$. This parameter is key to the definition of almost all OR metrics. Therefore, the parameter is detailed in section \ref{Sec:Scrutiny}. While not always true, in the literature, links are usually considered to be symmetrical, which means that $p_{ij}=p_{ji}$. Many other important parameters are also derived from this parameter.

$P_{\boldsymbol F_i}$ is defined as the probability that at least one member of ${\boldsymbol F_i}$ successfully receives the packet transmitted by node $i$:
\begin{equation}
	\label{eq01}
	P_{\boldsymbol F_i}= 1 - \displaystyle\prod_{j \in \boldsymbol{F}_i} {(1- p_{ij})}~~~.
\end{equation}
$P(i,k)$ is the probability that the $k$-th candidate (or forwarding) node receives the transmitted packet from node $i$ while all other higher-priority nodes fail to receive the packet:
\begin{equation}
	\label{eq02}
	P(i,k)= p_{ik} \displaystyle\prod^{k-1}_{j=1} {(1- p_{ij})}~~~.
\end{equation}
In eq.~(\ref{eq02}), the nodes in the forwarder list are assumed to be descendingly indexed in terms of priority. $P_{\boldsymbol F_i}$ is the opportunistic reception probability when the sender is $i$ and the receivers are members of $\boldsymbol F_i$. $P(i,k)$ is the opportunistic reception probability for the sender $i$ and a single opportunistic receiver (i.e., $k$).\\ For ease of reference, Table \ref{table:tab11} lists the most common notations appearing in the definition of OR metrics along with their descriptions.
  
\begin{table*}[htbp]
	\centering
	\caption{Unified notations}
	\label{table:tab11}

	\resizebox{1.0 \textwidth}{!}{
	\begin{tabular}{ | c| c |c| }
             	\Xhline{1.5pt}
             {\bf Notation} & {\bf  Description}& \bf Variants/Note\\ \Xhline{1.8pt}
              $i,j,h/src/dst$  & General node/Source node/Destination node & \\ \hline
              	$\boldsymbol{C}_i$ / ${\boldsymbol{F}_i}$& Candidate forwarder set/Forwarder list for node $i$& \\ \hline 
              $c$ & Cost& $c_{com}$, $c_{rc}$, $c_{ij}$ \\ \hline
               $B$ & Bandwidth& $B_{cur_i}$, $B_{new}$, $B_{tot_i}$, $B_{loc}$ \\ \hline
                $E$ & Energy&    $E_{new}$, $E_{tot_i}$, $E_{res}$, $E_b$, $E_{prb}$, $E_{wak}$, $E_{cons}$\\ \hline
                $R$ &Communication range & \\ \hline

                  $d$ & Geographical advancement& $d_{ij}\equiv d(i,j)$ = $d(i,dst)-d(j,dst)$\\ \hline
                  $Trust/ST$ & Trust/Social Tie&  $Trust_{ij}/ ST_{ij}$ \\ \hline
                   $T$ / $t$ & Total delay or time/Partial delay or time& $T_{stl}$, $T_{TX}$ , $t_i$, $t_{stl_i}$, $t_r$, $T_{hop}$, $T_{Crd}$, $T_{MAC}$, $T_{Crd}$   \\ \hline
                    $L$ & Packet size& $L$ , $L_i$ \\ \hline
                $r$ & Transmission rate& $r$, $r^*$, $r_i$, $r_k$\\ \hline
                 $p_{ij}$ & Packet deliver probability from node $i$ to node $j$ & $p_{i,j}$\\ \hline
                 	$P_{\boldsymbol F_i}$	& Probability that at least one member of $\boldsymbol F_i$ receives node $i$'s packet & $P_{\boldsymbol F_i}= 1 - \displaystyle\prod_{j \in \boldsymbol{F}_i} {(1- p_{ij})}$ \\ \hline
                 	$ P(i,k)$  &Probability that the $k$-th candidate is the first in $\boldsymbol F_i$ to receive  node $i$'s packet& $P(i,k)= p_{ik} \displaystyle\prod^{k-1}_{j=1} {(1- p_{ij})}$ \\ \Xhline{1.5pt}
	\end{tabular}
}
\end{table*}

\par
Figure~\ref{fig:fig5} illustrates the four classes of different computation scope/computation method combinations.  
\subsubsection{\bf Single-Path/End-to-End OR Metrics:  }  
In this class, metrics are calculated recursively along a single path (best path) from the destination backward. The following relevant metrics are presented in chronological order.
\begin{itemize}
\item {\it Expected Transmission Count (ETX) \cite{P5}: }
The ETX of a link is defined as the average number of packet transmissions required to deliver a packet over the link successfully. ETX is calculated using the forward and reverse delivery ratios of the link. If the forward and reverse delivery ratios of a link are $p_{ij}$ and $p_{ji}$, respectively, the ETX is:
\begin{equation}
\label{eq111}
ETX(i,j)=\frac{1}{p_{ij} . p_{ji}}~~~.
\end{equation}
The forward delivery ratio, $p_{ij}$, accounts for the successful delivery of the subject packet, and the reverse delivery ratio, $p_{ji}$, accounts for the successful delivery of the corresponding acknowledgment. The ETX of a route is calculated as the sum of its links' ETXs. A sample proposal for distributing ETX information among network nodes is to use a designated node. Suppose that each node has calculated the ETX metric for all the links between itself and its immediate neighbors. Then, the node sends this information throughout the network. After receiving all the one-hop ETXs, the designated node can calculate the ETX of the path between any two nodes (any potential source-destination pair) and distribute them across the network upon request. \\
Clearly, the ETX metric is entirely different from the hop count metric, which does not account for link quality. 

\item {\it {modified ETX (mETX), Effective Number of Transmissions (ENT) \cite{P_2017_5}:} }
As discussed previously, the ETX implicitly presumes i.i.d. bit errors in a single packet transmission and also in successive packet transmissions. mETX relaxes this assumption by considering the variability of the channel at the packet timescale and its probable subsequent retransmissions. mETX is defined as: 
\begin{equation}
\label{eq2017_1}
mETX = exp (E[\textstyle\sum] + \frac{1}{2} Var\small[\sum])~~~,
\end{equation} 
where $\textstyle\sum$ is the logarithm of the required number of transmissions over a link, the first term in the exponent on the right represents the average level of channel bit error probability over long periods of time, and the second term in the exponent accounts for the packet-to-packet variability of the bit error probability. From eq.~(\ref{eq2017_1}) it is obvious that $mETX \ge ETX$. Similar to the ETX metric, mETX of a path is equal to the summation of its constituent links' mETXs.\\
ENT attempts to improve the aggregate throughput by limiting the number of link-layer retransmissions and,  if needed, allowing higher layers to handle correct receptions. ENT is defined as:
\begin{equation}
\label{eq2017_2}
ENT = exp (E[\textstyle\sum] + 2 \delta Var\textstyle[\sum])~~~.
\end{equation} 
 ENT is very similar to mETX with the difference of an extra degree-of-freedom $\delta$, which is related to the maximum number of allowed link-layer retransmissions. 
 
\item {\it Markovian metric (M-Markov)\footnote{The abbreviation is from the authors of this survey for better reference.} \cite{P_2019_2_181359}:}
Conventional routing metrics generally consider independent links. The M-Markov metric \cite{P_2019_2_181359}, which is more of a concept rather than a concrete metric, introduces some context information into the link cost modeling. This metric calculates the cost of a path connecting node $i$ to $dst$ as the summation  of the constituent links' costs, each conditioned on its previous links cost:
\begin{equation}
	c(i,dst) = c(i,i+1) + \sum_{j=i+1}^{dst-1}c (j \rightarrow j+1 | j-1 \rightarrow j ).
\end{equation}
Considering independent links makes the conventional routing metrics as special cases of the M-Markov metric.

\item{\it Opportunistic Link Transmission (OLT) \cite{P_60_12}:}
OLT defines a multi-hop and single-path metric in slotted Cognitive Radio (CR) networks by considering transmission, queuing, and access and by excluding processing and propagation delays: 
\begin{equation}
OLT^{(N)} = \sum_{i=0}^{N}\frac{1}{\mu_i - \lambda_i}~~~,
\end{equation}
where $\mu_i$ and $ \lambda_i$ denote the processing and arrival rates at node $i$. Thereafter, node $i$'s merit for forwarding a packet is measured by:  
\begin{equation}
\resizebox{0.5\hsize}{!}{$%
	\begin{aligned}
	\min_{1 \leq j \leq M_i} &\{ OLT,(OLT^{(2)}_1, ...,OLT^{(2)}_{M^{(2)}_i}),\\ &...,(OLT^{(N)}_1 ,..., OLT^{(N)}_{M^{(N)}_i}) \}~~,%
	\end{aligned}
	$ %
}
\end{equation}
where $N$ is equal to the maximum number of hops over all opportunistic paths from node $i$ to the destination, and $M_i^{(N)}$ denotes the number of $N$-hop opportunistic paths from node $i$ to the destination.

\item {\it Energy Shortage Cost (ESC) \cite{P_61_21}:}
ESC concerns mobile WSNs, where sensing nodes occasionally meet sinks. If sinks are encountered frequently, nodes will transmit their data directly to them. Otherwise, the transmissions are done through other forwarding nodes with better links to the sinks. ESC is defined by:   
\begin{equation}
ESC(i,sink_i) = \frac{E_{cons}(i,sink_i)}{E_{res_i}}~~~,
\end{equation}
where $E_{cons}(i,sink_i)$ represents the energy required by node $i$ to deliver a packet (message) to its corresponding sink, $sink_i$. $E_{cons}(i,sink_i)$ is inversely proportional to the delivery probability towards $sink_i$, and $E_{res_i}$ is the remaining energy of node $i$. Although the scope of the metric appears to be end-to-end, the details of the computation method are unmentioned and unclear in the paper.

\item {\it Metric of the Opportunistic Routing Admission Control protocol (M-ORAC)\footnote{The abbreviation is from the authors of this survey for better reference.} \cite{P_40_19}:}

M-ORAC is an admission-based OR scheme that prunes candidates that are not able to fulfill the bandwidth and energy requirements of a new flow. A forwarding node $i$ admits a new flow that satisfies the following conditions: 
\begin{equation}
\label{eq36}
\left\{
\begin{array}{ll}
\rho_{_{B_{i}}} = \frac{B_{cur_i} + B_{new}}{B_{tot_i}} < 1~~~,\\
\rho_{_{E_{i}}} = \frac{E_{res_i} + E_{new}}{E_{tot_i}} < 1
\end{array}
\right.
\end{equation}

where  $\rho_{_{B_{i}}}$/$\rho_{_{E_{i}}}$ is the bandwidth/energy measure index, $B_{cur_i}$/$B_{new}$ shows the bandwidth requirement of the accepted flows/new flow, and $E_{res_i}$/$E_{new}$ denotes rough estimation of the energy consumption of the accepted flows/new flow. Finally, $B_{tot_i}$ and $E_{tot_i}$ represent the bandwidth of node $i$ to service incoming traffic flows and total energy budget of node $i$, respectively. After finding eligible candidates, a new metric, called M-ORAC, is used for prioritization. M-ORAC is defined as follows:
\begin{equation}
\label{eq361}
\begin{aligned}
\text{\it M-ORAC}(i,dst) =& \alpha ~ \rho_{_{B_i}}+ \beta ~ \rho_{_{E_i}}+ \gamma ~ ETX(i,dst) \\ & + \varphi  ~ \frac{d(src,dst)- d(i,dst)}{d(src,dst)}~~.
\end{aligned}
\end{equation}

Clearly, M-ORAC makes a trade-off between the congestion control, routing quality, and energy consumption by applying the corresponding coefficients $\alpha$, $\beta, \gamma$, and  $\varphi$.

\item {\it  M-SNR \footnote{The abbreviation is from the authors of this survey for better reference.} \cite{P_2017_3}:}
The cost metric $c(i,dst)$ represents the SNR-related cost from node $i$ to the destination $dst$. For node $i$, this cost is updated according to the information in the header of the received update messages flooded over the network as follows:
\begin{equation}
\resizebox{0.95\hsize}{!}{$%
\label{eq2017_10}
c(i,dst)= \left\{
\begin{array}{lll}
c(i+1,dst) + 1 ~~~~~~~~~~~~~~~~~~~~ ;~ SNR_{i+1}^{dst} > SNR_{THRSLD}\\
c(i+1,dst)+ ~~~~\\
 \lceil SNR_{THRSLD}+1-SNR_{i+1}^{dst}\rceil~~; ~~~~~ o.w.
\end{array}
\right.
$%
}
\end{equation}
where $SNR_{i+1}^{dst}$ and $SNR_{THRSLD}$ are the SNR of the signal from $dst$ at the node right after node $i$, i.e., node $i+1$, and a threshold representing the maximum observed SNR in practical Mobile Adhoc NETworks (MANET) environments, respectively. By this cost-update policy, higher packet forwarding probability is given to a node with high-SNR links to the destination.\\
The definition of $SNR_{THRSLD}$ in \cite{P_2017_3} requires clarification since it seemingly makes  the first condition in eq.~(\ref{eq2017_10}) irrelevant.

\item {\it Q-ETX \cite{P_2018_24}:}
This metric considers the backlog transmission queue of a relay node as well as its ETX and is defined as:

\begin{equation}
\label{eq2018_3}
\text{\it Q-ETX}(i,dst) = \alpha Q(i) + (1-\alpha) ETX(i,dst) ~~~,
\end{equation} 
where $Q(i)$ denotes the transmission queue length at node $i$. Since this metric is defined in the context of slotted CRNs, both $Q$ and ETX are calculated on the condition of channel availability and for the limited duration of a single time slot.

\end{itemize}
{\it Note}: The single-path/end-to-end OR metrics are not, by their very nature, opportunistic, since they do not benefit from the broadcast nature of the wireless medium. These metrics can be equally embedded in traditional and opportunistic routing protocols; for instance, ETX was initially introduced in the pre-OR era. However, their extensive use in OR protocols labels them as important OR metrics.

\subsubsection{\bf Multi-Path/End-to-End OR Metrics:  } 

As these are end-to-end metrics, they are calculated recursively along multiple paths from the destination backward. We intuitively expect this class of metrics, which employs more wireless path opportunities, to be more inclusive than its single-path counterpart. The relevant metrics are described in chronological order.
\begin{itemize}

\item {\it Expected Any-path transmissions (EAX)\cite{P13,P14}:}
 ETX ignores the broadcast nature of wireless media, and can only predict the behavior of the individual links that belong to a particular opportunistic path. However, OR aims to benefit from the broadcast nature of WNs to enhance network throughput. EAX was introduced to achieve greater consistency between OR and its employed cost metric.\\ 
As a pioneering opportunistic metric, EAX computes the transmission cost of each opportunistic path recursively to find the best opportunistic path. The $EAX(i,dst)$ metric is defined as the expected number of transmissions that $i$ should make for the successful delivery of the transmitted data packet at $dst$:  
\begin{equation}
\label{eq1}
EAX(i,dst)=\frac{1+\displaystyle\sum_{j \in \boldsymbol F_{i}}P(i,j) EAX(j,dst) }{P_{\boldsymbol F_{i}}}~~.
\end{equation}

Note that in eq.~(\ref{eq1}), the numerator is the expected number of transmission attempts that $i$ and the members of $\boldsymbol F_{i}$ must make for the successful delivery of the data packet at $dst$ (\enquote*{1} corresponds to the $i$'s attempt and the summation corresponds to the number of attempts made by the members of $\boldsymbol F_{i}$). The denominator is the probability that at least one member of $\boldsymbol F_{i}$ receives the packet transmitted by $i$. $P(i,j)EAX(j,dst)$ is the EAX cost of node $j$ given that it acts as a forwarding node. EAX is a multiple-hop, symmetric, and multi-path metric that considered the anypath forwarding scheme for the first time. 
Among the many works that have adopted EAX as their main metric are \cite{P17},\cite{P15,p_9_11}, and \cite{P18}.

 \item {\it EOTX\cite{P_60_25}:}
EOTX, the multi-path generalization of ETX, computes the minimum expected number of transmissions needed to deliver a packet from a sender to the destination using all the available paths:  
\begin{equation}
	\resizebox{0.7\hsize}{!}{$%
\label{eq189}
EOTX(i,dst)=\left\{
\begin{array}{@{}lll@{}}
0 & ; &\text{\it i=dst} \\
1+\sum_{j} p_{\boldsymbol F_i} \operatorname*{min}_{j \in \boldsymbol{F}_i}EOTX(j,dst)& ;&\text{\it o.w.}
\end{array}\right.	$ %
}~~,
\end{equation}
where $p_{\boldsymbol F_i}$ denotes the probability that all nodes in set $\boldsymbol F_i$ receive a packet broadcast by node $i$ (therefore, it is different from $P_{\boldsymbol F_i}$, which represents the probability of at least one node in $\boldsymbol F_i$ receiving the packet). This metric resembles $z(i,dst)$ in eq.~(\ref{eq40}), which is used for the credit calculation of the forwarding nodes in the MORE routing protocol \cite{P9}.

\item {\it Expected Any-path Communication Time (ExACT)\cite{P20}:}
The network nodes might not all use the same bit rate when forwarding an opportunistic flow due to different SNR values, inter-node distances, and interference levels. Using the same bit rate on all the links results in overutilization (under-utilization) for long (short) links. A high bit rate would disconnect the network since some links may not work properly due to the short radio range. By contrast, low bit rates will increase the number of network links and spatial diversity. ExACT is an extended version of EAX with the bit rate selection feature. ExACT is a recursively defined metric that minimizes the total transmission time at the specified bit rate $r$. ExACT starts from $dst$ and finds the minimum ExACT value for the previous 1-hop candidates set and determines their ideal transmission rate, $r^*$. This procedure is performed recursively until reaching node $i$.

\begin{equation}
\label{eq8}
\resizebox{0.7\hsize}{!}{$%
	ExACT(i,dst,r)=\frac{{\displaystyle\frac {1}{r}}+
		\displaystyle\sum_{ j \in \boldsymbol F_{i} } P^{r}(i,j) ExACT(j,dst,r_j^*)  }       {\displaystyle {P^{r}_{\boldsymbol F_i}}} %
	$ %
}~~.
\end{equation}

In eq.~(\ref{eq8}), $\displaystyle\frac{1}{r}$ represents a single packet's transmission time. The correct formulation should use $\displaystyle\frac{L}{r}$ to account for the packet length (for instance, see eq.~(\ref{eq112}) for EATT). in eq.~(\ref{eq8}), the use of $r^*$ instead of $r$ in $ExACT(j,dst,r_j^*)$ emphasizes  the different bit rates of the links.

\item {\it Expected Medium Time (EMT):}
EMT \cite{P33} is another recursive opportunistic metric that shares several similarities with other opportunistic metrics such as EAX \cite{P13}, EATT \cite{P19} and ExACT \cite{P20}. EMT models the expected transmission time from node $i$ to $dst$ as follows:

\begin{equation}
\label{eq25}
EMT(i,dst)=\frac{ \displaystyle\sum_{{j \in \boldsymbol{F}_i}} P(i,j)EMT(j,dst) }       {P_{\boldsymbol F_i}}~~~.
\end{equation}

The lack of an additive term like the one in similar metrics above, which accounts for the source's transmission, needs further verification.
\item {\it Expected Anypath Transmission Time (EATT)\cite{P19,P119}:}
EATT, which is very similar to ExACT, selects an appropriate flow rate to simultaneously maximize network throughput and minimize transmission time. EATT is defined as follows:
\begin{equation}
\resizebox{0.7\hsize}{!}{$%
\label{eq112}
EATT(i,dst,r)=\frac{{\displaystyle\frac {L}{r}}+\displaystyle\sum_{j \in \boldsymbol F_{i}} P^{r}(i,j) EATT(j,dst,r) } {\displaystyle {P^{r}_{\boldsymbol F_i}}}~~.%
$%
}%
\end{equation}
where $L$ is the packet length, and $r$ is the bit rate. $r$ is selected to minimize $EATT(i,dst,r)$.

\item {\it Opportunistic routing Residual Expected Network Utilities (OpRENU):}
The OpRENU \cite{P37,P37_2}\cite{P25} metric is designed for a utility-based routing paradigm where a positive value rewards the successful routing of a data packet after reducing the transmission cost. OpRENU is defined recursively with the same features of STR except for the addition of the transmission cost $c$:
\begin{equation}
\resizebox{0.7\hsize}{!}{$
\label{eq31}
OpRENU(i,dst)={{\displaystyle\sum_{j \in \boldsymbol{F}_i}{P(i,j)OpRENU(j,dst)}}} - c~~.
$%
}
\end{equation}

\item {\it Successful Transmission Rate (STR)\cite{P22}:}
STR is another recursively defined metric that computes the successful transmission rate for every candidate node. The calculation of STR for an immediate neighbor of $dst$ is straightforward: it equals the available transmission rate of the link between that neighbor and $dst$. For other candidate nodes, the following formula is used: 
\begin{equation}
\resizebox{0.7\hsize}{!}{$
\label{eq10}
STR(i,dst)={{p_{i1}}}{STR(1,dst)+{\displaystyle\sum_{j \in \boldsymbol{F}_i}{P{(i,j)}}}{STR(j,dst)}}.
$%
}
\end{equation}
The first term accounts for the transmission contribution
of the highest-priority candidate, and the second term accounts for the other candidates' contributions.

\item {\it Opportunistic End-to-end Cost (OEC) \cite{P_50_17,P_50_18}:}
OEC reflects the energy cost of using an opportunistic path and is defined as follows:
\begin{equation}
\resizebox{0.65\hsize}{!}{$
\label{eq375}
\begin{aligned}
OEC(i,dst)&=\displaystyle\frac{\frac{E_{cons-TX}}{E_{res_{i}}} + \sum_{j \in \boldsymbol F_{i}} \frac{E_{cons-RX}}{E_{{res}_j}} } {P_{\boldsymbol F_{i}}} \\ &+ \displaystyle\frac{ \sum_{j \in \boldsymbol F_{i}} P(i,j)OEC(j,dst)}{P_{\boldsymbol F_{i}}}~~,
\end{aligned}
$%
}
\end{equation}
where $E_{res}$ is the remaining energy level of a node, and $E_{cons-TX}/E_{cons-RX}$ is the consumed transmission/reception energy of a wireless node. Therefore, the first term in the numerator denotes the energy efficiency of the sender when sending a packet to $\boldsymbol F_{i}$. The second term in the numerator represents the energy efficiency of the members of $\boldsymbol F_{i}$ when receiving the transmitted packet.

\item {\it Expected number of Duty-Cycled wake-ups (EDC) \cite{P21,P211}:}
The main goal of EDC is to minimize the node wake-up
 interval in WSNs. The lower the total duration of the wake-up intervals of a node is, the lower its energy consumption. Clearly, node $i$ has two main concerns when selecting $\boldsymbol{F}_i$ members. First, node $i$ should deliver the packet to at least one member of $\boldsymbol{F}_i$, which implies that the cardinality of $\boldsymbol{F}_i$ (i.e., $|\boldsymbol{F}_i |$) should be increased. Second, the overall progress of the packet toward the destination ($dst$) should be maximized. If the selected $\boldsymbol{F}_i$ members are not optimal, the overall progress is deteriorated. In other words, from an overall progress perspective, high-quality forwarding members are favored. Hence, EDC should restrict the number of forwarding members. EDC attempts to make a fair trade-off between these two contradicting requirements as follows:

\begin{equation}
\label{eq9}
EDC(i,dst)={\frac{1}{\displaystyle\sum_{j \in \boldsymbol{F}_i}{p_{ij}}}}+{\frac{\displaystyle\sum_{j \in \boldsymbol{F}_i}{p_{ij}} . EDC(j,dst)}{\displaystyle\sum_{j \in \boldsymbol{F}_i}{p_{ij}}}}~~~.
\end{equation}
The first part of eq.~(\ref{eq9}) is a rough estimation of the number of transmission attempts by node $i$ to deliver a packet to a member of $\boldsymbol{F}_i$. In this formulation, the order of $\boldsymbol{F}_i$ members and their wake-up status are ignored. In the second part of the equation, the required wake-up duration of each $\boldsymbol{F}_i$ member, in time interval $T$, for the successful delivery of the packet at $dst$ is roughly estimated by computing the expected EDC values over all members of $\boldsymbol{F}_i$. The first part of the equation improves with $|\boldsymbol{F}_i |$ while the second part improves with the quality of the selected forwarding nodes. Minimizing the EDC metric reduces energy consumption. Note that each node will resend an undelivered data packet until a forwarding node has acknowledged it.

\item {\it Expected Transmission Cost (ETC), Expected Available Bandwidth (EAB), Bandwidth-Cost Ratio (BCR):}
BCR \cite{P36,P36_2_2} is proposed to estimate the available bandwidth while considering the expected transmission cost (ETC) for an opportunistic flow. ETC is defined in the same way as EAX:
\begin{equation}
\label{eq28}
ETC(i,dst)=\frac{1+\displaystyle\sum_{{j \in \boldsymbol{C}_i}} P(i,j)ETC(j,dst) }  {P_{\boldsymbol C_i}}~~~.
\end{equation}

ETC considers the cost from node $i$ to $\boldsymbol{C}_i$ and the cost from $\boldsymbol{C}_i$ to $dst$ as usual. Then, the expected available bandwidth (EAB) for node $i$ is defined as:
\begin{equation}
\label{eq29}
EAB(i)=\min \{ B_{loc}(i) , B_{opp}(i)\}~~~,
\end{equation}

where $B_{loc}(i)$ and $B_{opp}(i)$ are the local and opportunistic available bandwidths of node $i$, respectively. The channel capacity, channel idle time, packet length, average backoff time, SIFS, DIFS and the time required for sending an ACK packet must be considered to obtain an accurate estimation of $B_{loc} (i)$. $B_{opp}(i)$ is estimated in the same manner as ETC:  

\begin{equation}
\label{eq30}
B_{opp}(i)=\frac{\displaystyle\sum_{{j \in \boldsymbol{C}_i}} P(i,j)EAB(j) }  {P_{\boldsymbol C_i}}~~~.
\end{equation}

$B_{loc}(i)$ is not required in the definition of $B_{opp}(i)$ since we have considered its effect in eq.~(\ref{eq29}). Finally, the BCR metric is defined as the ratio of ETC to EAB.

\item {\it Correlation-aware EAX (cEAX):}
Link correlation is caused by (i) cross-network interference under a shared medium and (ii) correlated fading introduced by highly dynamic environments. This phenomenon leads to correlated packet receptions among receivers that are closely located and might result in suboptimal forwarder selection, thereby reducing network performance. Wang et al. \cite{P_50_22,P_50_23,P_50_24,P_50_25,P_50_26} studied the effect of link correlation on the performance of OR and proposed the correlation-aware EAX (cEAX) metric, which is defined in a same way as  EAX:
\begin{equation}
\label {eq385}
\resizebox{0.7\hsize}{!}{$
cEAX(i,dst )= \frac{\displaystyle 1+ \displaystyle\sum_{j \in \boldsymbol F_{i}} JPRP(i,j)cEAX(j,dst)}{\displaystyle 1-JPLS(\boldsymbol F_{i})}~~,
$%
}
\end{equation}

where $JPRP$ and $JPLS$ stand for the joint packet reception and joint packet loss probabilities, respectively. $JPRP(i,j)$ represents the probability that $j$ receives the packet transmitted by node $i$ when higher-priority forwarding nodes fail to do so in the presence of link correlations. $(1-JPLS(\boldsymbol F_{i} ))$ means that at least one member of $\boldsymbol F_{i}$ has successfully received the transmitted packet.

\item {\it Multi-channel Expected Anypath Transmission Time (MEATT):}
MEATT \cite{P38} is an opportunistic metric that extends the ETT metric for multi-channel multi-radio mesh networks in the same manner as EATT. MEATT is designed so that the proper channel is implicitly selected by the current forwarder when estimating the expected anypath transmission time. MEATT attempts to minimize the expected transmission time by searching available channels $(\boldsymbol{K}_i)$ for a proper channel number at each forwarding node, as follows: 
\begin{equation}
	\resizebox{0.7\hsize}{!}{$
	\begin{aligned}
\label{eq32}
 & MEATT(i,dst)=\operatorname*{arg\,min}_{k \in \boldsymbol{K}_i} MEATT(i,dst,k)=\\
&= \operatorname*{arg\,min}_{k \in \boldsymbol K_i}\frac{{\displaystyle\frac {L}{r_k}}+\displaystyle\sum_{j \in \boldsymbol F_{i} } \alpha_j P(i,j,k)MEATT(j,dst,k)}    {P_{\boldsymbol F_{i}}(k)}~~.
\end{aligned}
$%
}
\end{equation}
In eq.~(\ref{eq32}), $L$ is the packet length, $r_k$ is the bit rate in channel $k$, $\alpha_j$ is a tunable parameter to account for channel diversity. $P(i,j,k)$ and $P_{\boldsymbol F_{i}}(k)$ have the same meaning as before, except for the inclusion of channel $k$.

\item {\it Expected Anypath Delay (EAD)\cite{P_40}:}
The EAD metric is proposed for establishing a minimum-delay OR route for an opportunistic flow in a multi-channel scenario. The data delivery delay comprises both the transmission time ($T_{TX}$) and the stall time ($T_{stl}$) resulting from intra-flow interference. EAD uses the path established by the EAX \cite{P13} metric as its primary path and estimates the data delivery delay as follows:
\begin{equation}
	\resizebox{0.7\hsize}{!}{$
\label{eq33}
EAD(i,dst)=T_{stl_i} + T_{TX_i}=\displaystyle \sum_{j \in path(i\rightarrow dst)}(t_{stl_j}+t_{TX_j})~~~.
$%
}
\end{equation}
In eq.~(\ref{eq33}), the summation is over the set of nodes on the EAX path between node $i$ and $dst$, and $t_{TX_j}$ is the required transmission time for node $j$. When two direct neighbors attempt to use the same wireless channel simultaneously, EAD prioritizes the neighbor that has a shorter transmission time. Therefore, the stall time of node $j$ ($t_{stl_j}$) can be estimated as the minimum transmission time among node $j$ and its direct neighbors on the ${i \rightarrow dst}$ path. EAD estimates $t_{TX_j}$ in the same manner as other opportunistic metrics with minor differences. EAD first defines $N(j \rightarrow \boldsymbol{F}_j)$ as the average number of packet transmission attempts by node $j$ toward $\boldsymbol{F}_j$ to deliver a data packet from $j$ to $dst$: 
\begin{equation}
\label{eq34}
N(j \rightarrow \boldsymbol{F}_j)= \displaystyle \sum_{j \in \boldsymbol{F}_h}{\frac{N (h \rightarrow \boldsymbol{F}_h)}{P_{\boldsymbol F_h}}}.P(h,j)~~.
\end{equation}
In eq.~(\ref{eq34}), $\frac{N (h \rightarrow \boldsymbol{F}_h) }{P_{\boldsymbol F_h}}$ is the expected number of packets node $h$ should send to successfully deliver a data packet to its forwarder list $\boldsymbol{F}_h$. We can estimate $t_{TX_j}$ by multiplying the time required for transmitting a single data packet $\displaystyle\frac{L}{r} $ and the expected number of transmissions node $j$ should make: 
\begin{equation}
\label{eq35}
t_{TX_j}=  {\frac{N(j \rightarrow \boldsymbol{F}_j)}{P_{\boldsymbol F_j}}} .{\frac{L}{r}}~~.
\end{equation}

\item{\it Expected Weight of Anypath Transmissions (EWATX) \cite{P_60_7}\cite{P_60_24}:} EWATX introduces a multi-form (but not multi-dimensional) anypath OR metric that can consider different quality parameters one at a time:
\begin{equation}
\label{eq202}
\resizebox{0.7\hsize}{!}{$%
	\begin{aligned}
	&{EWATX}_k(i,dst)=\\&\displaystyle\frac{\displaystyle w_k(i)+\displaystyle\sum_{{j \in \boldsymbol{F}_i}}P(i,j) EWATX_k(j,dst) }       {\displaystyle P_{\boldsymbol F_i}};~ 1 \leq k \leq K ~.%
	\end{aligned}
	$ %
}
\end{equation}
${EWATX}_k(i,dst)$ in eq.~(\ref{eq202}) represents the weighed average of the anypath weights from node $i$ to the destination through a forwarder list $F_i$, $w_k(i)$ is the $k$-th quality parameter value of node $i$, and $K$ is the number of different quality parameters to be considered. For instance, 
when $K=1$, $w_1(i)$ may represent the average time required for node $i$ to complete one transmission. When $K=2$, $w_1(i)$ may be as above, and $w_2(i)$ may represent the energy consumed by node $i$ to complete one transmission.
This metric is very similar to the EATT metric.

\item {\it Metric of the Long Lifetime Any-path protocol (M-LLA)\footnote{The abbreviation is from the authors of this survey for better reference.}\cite{P32}:}
M-LLA is a dynamic opportunistic metric that addresses the link stability issues in VANETs. First, M-LLA estimates the stability of a dynamic wireless link established between two moving vehicles. Then, the best opportunistic path between $src$ and $dst$ is found using the M-LLA metric. Assume that two vehicles ($i$ and $j$) are connected by a wireless link at time $t_{0}$ and that the distance between them is denoted by $d_{ij}(t_0)$. Assume also that these vehicles remain connected after $\Delta t$ if the distance between them $(d_{ij} (t_0+\Delta t))$ does not exceed $R$, i.e., the wireless coverage range. We can decompose $d_{ij} (t_0+\Delta t)$ into $d_{ij} (t_0)+\Delta d_{ij}$, where $\Delta d_{ij}$ is the relative increase in the geographical advancement between node $i$ and node $j$. The stability index is defined as:
\begin{equation}
\label{eq19}
s_{ij} =1- \frac{\min{( \Delta d_{ij} ,R )}}{R}~~.
\end{equation}
The best value of $s_{ij}$ is 1. This value is assigned to the link when both vehicles have equal relative movement vectors. The worst value of $s_{ij}$ is 0. This value is assigned to the link when the distance between two nodes exceeds $R$ (communication range) in a time period $\Delta t$. We can modify the link cost definition to include the stability index as follows: 

\begin{equation}
\label{eq20}
c_{ij}= \frac{1}{p_{ij}s_{ij}}~~.
\end{equation}

The opportunistic cost between node $i$ and $dst$ is defined as: 

\begin{equation}
\label{eq21}
\text{\it M-LLA}(i,dst)=\text{\it M-LLA}(i,\boldsymbol{F}_i)+ \text{\it M-LLA}(\boldsymbol{F}_i,dst)~~,
\end{equation}

where $\text{\it M-LLA}(i,\boldsymbol{F}_i)$ is the cost of delivering a packet from node $i$ to at least one member of $\boldsymbol{F}_i$, and $\text{\it M-LLA}(\boldsymbol{F}_i,dst)$ is the opportunistic path cost from $\boldsymbol{F}_i$ to $dst$. M-LLA is derived as follows: 
\begin{align}
	\label{eq22}
	\notag \text{\it M-LLA}(i,dst)&= \frac{1}{1- \displaystyle \prod_{j \in \boldsymbol{F}_i}(1- p_{ij}s_{ij})}+ \\ 
	&\displaystyle \sum_{j \in \boldsymbol{F}_i} \frac{p_{ij}s_{ij}\text{\it M-LLA}(j,dst) \displaystyle \prod_{k=1}^{j-1}{(1-p_{ik}s_{ik})}}   {1-\displaystyle \prod_{j \in \boldsymbol{F}_i}(1-p_{ij}s_{ij})}
\end{align}

or equivalently:
\begin{equation}
	\resizebox{0.75\hsize}{!}{$%
\label{eq22_2}
\text{\it M-LLA}(i,dst)= \frac{1+\displaystyle \sum_{j \in \boldsymbol{F}_i} \text{\it M-LLA}(j,dst)p_{ij}s_{ij} \displaystyle \prod_{k=1}^{j-1}{(1-p_{ik}s_{ik})}}{1- \displaystyle \prod_{j \in \boldsymbol{F}_i}(1- p_{ij}s_{ij})}~.
	$ %
}
\end{equation}

The only difference between eq.~(\ref{eq21}) and its counterpart in \cite{P19,P119} is the inclusion of a stability index. Additionally, eqs.~(\ref{eq22_2}) and (\ref{eq1}) are closely related. Being similar to EAX, except for the inclusion of the node speed and stability index, \text{\it M-LLA} has an identical feature set.

\item {\it Opportunistic Energy Cost with Sleep-wake schedules (OECS) \cite{P_50_20}:} The OECS metric reflects the energy cost of an opportunistic path:
\begin{equation}
	\begin{aligned}
	\label {eq384}
	  OECS(i,dst )=& \displaystyle\frac{\frac{{E}_{prb}+{E}_{wak}+E_{cons-TX}}{E_{res_i}}}{P_{\boldsymbol F_i}}+ \\ &\frac{\sum_{j \in \boldsymbol F_i}P(i,j)OECS(j,dst)}{P_{\boldsymbol F_i}}~.
	\end{aligned}
\end{equation}

OECS and OEC are similar except that the former models the required energies for broadcasting and receiving beacons.

\item {\it Expected Transmission Cost (ETC*)\footnote{The * superscript is intended for distinction with the ETC metric in eq.~(\ref{eq28}).} \cite{P_50_10, P_2019_10_041028}:}
This metric is proposed  in heterogeneous duty-cycled WSNs. In duty-cycled networks, the transmission cost associated with finding at least one awake forwarding node is called the rendezvous cost, $c_{rv}$. This cost is stated to be proportional to the wake-up ratio of all the forwarding nodes. The total cost of node $i$ sending a packet over a single hop is then:  
\begin{equation}
{ETC}^{*}(i,\boldsymbol F_i) = \frac{c^{i}_{rv} + c^{i}_{com}}{T_{cycle}}~~,
\end{equation}
where $c_{com}$ and $T_{cycle}$ are the packet transmission cost and cycle duration, respectively. Thereafter, the expected multi-hop transmission cost or metric is calculated as:
\begin{equation}
\label{eqetc}
{ETC}^*(i,dst) = {ETC}^{*}(i,\boldsymbol F_i) + \sum_{j\in \boldsymbol F_i} \frac{ETC^*(j,dst)}{|\boldsymbol F_i|}~~.
\end{equation}

\item {\it Expected Transmission Cost over a multi-hop Path (ETCoP) \cite{P_50_19}:} The ETCoP metric reflects the total expected number of transmissions required for the successful delivery of a packet on the path between two end nodes. If the nodes on the path are numbered-labeled, the path between node $i$ and $dst$ consists of $(dst-i+1)$ nodes, then, we define:
\begin{equation}
\label {eq382}
c(i,dst)=(dst-i)\prod_{j=i}^{dst-1}p_{j,j+1} + c(i,dst-1) (1-p_{dst-1,dst}).
\end{equation}

The first term on the RHS of eq.~(\ref{eq382}) is the success probability of successive transmissions from node $i$ to $dst$. The second term on the RHS represents the cost of a situation where $dst-1$ fails to deliver the packet to $dst$. ETCoP is then defined as:
\begin{equation}
\label {eq383}
ETCoP(i,dst)=\frac{c(i,dst)}{\prod_{j=i}^{dst-1}p_{j,j+1}(i,dst)}~~.
\end{equation}
ETCoP prefers a path whose link component's quality gradually increases towards the destination. 

\item {\it Expected Energy Consumed Along the Path (EEP) \cite{P_2018_30}:}
EEP was originally introduced as an anycast routing metric and was used by EDAD \cite{P_2018_31}, an energy-centric cross-layer data collection protocol designed for anycast communications in asynchronously duty-cycled WSNs, and not in the context of OR. EEP was later used as an OR metric by \cite{P_2018_30} for opportunistic many-to-many multicasting in duty-cycled WSNs. $EEP_i$ denotes the expected energy required to deliver a packet from node $i$ to the destination and is defined as:

\begin{equation}
\label {eq2018_24}
EEP (i,dst) =\displaystyle \frac{\displaystyle \sum_{j \in \boldsymbol{F}_i} EEP(j,dst) + \frac{2}{p_{ij}}}{|\boldsymbol{F}_i|}  +  \frac{\frac{T_w}{T_F}}{|\boldsymbol{F}_i| +1}~~.
\end{equation}
The first part of the RHS in eq.~(\ref{eq2018_24}) is an average of the energies required by each member of $\boldsymbol{F}_i$ for delivery to the destination (the first term of the summation), plus the energy consumed for the packet transaction between node $i$ and node $j$ (the second term of the summation). The second part of the RHS in eq.~(\ref{eq2018_24}) accounts for the average energy consumed until one member of the CFS wakes up to respond, as the network is duty-cycled.  $T_w$ and $T_F$ are the average sleep time of each node and the time duration of each data frame transmission, respectively. 

\item {\it Expected End-to-end Latency (EEL)\cite{P_50_15}:} $EEL$ is proposed for Underwater Wireless Sensor Networks (UWSN). The goal of $EEL$ is to maximize goodput. $EEL(i,dst)$ is defined as the constrained end-to-end delay from node $i$ to $dst$:
\begin{equation}
\resizebox{0.6\hsize}{!}{$
	\label{eq372}
	\begin{aligned}
		EEL (i,dst) &=\displaystyle \frac{T_{MAC_{\boldsymbol F_i}} + T_{Crd_{\boldsymbol F_i}} }{P_{\boldsymbol F_i}}\\ & + \frac{ \sum_{j \in \boldsymbol F_i}P(i,j)EEL(j,dst)}{P_{\boldsymbol F_i}}~.
	\end{aligned}
	$%
}
\end{equation}

$EEL(i,dst)$ must be lower than a threshold. The parameter $T_{MAC_{\boldsymbol F_i}}$ is composed of four parts, the MAC contention time, the packet transmission delay, the propagation delay, and the ACK delay. $T_{Crd_{\boldsymbol F_i}}$ is the coordination delay among the members of $\boldsymbol F_i$. In $EEL$, each forwarding node waits for higher-priority nodes to respond by sending an ACK packet. 

\item {\it End-to-end transmission Cost (EC) \cite{P_60_11}:}
This metric is equal to the energy cost required for delivering an $L$-bit packet from node $i$ to the final destination through an optimally selected cluster-parent-set ($CPS_i$) with respect to $R$, the communication range, which we call it $\boldsymbol F_{i,R}$ in our own notation:   
 \begin{equation}
 \label{eqEC}
{EC}(i,dst) = \min_{R} ({AEC}(i,\boldsymbol F_{i,R}) + {REC}(\boldsymbol F_{i,R},dst))~.
\end{equation}
 If we denote the energy cost of transmitting an $L$-bit packet from node $i$ with $E_i(L,R)$, then the first term on the right of eq.~(\ref{eqEC}) represents the average energy cost of delivering an $L$-bit packet from node $i$ to $\boldsymbol F_{i,R}$ in which :
\begin{equation}
AEC(i,\boldsymbol F_{i,R} )= \frac{E_i(L,R)}{P_{\boldsymbol F_{i,R}}}~~,
\end{equation}
 and the second term on the right shows the energy cost of delivering the $L$-bit packet from $\boldsymbol F_{i,R}$ to the final destination:
 \begin{equation}
 REC(\boldsymbol F_{i,R},dst)=\frac{ \displaystyle\sum_{{j \in \boldsymbol F_{i,R}}} P(i,j)EC(j,dst) }  {P_{\boldsymbol F_{i,R}}}  ~~.
 \end{equation}
\item {\it Expected Cognitive Anypath Transmissions (ECATX) \cite{P_61_20}:}
Defined in cognitive WNs, ECATX greatly resembles M-LLA and EATT structurally and also in the sense that all these three metrics revise the link probability by one factor of interest. ECATX revises the link probability by the factor of link quality:
\begin{equation}
ECATX (i,\boldsymbol F^k_i) = \frac{1}{{P^k_{\boldsymbol F_i}}.{q^k_i}}~~~,
\end{equation}
in which the revision factor $q^k_i$ denotes the quality of channel $k$ of node $i$ in the sense of its average OFF-time duration compared to that of the other channels. A larger $q$ indicates a greater eligibility for forwarding packets. Therefore, the end-to-end metric is defined as:
\begin{equation}
	\resizebox{0.7\hsize}{!}{$%
\label{eq300}
ECATX (i,dst)= ECATX(i,\boldsymbol F^k_i)+ ECATX(\boldsymbol F^k_i,dst).
	$ %
}
\end{equation}
 
\item {\it Expected Delivery Ratio (EDR), Expected End-to-End Delay (EED) \cite{P_2019_3_280407}:}
\cite{P_2019_3_280407} proposes two OR algorithms for connecting Narrowband-Internet-of-Things (NB-IoT) users to a Base Station (BS) in cellular communications paradigm, via a set of duty-cycled D2D relays. The basic assumption, herein, is that the routings occur in two hops. The authors introduce two OR metrics, EDR, which tries to maximize the expected end-to-end packet delivery probability, and EED, which aims at minimizing the expected total transmission delay. The two metrics can be defined in the form of simplified optimization formulations as follows:
\begin{equation}
\begin{aligned}
	& \operatorname*{max}_{relay,slot} EDR(src,BS)= \\ &\sum_{slot} \sum_{relay}PDP(src,relay,slot). PDP(relay,BS,slot)
	\end{aligned}
\end{equation}
and
\begin{equation}
\resizebox{0.7\hsize}{!}{$
\begin{aligned}
	&\operatorname*{min}_{relay,slot} EED(src,BS)=\\ 
	&{(EDR(src,BS))}^{-1} \times\displaystyle\sum_{slot} \sum_{relay}PDP(src,relay,slot)\times\\ & PDP(relay,BS,slot)\times T_{TX}(src,relay,slot)~~,
	\end{aligned}
		$%
	}
\end{equation}
wherein the $PDP$s and the $T_{TX}$ are defined between $src/relay$ and $relay/BS$ in specific time slots. The optimization problems boil down to finding the optimum relay set and the transmission time slot.

\item {\it {Stability-Quality-Advancement (SQA) \cite{P_2017_4}:} }
This metric, introduced particularly for VANETs with vehicles in highways moving in the same direction, is very similar to the M-LLA metric with a modified cost: 
\begin{equation}
\label{eq525}
SQA(i,j)= p_{ij} . s_{ij} . a_{ij}~~~.
\end{equation}
Compared to M-LLA, the only modification is the packet relative advancement, $a_{ij}$, defined as:   
\begin{equation}
\label{eq2017_11}
a_{ij}= \left\{
\begin{array}{ll}
\frac{\displaystyle d_{ij}}{R}~~;~~ 0<d_{ij}\le R\\
0~~~~~; ~~~~~~ o.w.~~~~~~~~~~.
\end{array}
\right.
\end{equation}
The double consideration of the geographical advancement through $s_{ij}$ (the stability index of the link connecting node $i$ and node $j$ as in eq.~(\ref{eq19})) and $a_{ij}$ in the cost function of eq.~(\ref{eq2017_11}) may require further justification.

\item {\it {Expected Transmission Direction (ETD) \cite{P_2018_40}: }}
ETD is defined in the context of data transmission in IoT applications with multiple unreliable gateways. In this application, it is assumed that all sensor nodes, as well as gateways, are duty-cycled. So, not only the PDP of the individual links belonging to a specific path but the extent of duty-cycle overlaps along those links are also important. The ETD metric is defined as the cost of delivery at the minimum-cost gateway among the candidate gateways corresponding to a specific destination node, within a flexible budget of energy and delay. Assuming a number-labeled set of nodes along the path from node $i$ to $dst$ (i.e., the minimum cost gateway), the metric is formulated as:
\begin{equation}
\resizebox{0.9\hsize}{!}{$
ETD(i,dst)= \displaystyle\sum_{j=i}^{dst-1} [E_{cons}(j\rightarrow j+1)\displaystyle\sum_{0\leq m<ReTX}m.(1-p_{j,j+1})^{m-1}p_{j,j+1}]  ,
	$%
}
\end{equation}
wherein the inner summation accounts for the transmission cost over all links ({\it j$\rightarrow$j+1}) considering $ReTX$ number of retransmissions allowance, and  $E_{cons}$ denotes a link's consumed transmission energy. It should be emphasized that \cite{P_2018_40} treats the cost to the selected gateway as the cost to the destination.

	\item {\it {Sigment's Packet Delivery Ratio (SPDR), Packet Delivery Advancement (PDA) \cite{P_2018_41}:}}
	In CRNs, the condition and the availability of links (due to spectrum variation) are changed dynamically. To overcome these limitations regarding optimal routing, \cite{P_2018_41} divides the path between $src$ and $dst$ into a set of smaller and more stable route segments by introducing temporary source-destination pairs. The temporary source (${src}^{'}$) and destination (${dst}^{'}$) are selected so that they can communicate through a single candidate set. SPDR is then defined as the classical delivery cost from ${src}^{'}$ to ${dst}^{'}$ through the candidate set using just the link probabilities as follows: 
	\begin{equation}
	SPDR ({src}^{'}, {dst}^{'}) = \sum_{i=1}^{|{\boldsymbol F_{{src}^{'}}}|} p_{{src}^{'}i} . P(i,{dst}^{'}) ~~~.
	\end{equation}
	In the continuation, to make a trade-off between longer geographical advancement and shorter forwarder candidate list, \cite{P_2018_41} further amends SPDR to account for geographical progress: 
	\begin{equation}
	PDA({src}^{'}, {dst}^{'}) = SPDR ({src}^{'}, {dst}^{'}) . d({src}^{'}, {dst}^{'})~~.
	\end{equation}

\item {\it Expected Dynamic Transmission Cost (EDTT) \cite{P_2020_03_280059}:}
Dynamic transmission power strategy is sometimes adopted to serve energy-critical networks such as Energy-Harvesting Wireless Sensor Networks (EH-WSNs) more efficiently. However, this provides the communication links with time-varying qualities, whose information to be updated and disseminated regularly. The EDTT represents the end-to-end transmission time-cost, in the time-slotted underlying network, with the familiar two-step form (e.g., see ETC${}^*$ in eq.~(\ref{eqetc})) of:  

\begin{equation}
	EDTT(i,dst,t) = EDTT(i,j,t)+\frac{\sum_{j\in \boldsymbol{F}_i}EDTT(j,dst,t)}{|\boldsymbol{F}_i|}
\end{equation}
where its single-hop counterpart is defined as:
\begin{equation}
	EDTT (i,j,t)= \frac{1}{p_{ij}}.\vv{T_{com}}.\vv{T_{wait}}(t)
\end{equation}
$\vv{T_{com}}$ is the average transmission time-cost and $\vv{T_{wait}}$ denotes the average waiting duration at time $t$, throughout each time slot.  We suspect that the original formulation of EDTT in eq.~(7) of \cite{P_2020_03_280059}  might be erroneous, since to be opportunistic, the single-hop progress should target a forwarder list rather than a single node, which we corrected herein.
\end{itemize}

\subsubsection{\bf Multi-Path/Per-Hop OR Metrics:} 
These metrics evaluate progress in a short-sighted fashion. In contrast to the two previous classes, the relevant metrics are calculated non-recursively. The metrics are presented in the chronological order of their appearance in the literature.
\begin{itemize}

\item {\it Expected One-hop Throughput (EOT)\cite{P27}:}
The authors of the OEE metric \cite{P23} proposed another metric, known as EOT, which is redefined in several papers with different names and optimization goals\cite{P26,P28,P29,P_40_22,P_40_23}. In EOT, each forwarding node $i$ selects $\boldsymbol F_i$ so that 1) the geographical (physical) progress of the packet toward the destination is maximized and 2) the time required for the transmission of the packet from node $i$ to node $j \in \boldsymbol{F}_i$, $t_{TX_i}$, is minimized simultaneously. The authors decompose $t_{TX_i}$ to several components, such as the channel contention time, the data transmission time, the propagation delay, and the forwarding node coordination delay. EOT is defined as follows:
\begin{equation}
\label{eq15}
\resizebox{0.7\hsize}{!}{$%
	\begin{aligned}
	& EOT(i,\boldsymbol{F}_i)=\\&{{\displaystyle\sum_{j \in \boldsymbol{F}_i}{d_{ij}}P(i,j)}} .\frac{\displaystyle\huge L}{\displaystyle t_{TX_i}^{'} {\displaystyle\prod_{j \in \boldsymbol{F}_i}  {(1-p_{ij})} }+{\displaystyle\sum_{j \in \boldsymbol{F}_i}{t_{TX_i}}P(i,j)}} ~.%
	\end{aligned}
	$ %
}
\end{equation}
Note that $\displaystyle\sum_{j \in \boldsymbol{F}_i}{t_{TX_i}}P(i,j)$ is the expected delay for the successful transmission of the packet from node $i$ to the members of $\boldsymbol{F}_i$. Although not determined in the paper exactly, $ t_{TX_i}^{'}$ appears to be the retransmission time of the packet, which is multiplied by ${\displaystyle\prod_{j \in \boldsymbol{F}_i}  {(1-p_{ij})} }$, i.e., the probability that none of the members of $\boldsymbol{F}_i$ receive the packet successfully. Therefore, the denominator is the excepted one-hop delay of the packet from node $i$ to $\boldsymbol{F}_i$ considering both opportunistic forwarding and retransmission delays.\\
The first multiplicand on the RHS of eq.~(\ref{eq15}) was later named as a new OR metric, EPA \cite{P23}, by the same authors.

\item {\it Expected Advancement Rate (EAR):}
EAR \cite{P33} is an extension of EPA \cite{P23} proposed by the same authors. EAR models the expected bit-advancement per second by incorporating the node rate:
\begin{equation}
\resizebox{0.7\hsize}{!}{$
	EAR(i,\boldsymbol{F}_i)= r_i . {{\displaystyle\sum_{j \in \boldsymbol{F}_i}{d_{ij}}P(i,j)}}= r_i . EPA(i,\boldsymbol{F}_i)~~, %
	$%
}
\end{equation}
where $r_i$ is the broadcast rate of node $i$. 
	
	\item {\it Opportunistic Expected Transmission Time (OETT):}
	In \cite{P33}, the ETT \cite{P34} unicast metric is generalized for opportunistic routing scenarios by considering the hidden node problem. OETT is defined for the hyper opportunistic link between node $i$ and its forwarding candidate set (i.e., $\boldsymbol{F}_i$) as follows: 
	
	\begin{equation}
	\label{eq24}
	{OETT}(i,{\boldsymbol F_i})= \frac{L}{r_i . P_{\boldsymbol F_i}}~~,
	\end{equation}
	
	where $L$ is the packet length, and $r_i$ is the broadcast rate of node $i$. OETT measures the expected transmission time to send a packet from node $i$ to any node in $\boldsymbol{F}_i$.

\item {\it Energy Distance Ratio per bit ($EDRb$), Delay Distance Ratio ($DDR$):}
\cite{P_61_24,P_2019_151} introduces two metrics related to the joint geographical progress/consumed energy and the joint geographical progress/delay. Thereafter, a routing protocol that employs Pareto optimization with a reliability constraint of these two metrics is proposed. The joint geographical progress/consumed energy metric, called EDRb, is defined as:

\begin{equation}
\label{eq1122}
EDRb (i,\boldsymbol F_i)=\frac{E_{b}}{EPA(i,\boldsymbol F_i)}~~~,
\end{equation}
and the joint geographical progress/delay metric, called DDR, is defined as:
\begin{equation}
\label{eq1122_2}
DDR(i,\boldsymbol F_i) =\frac{T_{hop}}{EPA(i,\boldsymbol F_i)}~~~.
\end{equation}
In eq.~(\ref{eq1122}), $E_{b}$ is the energy consumption per bit, and in eq.~(\ref{eq1122_2}), $T_{hop}$ denotes the delay of a packet being transmitted over one hop.

\item {\it Multicast Expected Advancement Rate (MEAR), I-MEAR:}
MEAR \cite{P35} is a naive extension of the EAR metric to multicast applications. The MEAR of node $i$ with respect to all multicast destinations set $\boldsymbol{DST}$ is defined as: 
\begin{equation}
\label{eq26}
MEAR(i,\boldsymbol{DST})=\operatorname*{max}_{r_i^h} \displaystyle \sum_{h \in \boldsymbol{DST}}EAR(i,\boldsymbol F_i^h)~~.
\end{equation}
Based on eq.~(\ref{eq26}), MEAR of a node $j$ is calculated by finding the best rate that yields the largest MEAR. The authors argue that because of the tree backbone, the destinations of node $i$ will depend on its location. Hence, they propose a modified version of MEAR named I-MEAR: 

\begin{equation}
\label{eq27}
\text{\it I-MEAR}(i,\boldsymbol{DST_i})=\operatorname*{max}_{r_i^h} \displaystyle \sum_{h \in \boldsymbol{DST_i}}EAR(i,\boldsymbol F_i^h)~~.
\end{equation}

\item {\it Expected Packet Advancement (EPA), One-hop Energy Efficiency (OEE)~\cite{P23}:}
OEE is an energy-aware metric proposed for WSNs using the physical location of sensor nodes. Node $i$ selects members of $\boldsymbol{F}_i$ so that, simultaneously, 1) the geographical progress of the packet toward the destination is maximized and 2) the energy consumed for the successful transmission of the packet (denoted by ${E}_{or}$($\boldsymbol{F}_i$)) is minimized. ${E}_{or}(\boldsymbol F_i)=E_i+E_{\boldsymbol F_i}+E_o$ is composed of three parts: 1) $E_i$, the expected energy consumed by node $i$ for transmitting the packet, 2) $E_{\boldsymbol F_i}$, the expected energy consumed by the members of $\boldsymbol{F}_i$ when dealing with the packet, and 3) $E_o$, the expected energy consumed by other neighbors of node $i$. Each of these parts is further decomposed into the power consumption required for data reception, data transmission, and idle listening. Then, the expected packet advancement (EPA) of a packet is defined as: 

\begin{equation}
\label{eq11}
EPA(i,{\boldsymbol{F}_i})={{\displaystyle\sum_{j \in \boldsymbol{F}_i}{d_{ij}}P(i,j)~~,
}}
\end{equation}
where $d_{ij}$ is the geographical advancement of the packet toward $dst$ from node $i$ to node $j$: 
 
\begin{equation}
\label{eq12}
d_{ij}=d(i,dst)-d(j,dst)~~. 
\end{equation}

Finally, the OEE metric is defined as follows: 

\begin{equation}
\label{eq13}
  OEE(i,\boldsymbol{F}_i^*) = \operatorname*{max}_{\substack {\forall \boldsymbol{F}_i \subset \boldsymbol{C}_i}}\bigg\{  \frac{EPA{(i,\boldsymbol{F}_i)}}  {{P_{\boldsymbol{F}_i}}{.{{E}_{or}   {(\boldsymbol{F}_i)}}}}\bigg\}. L~~.
\end{equation}

In eq.~(\ref{eq13}), $\frac {EPA(i,\boldsymbol{F}_i )} {P_{\boldsymbol{F}_i }}$ is the expected packet advancement for a successful opportunistic forwarding considering retransmissions, and $\boldsymbol{F}_i^*$ is a subset of $\boldsymbol{C}_i$ which maximizes OEE.

\item {\it Expected Distance Progress (EDP)\cite{P24}:}
EDP is a recursive opportunistic metric based on the Distance Progress OR (DPOR) \cite{P25}, an earlier work from the same authors, which considers the distance progress of the data packet toward $dst$. The EDP's definition is similar to the EPA's definition given in OEE \cite{P23}. 
\item{ \it Cognitive Transport Throughput (CTT):}
\cite{P_60_14} introduces an OR metric in multi-hop CR networks, although the metric itself is calculated over a one-hop relay:
\begin{equation}
CTT (i,\boldsymbol F^{k}_i) =\sum_{j\in \boldsymbol F^{k}_i} p^{k}(i,j) \frac{L.d_{ij}}{T_{RX}(j)}~~~,
\end{equation}
wherein $k$ and $T_{RX}(j)$ are respectively the channel index and the reception delay of node $j$. The metric measures the expected bit advancement per second for the one-hop relay of an $L$-bit packet in channel $k$. \cite{P_60_14} states that maximizing the one-hop relay performance contributes to the end-to-end performance improvement. 

\item {\it  Forwarder Score (FS) \cite{P_61_22}:}
This metric is a direct derivative of EDC that provides load balancing through the consideration of the forwarding node's residual energy in duty-cycled WSNs. FS is defined as: 

\begin{equation}
	\label{eq_2018_11_1}
	FS(i,\boldsymbol F_i) = \displaystyle\frac{1}{E_{res_i}^\alpha (|\boldsymbol F_i| + 1)} + \frac{\sum_{j \in \boldsymbol F_i} FS_j}{|\boldsymbol F_i|}~~~,
\end{equation}
where $E_{res_i}$ denotes the scaled residual energy ratio of node $i$, and $\alpha$ is a weight for adjusting the effect of the residual energy.\\   
According to \cite{P_61_22}, the probability that a node overhears another neighboring node's transmission decreasingly depends on the overheard node's number of neighbors. However, this statement is not entirely justified.

\item {\it Expected single-hop packet speed (espeed) \cite{P_61_14}:}
QoS-aware geographic opportunistic routing in WSNs, introduced in \cite{P_61_14}, considers both end-to-end reliability and delay constraints and formulates the routing as a multi-objective multi-constraint optimization problem. Regarding the forwarder list formation, let $\pi_j (\boldsymbol F_i)=\{j_1, j_2,...,j_n \}$ be one ordered permutation of nodes in $\boldsymbol F_i \subseteq \boldsymbol C_i$ with priority $(j_1>j_2>...>j_n)$, then the $espeed$ metric is defined as the solution to the following maximization problem:
\begin{equation}
\label{eq5455}
 espeed(i, \boldsymbol F_i)=  
  \displaystyle\operatorname*{max}_{\pi_j (\boldsymbol F_i)} ~ \frac{EPA(i,\pi_j (\boldsymbol F_i))}{ed(i,\pi_j (\boldsymbol F_i))} ~~~~,
\end{equation}
wherein EPA and $ed$ represent the expected packet advancement per-hop mentioned previously and the expected single-hop media delay, respectively, and both are limited to corresponding predetermined thresholds. Also, it is assumed that all members of the forwarder list can hear each other.

\item{ \it Patterned Synchronization Tendency Metric (PSTM) \cite{P_61_23}:}
To prevent the overuse of nodes due to the patterned synchronization effect \cite{P_61_23} in duty-cycled WSNs, the value of PSTM at each forwarding node is compared against a threshold. The results of the comparisons are then used to adjust the duty-cycle of the node. PSTM is defined as:
\begin{equation}
PSTM(i,S_i) = \frac{ \displaystyle\sum_{j \in S_i} P(i \leftarrow j)}{|S_i|}~~~~~~. 
\end{equation}
$ P(i \leftarrow j)$ denotes the probability of node $j$ to select the particular upstream node $i$ as the next-hop forwarder, and $S_i$ is the set of nodes to which node $i$ acts as the upstream node. Node $i$ is informed of $P(i \leftarrow j)$, which is calculated at node $j$, through the header field of the transmitted packets from node $j$ to node $i$.

\item {\it Distance Utility Energy Ratio (DUER)\cite{P_50_14}:}
The DUER metric, which is similar to its predecessors EDRb and DDR \cite{P_61_24}, is a normalized version of EPA aiming to maximize the network's lifetime. DUER is defined as:
\begin{equation}
\label{eq371}
DUER(i, \boldsymbol F_i)=\frac{EPA(i,\boldsymbol F_i)}{UE_{tot}(i,\boldsymbol F_i)}~~~.
\end{equation}
Herein, $UE_{tot}$ is a utility function that adds up the energy consumed by node $i$ and members of $\boldsymbol F_i$ for packet transmission and reception.
 
\item {\it Forwarding Efficiency (FE)\cite{P_50_16}:} FE is a multi-flow metric. Here, we present FE for a single flow: 
\begin{equation}
\label{eq373}
FE(i,\boldsymbol F_i,r)=EPA^{*}(i,\boldsymbol F_i,r) \frac{r}{P_{\boldsymbol F_i}}~~,
\end{equation}
where $r$ is the transmission rate, and $EPA^{*}$\footnote{The * superscript is intended for distinction with the EPA metric in eq.~(\ref{eq374})} is defined as follows:
\begin{equation}
\label{eq374}
EPA^{*}(i,\boldsymbol F_i,r)=\sum _{j \in \boldsymbol F_i} d_{ij}p_{ij}^{r}\Delta_{j} \prod_{k=0}^{j-1}(1-p_{ik}^{r}\Delta_k)~.
\end{equation}
In eq.~(\ref{eq374}), $\Delta_j$ is the probability that the coded packet is decodable at the $j$-th forwarding node.

\item {\it Expected Energy and Packet Advancement (EEPA) \cite{P_2019_3_240848}:}
In UWSNs, which feature a 3D topology, high possibility of void regions, and the underwater acoustic propagation model, this metric aims at compromising  between the normalized reliability (associated with the first term on the RHS of the equation below) and the normalized energy consumption (associated with the second term) and is defined as:
\begin{equation}
\label{eq_2019_24_01}
	EEPA(i,\boldsymbol{F}_i)= \alpha \frac{EPA(i,\boldsymbol{F}_i)}{EPA(i,\boldsymbol{C}_i)} - \beta \frac{E(i,\boldsymbol{F}_i)}{E(i,\boldsymbol{C}_i)}~~,
\end{equation}
wherein $\boldsymbol{F}_i$ is a subset of $\boldsymbol{C}_i$ which maximizes the above expression. EPA, the expected packet advancement, is a variant of the EPA metric in eq.~(\ref{eq11}) which assigns a higher priority to a forwarding node with lower underwater depth. $E$ denotes the consumed energy of the forwarder list in the packet forwarding process. $\alpha$ and $\beta$ are the weighting coefficients to favor one factor to another. Since both fractions in eq.~(\ref{eq_2019_24_01}) are normalized and between zero and one, though not mentioned in the original paper, we suspect that most probably the minus operator should be an addition and $\alpha + \beta = 1$ ($0 \leq \alpha , \beta \leq 1$).

\item {\it Forwarder Score ($FS^*$)\footnote{The * superscript is intended for distinction with the FS metric in eq.~(\ref{eq_2018_11_1}).} \cite{P_2019_3_231559}:}
This metric, as its identically-titled predecessor \cite{P_61_22}, aims at providing load balancing in duty-cycled WSNs with minor differences. It is defined as:
\begin{equation}
\label{eq2019_3_22}
	FS^{*}(i,\boldsymbol{F}_i)=\displaystyle\frac{1}{{\big\lceil \frac{E_{res_i}}{E_{tot_i}}.g \big\rceil}^\alpha . (|{\boldsymbol F}_i|+1)} +  \frac{\sum_{j \in {\boldsymbol F}_i} FS^*_j}{|{\boldsymbol F}_i|}~~,
\end{equation}
where, compared to the definition of eq. (\ref{eq_2018_11_1}), $E_{res_i}$ has been replaced by $\lceil \frac{E_{res_i}}{E_{tot_i}}.g \rceil$ as the energy-survival indicator. $E_{res_i}$, $E_{tot_i}$ , and the parameter $g$ are the node $i$'s residual energy, total energy, and the energy quantization granularity, respectively.

\item {\it Enhanced Expected Packet Advance ($EEPA^{*}$)\footnote{The * superscript is intended for distinction with the EEPA metric in eq.~(\ref{eq_2019_24_01})} \cite{P_2018_12}:}
There is another version of EPA, called EEPA${}^{*}$, that also considers, in a more realistic scenario, the correlation between contributing links in packet delivery from a node to a specific forwarder list. Specifically, EEPA${}^{*}$ revises the term $P(i,j)$ (see eq.~(\ref{eq02})) in eq.~(\ref{eq11}) as:

\begin{equation}
\label{eq475}
 P_c(i,j)=p_{ij}\displaystyle\prod^{j-1}_{m=0} {(1- p_{im}\rho(i,j,m))}~~,
\end{equation}
where $\rho(i,j,m)$ denotes the correlation between the $i~\rightarrow~j$ and $i~\rightarrow~m$ links. Therefore, the final derivation for EEPA${}^{*}$ becomes:   

\begin{equation}
\label{eq476}
EEPA^{*}(i,{\boldsymbol{F}_i})={{\displaystyle\sum_{j \in \boldsymbol{F}_i}{d_{ij}}P_c(i,j)
	}}~~~.
	\end{equation}

\item {\it Radio Signal Strength Indicator (RSSI)-based Metric (M-RSSI)\footnote{The abbreviation is from the authors of this survey for better reference.} \cite{P_2018_43}:}
This over-link metric is introduced in \cite{P_2018_43} for a duty-cycled network-coded WSN to represent the link quality. The metric, which is based on the RSSI between two nodes, is a compromise between the path length and the PDP of the link connecting them. The M-RSSI metric is calculated by:
\begin{equation}
\resizebox{0.7\hsize}{!}{$%
\text{\it M-RSSI}(i,j)= \left\{
\begin{array}{ll}
1+\gamma ~~&;~ RSSI_{ij} < RSSI_{THRSLD}\\
1 ~~~~&; ~ o.w.
\end{array} \right.%
$%
}
\end{equation} 
wherein the parameters $RSSI_{THRSLD}$ and $\gamma$ are tuned experimentally. The degradation of RSSI below $RSSI_{THRSLD}$ is penalized by $\gamma$ in the metric's value to avoid long links.

\item {\it Dempster-Shafer Theory (DST)-based Trust (DSTT)\footnote{The abbreviation is from the authors of this survey for better reference.} \cite{P_2019_1}:}
This metric, presented in the context of a UAWSN application, combines three pieces of environmental evidence. Each evidence represents a single-objective metric based on DST. The target objectives are the residual energy, the PDP, and the geographical advancement of the next hop. Using the Basic Probability Assignment (BPA) function, a mass value is assigned to every node $j \in {\boldsymbol C}_i$, denoting its strength as a piece of evidence. The mass values corresponding to the above three metrics are then defined as:  
\begin{equation}
\resizebox{0.5\hsize}{!}{$%
	 \left\{
	\begin{array}{ll}
	m_1 (i,j)= \frac{\displaystyle E_{res}(j)}{\displaystyle\sum _{{h \in \boldsymbol C}_i}E_{res}(h)}~~,&\\ \\

m_2 (i,j)= \frac{\displaystyle p_{ij}}{\displaystyle\sum _{{h \in \boldsymbol C}_i}p_{ih}}~~,&\\ \\

m_3 (i,j)= \frac{\displaystyle d(i,j)}{\displaystyle\sum _{{h \in \boldsymbol C}_i}d(i,h)}~~.&\\

	\end{array} \right.%
	$%
}
\end{equation}
 
\cite{P_2019_1} uses the Dempster's rule of combination from multiple metrics to get the trust value of node $j$ as a forwarding node for node $i$ as follows:
\begin{equation}
	DSTT(i,j)=\frac{m_1(i,j) . m_2(i,j) . m_3(i,j)}{1-K}~~,
\end{equation}
where $K$ is a normalization factor used to measure the conflict between BPA functions of different metrics. The nodes from ${\boldsymbol C}_i$ are added to ${\boldsymbol F }_i$ (node $i$'s forwarder list) and prioritized in decreasing-trust order. The addition of nodes to ${\boldsymbol F }_i$ continues until $P_{{\boldsymbol F }_i}$ (see Table~\ref{table:tab11}) reaches a predetermined minimum threshold.

\item {\it Effective Forwarding Rate (EFR)\cite{P_2019_2}: }
\cite{P_2019_2} introduces EFR in the context of Cognitive Radio Ad-hoc Networks (CRAHN) wherein the authors address the volatility in link states by their effective forwarding rates. The EFR metric is defined as:
\begin{equation}
	EFR (i,{\boldsymbol F }_i,r_i)= r_{i} \sum_{j=1}^{|{\boldsymbol F }_i|}  p_{ij} P(i,j)~~,
\end{equation}
in which $r_{i}$ denotes the transmission rate emanated by node $i$. Though EFR might seem like a new metric, aside from the multiplicative factor $r_{i}$ which is independent of the forwarding links' states, it is completely similar to the classical PDP-based metrics.

\item {\it Distance Progress (DP), Expected Distance Progress ($EDP^{*}$)\footnote{The * superscript is intended for distinction with the EDP metric.} \cite{P_2019_2_192147}:}
Assuming an Underwater Optical Wireless Network (UOWN) environment where each node is aware of its location information as well as its one-hop neighbors', \cite{P_2019_2_192147} introduces the following geographical-advancement metric: 
\begin{equation}
DP(i , j) = ||d_{src}-d_{dst}|| - || d_{j} - d_{dst}||, ~\forall j \in \boldsymbol{F}_{i}~~ , 
\end{equation} 
to prioritize the members of node $i$'s forwarder list along the transmission from $src$ to $dst$. Later in \cite{P_2019_2_192147}, $DP(i,j)$ is amended by averaging over all members of the node $i$'s forwarder list, $\boldsymbol{F}_{i}$:
\begin{equation}
EDP^{*}( i, F_i ) = \sum_{j \in F_{i}} DP (i,j) P(i,j)~~~.
\end{equation}
DP and EDP${}^{*}$ are very much similar to, with minor differences, the geographical advancement definition in eq.~(\ref{eq12}) and the EPA metric in eq.~(\ref{eq11}), respectively.

\end{itemize}

\subsubsection{\bf Single-Path/Per-Hop OR Metrics:} 
This class of metrics does not benefit from the broadcast nature of the wireless medium in their definitions. Consequently, they can be equally embedded in traditional and opportunistic routing protocols. The per-hop attribute of this class of OR metrics limits the span to two neighboring nodes. As an example, traditional definitions of the trust involve mutual transactions between two nodes, so, in the context of wireless networks, it can only be defined between two wirelessly-neighbor nodes. As a general statement, it is fair to say that the trust-based metrics are most likely single-path/per-hop metrics.
\begin{itemize}

\item{ \it Metric of the Most Probable Path protocol (M-MPP)\footnote{The abbreviation is from the authors of this survey for better reference.} \cite{P_60_17}:}
This metric measures the degree to which a link capacity can satisfy a connection request in bits/second. For a given CR connection request of rate demand $D$ (in bits/second), the probability that channel $k$ can support this demand, using Shannon theorem, is given by:
\begin{equation}
	\label{eq196}
	P[C^{k}_{ij} \ge D] = P\bigg[S^{(k)}_{I,j} \leq \frac{S^{(k)}_{RX,j}}{2^{\frac{D}{B^k}} -1} - G_0 \bigg]~~~,
\end{equation}
where $B^k$ is the channel $k$'s bandwidth, $S^{(k)}_{RX,j}$ and $ S^{(k)}_{I,j}$ represent the received signal power from node $i$ at node $j$ and the total primary-to-CR interference power at CR node $j$, respectively, and $G_0$ denotes the white Gaussian noise power spectral density.  
	
\item {\it Trustworthiness and ETX (E2TX)\cite{P30}:}
The ETX metric estimates the PDP of a wireless link using periodic measurements and then trusts these measurements for the next 5 or 10 minutes. In E2TX, the trustworthiness of the ETX metric is revisited. The trust index of node $j$ assigned by node $i$ after the n-th topology update cycle is defined as follows:  
	\begin{equation}
	\label{eq16}
	Trust(i,j,n)= \frac{\theta_{ij}(n)}{N_{ij}(n)}~~~,
	\end{equation}
		where $N_{ij}(n)$ is the number of packets forwarded by node $i$ toward node $j$ in time slot $n$. $\theta_{ij}(n)$ is the number of packets correctly received by node $j$ in the same time slot. After each new observation in the $(n+1)$-th topology update cycle, $Trust(i,j,n)$ is calculated by using a moving average window model similarly to the RTT computation in TCP (RFC2988 \cite{P311}): 
	\begin{equation}
	\resizebox{0.90\hsize}{!}{$
		\label{eq17}
		Trust(i,j,n+1) \longleftarrow  {\alpha . Trust(i,j,n)}+ {(1- \alpha) . Trust(i,j,n+1)}~~,	$%
	}
	\end{equation}
	where $\alpha$ is a positive weighting factor smaller than 1. Finally, the E2TX metric is defined as follows: 
	\begin{equation}
	\label{eq18}
	E2TX(i,j,n) = (1-Trust(i,j,n)) . ETX(i,j,n)~~~.
	\end{equation}
	In this equation, the E2TX for a trusted link is less than the original link's ETX. Trust opportunistic multicast Routing (TOMR) \cite{P31} has used the E2TX metric to design a secure multicast protocol.

	\item{ \it Probability of Success (PoS) \cite{P_60_13,P_70_2}:}
	PoS defines the probability of the channel availability period being greater than or equal to the required transmission period between two neighboring nodes in a specific channel. For two neighboring nodes $i$ and $j$ communicating in channel $k$, the metric is equal to:         
	
	\begin{equation}
		PoS^k(i,j) = P[ T_{avl}^k(i,j) \ge T_{TX}^k(i,j) ]~~~.
	\end{equation}
	wherein $T_{TX}$ denotes the transmission time and the channel availability time period, $T_{avl}$, is assumed to follow an ON/OFF state model.

	\item {\it Dynamic Forwarding Delay metrics (DFD${}_1$ \cite{P_2019_4_010034}\cite{P_2019_4_010059}, $DFD_2$ \cite{P_2019_4_010041}, $DFD_3$ \cite{P_2019_4_010053}, and $DFD_4$ \cite{P_2019_4_011450}) \footnote{The index is intended for distinction between different DFD metrics.} :} These are a family of very similar OR metrics introduced in different wireless network types by almost common authors. The original forms of the metrics suffer from the poor choice of notation, and we, herein, reformulate them into as consistent as possible forms. The objective of these metrics is to minimize the forwarding delay by considering a linear combination of multiple factors, the link quality, the geographical advancement, the residual energy, and the queue population of the forwarding relays. \\
	In the context of WANETs \cite{P_2019_4_010034} and IoT \cite{P_2019_4_010059}, DFD${}_1$ is defined as:
	\begin{equation}
	\label{eq-DFD11}
	\begin{aligned}
	DFD_1(i,j)=&( \alpha_1 f_1(E_{res_j})+ \alpha_2 f_2(LQI(i,j))+\\ & \alpha_3 f_3(d(j,dst))).DFD_{max}~~~~~~~~,
	\end{aligned}
	\end{equation}
	in the context of MANETs \cite{P_2019_4_010041}, DFD${}_2$  is defined as:
	\begin{equation}
	\label{eq-DFD22}
	\begin{aligned}
	DFD_2(i,j)=&( \alpha_1 f_1(E_{res_j})+ \alpha_2 f_2(LQI(i,j))+\\ & \alpha_3 f_3(d(j,dst))+ \\ &\alpha_4 Live(i,j)).DFD_{max}~~~~~,
	\end{aligned}
	\end{equation}
	and in the context of WSNs \cite{P_2019_4_011450}, DFD${}_4$  is defined as: 
	\begin{equation}
	\label{eq-DFD4}
	\begin{aligned}
	DFD_4(i,j)=&( \alpha_1 f_2(LQI(i,j))+\\ & \alpha_2 f_3(d(j,dst))+ \\ &\alpha_3 EOF(j)).DFD_{max}~~~~~.
	\end{aligned}
	\end{equation}
	In the above equations, DFD${}_{max}$ is the predefined maximum delay allowed on each link, the sum of $\alpha_i$s totals 1, and the contributing functions are defined as follows:
	\begin{equation}
	f_1(E_{res_j})= \left\{
	\begin{array}{ll}
	\frac{E_{init}-E_{res_j}}{E_{init}};~~ E_{res_j}\ge E_{min}\\
	0~~~~~~~~~~~~; ~~E_{res_j}<E_{min}~~~~~~,
	\end{array}
	\right.
	\end{equation}
	\begin{equation}
	\begin{aligned}
	& f_2(LQI(i,j))=\\& \left\{
	\begin{array}{ll}
	1~~~~~~~~~~~~; ~~LQI(i,j)<LQI_{bad}~~\\
	\frac{LQI_{max}-LQI(i,j)}{LQI_{max}}; LQI_{bad}\le LQI(i,j)<LQI_{good}\\
	0~~~~~~~~~~~~; ~~LQI(i,j) \ge LQI_{good}~~,
	\end{array}
	\right.
	\end{aligned}
	\end{equation}
	\begin{equation}
	f_3(d(j,dst))= \left\{
	\begin{array}{ll}
	\frac{2R-d(j,dst)}{2R};~~ d(j,dst)\ge R\\
	0~~~~~~~~~~~~; ~~d(j,dst)<R~~~~~~,
	\end{array}
	\right.
	\end{equation}
	\begin{equation}
	LIVE(i,j)=\frac{1}{\frac{180-\theta(j)}{180}.T_{LV}(i,j)}~~,
	\end{equation}
	and
	\begin{equation}
	EOF(j)=\frac{EDR(j)}{E_{res_j}}~~,
	\end{equation}
	wherein $f_1$ and Energy Objective Function (EOF) reflect the effect of the residual energy, $f_2$ represents the link quality,  and $f_3$ and $LIVE$ represent the impacts of the geographic advancement and mobility. The parameters Link Quality Indicator ($LQI(i,j)$) \cite{P_2019_4_021403},  $\theta(j)$, $T_{LV}(i,j)$, and Energy Drain Rate ($EDR(j)$) denote the link $ij$'s  quality measure, the angle between the node $j$'s moving direction and the virtual line connecting the source and the destination, the stability period of link $ij$, and the rate at which node $j$ loses its energy, respectively.  \\
	In a similar way, except in the context of video dissemination over FANETs \cite{P_2019_4_010053} and with differently defined constituent functions, DFD${}_3$ is defined as:
	\begin{equation}
	\begin{aligned}
	DFD_3(i,j)=&(  \alpha_1 f_4(LQI(i,j))+\\ & \alpha_2 f_5(d(j,dst))+\\ &\alpha_3 f_6(Q(j))).DFD_{max}~~~~~,
	\end{aligned}
	\end{equation}
	in which the constituent functions $f_4$, $f_5$, and $f_6$ represent the link quality, the geographic advancement, and the relaying node's congestion status, respectively. The functions $f_4-f_6$ are:
	\begin{equation}
	f_4(LQI(i,j))=\frac{1}{1+e^{-c_1 (LQI(i,j)-c_2)}}~~,
	\end{equation}
	\begin{equation}
	f_5(d(j,dst))=\frac{1}{1+e^{-c_3 (d(j^{'},dst)-R)}}~~,
	\end{equation}
	\begin{equation}
	f_6(Q(j))=\frac{1}{1+e^{-c_4 (Q(j)-\frac{Q_{max}}{2})}}~~,
	\end{equation}
	in which $j^{'}$ is the intersection between the node $j$'s communication range boundary and the $src-dst$ virtual connecting line, $Q(j)$ is the node $j$'s queue population, the sum of $\alpha$s totals 1, and $c$s are the adopted models' constants.

	\item {\it M-SGOR\footnote{The abbreviation is from the authors of this survey for better reference.} \cite{P_2019_3_260013}:}
	This metric prioritizes neighboring nodes (i.e., $i \rightarrow j \in \boldsymbol{C}_i$) based on their geographical advancements toward the destination and total trust levels (a linear combination of direct and indirect trust), in insecure WSNs. The metric considers a linear combination of advancement and trust as follows:
	\begin{equation}
	\text{\it M-SGOR}(i,j)=\beta (1-\frac{d(j,dst)}{d(i,dst)}) + (1-\beta)Trust_{ij}~~,
	\end{equation}
	wherein varying $\beta$ from $0$ towards $1$ gradually favors the advancement over the trust.

\item {\it Reliable Trust-based OR (RTOR), TORDP, GEOTOR:} \cite{P_50_21} and \cite{P_60_10} consider the trustworthiness of a neighboring node as well as its link reliability by defining a new metric, $RTOR$. Each node $i$ calculates its neighboring node $j$'s RTOR according to the following:
\begin{equation}
\label{eq182}
RTOR (i,j)= Trust_{ij} ~ . ~ p_{ij}~~,
\end{equation}
where $Trust_{ij}$ denotes the trust value ($\in [0,1]$) of node $j$ from node $i$'s perspective. The trust value is updated periodically based on the number of uncooperative interactions between node $i$ and node $j$. This local metric is then used for selecting the candidate forwarder set~\cite{P_50_21}. Assuming that the geographical locations of all the nodes are available, $TORDP$ limits the candidates set's size to include only members that can contribute more than a predetermined minimum geographical advancement (distance progress), $MinDP$, towards the final destination. Therefore, $TORDP$ is defined as:
\begin{equation}
\label{eq183}
\resizebox{0.7\hsize}{!}{$%
	TORDP(i,j)=\left\{
	\begin{array}{@{}lll@{}}
	Trust_{ij} ~ . ~ p_{ij} &; &{DP(i,j,dst)} \geq MinDP \\
	0 &; & \text{otherwise}
	\end{array}\right.%
	$ %
}
\end{equation}
where $DP(i,j,dst)$ represents the additional geographical advancement towards the final destination obtained by going from node $i$ to node $j$.\\ 
Finally, $GEOTOR$ incorporates the distance progress, $DP(i,j,dst)$, into an OR metric, as below: 
\begin{equation}
\label{eq184}
GEOTOR (i,j)= Trust_{ij} ~ . ~ p_{ij}~.~{DP(i,j,dst)}~~~.
\end{equation}

\item {\it Energy and Trust Aware Opportunistic routing (ETOR) \cite{P_2018_13}:}
This metric is defined in terms of OR in slotted Cognitive Radio Social Internet of Things (CR-SIoT). ETOR comprises three factors, namely social tie, energy consumption, and trust. The metric measures the social tie and trust between a node $i$ and a member of its next-hop candidate forwarder set $j$ during slot time $T$. Moreover, ETOR accounts for the residual energy of that member, as follows:  
\begin{equation}
\label{eq478}
\resizebox{0.7\hsize}{!}{$
ETOR(i,j,T)= \alpha_1 Trust_{ij}(T) + \alpha_2 E_{res_{j}}(T) +\alpha_3 ST_{ij}(T)~.
$%
}
\end{equation}
In eq.~(\ref{eq478}), $Trust_{ij}$ and $ST_{ij}$ represent the total trust (direct and indirect) and the social tie between nodes $i$ and $j$, respectively. $E_{res_{j}}$ accounts for the residual energy of node $j$. The parameters  $\alpha_1$, $\alpha_2$ and $\alpha_3$ provide a weighted average over these three factors.

\item {\it Opportunistic Routing based Distance Progress (ORDP), Expected Packet Progress (EPP) \cite{P_2019_3_280633}:}
These metrics are proposed for UWSNs:
	\begin{equation}
	ORDP(i,dst)= d(src,dst)-d(i,dst)
	\end{equation}
	and
	\begin{equation}
	EPP(i,j)=ORDP(j,dst).E_{res_{i}}.p_{ij}~~~.
	\end{equation}
	The geographical progress of a node is inherently an end-to-end metric; However, there is nothing new about ORDP except that, as opposed to the traditional definition of geographical progress, it becomes bigger by the node getting closer to the destination. The second metric combines the geographical progress, the residual energy, and the PDP of the next hop link.

	\item {\it Four Distributions (4-D)\footnote{The abbreviation is from the authors of this survey for better reference.} \cite{P_2019_3_290326,P_2019_10_061532}:} 
	By adopting 4-D metric, the OR protocol in \cite{P_2019_3_290326}, and later in \cite{P_2019_10_061532}, try to assign higher priorities to the relay nodes which are geographically further away from the source, closer to the destination, vertically closer to the virtual line connecting the source to the destination, and finally have higher residual energy available. The 4-D metric achieves the latter by multiplying the appropriate probability distribution function corresponding to each factor:
	\begin{equation}
	\text{\it 4-D}(i,j)= \boldsymbol{d}(i,j).\boldsymbol{\theta}(i,j).\boldsymbol{dp}(i,j).\boldsymbol{E}(i,j);  ~\forall j \in \boldsymbol{C}_{i}~~,
	\end{equation}
	in which $\boldsymbol{d}(i,j)$, $\boldsymbol{\theta}(i,j)$, $\boldsymbol{dp}(i,j)$, and $\boldsymbol{E}(i,j)$ are the transmission-distance distribution, the direction distribution, the perpendicular-distance distribution, and the residual energy distribution, respectively. Despite of the claim that the 4-D metric is an end-to-end metric, it looks like to be otherwise.

	\item {\it Routing Metric (RM) \cite{P_2019_3_280030}:}
	In the context of WSNs for IoT, this metric, which apparently does not offer any specific novelty, linearly combines the one-hop geographical progress times the one-hop link trust, and the residual energy of the next-hop node as follows:
	\begin{equation}
	\begin{aligned}
	RM(i,j)=&\alpha (d(i,j).Trust_{ij}) +(1-\alpha)E_{res_i}~;\\ & ~\forall j \in \boldsymbol{F}_{i},~~ 0<\alpha<1~,
	\end{aligned}
	\end{equation}
	wherein the $Trust$ itself is a linear combination of the current and the historic trusts as in eq.~(\ref{eq17}).

	\item {\it Quality-Advancement-Density (QAD) \cite{P_2019_5_181545}:}
Very similar to the SQA metric and from the same authors, QDA is defined for VANETs as follows:
\begin{equation}
QAD(i,j)= \frac{p_{ij}. a_{ij}}{X_j} ~~~~,
\end{equation}    
wherein $a_{ij}$ denotes the relative packet advancement, and $X_j$, the density of neighboring nodes around node $j$, has replaced the stability index in SQA. QAD favors forwarders in dense neighborhoods. Due to frequent changes in VANETs, the QAD metric should be updated accordingly. 

\item {\it Energy-aware Coded Opportunistic Routing (ECOR) \cite{P_2019_12_051127}, Obstacle aware Opportunistic
	Data Transmission (OODT) \cite{P_2020_06_230000}:} This metric is defined in the context of CR-SIoT from almost the same authors who introduced ETOR in eq.~(\ref{eq478}). The metric considers the impact of social ties between the users of secondary wireless devices regarding OR in time-slotted CRNs. ECOR is the weighted summation of three components as:

\begin{equation}
	\begin{aligned}
ECOR(i,j,T)= & \alpha_1 ETX_{ij}(T) + \alpha_2 E_{cons}( i,C_i, T)+\\& \alpha_3 ST_{ij}(T)~~~,
\end{aligned}
\end{equation}
wherein $E_{cons}(i,C_i,T)$ is the energy consumed for successfully transmitting one packet from node $i$ to its downstream nodes, $C_i$, during the time slot $T$, and the other two components, $ETX_{ij}(T)$ and $ST_{ij}(T)$, are as described in ETOR. The ECOR metric is then used by an OR protocol, with the same name, to choose the primary and backup candidate sets, through an auction-model-based selection strategy. \\
The OODT \cite{P_2020_06_230000} metric, while claimed to be a new one, is almost equivalent to the ECOR metric from the same authors a year earlier. The only discrepancy seems to be in their final weighting relationships in which the last proportionality looks more reasonable in the earlier version.     

\item{Traffic-Differentiated Secure Opportunistic Routing (DSOR) \footnote{The abbreviation is from the authors of this survey for better reference.} \cite{P_2020_03_271947}:}
In the context of MANETs, TDSM quantifies a node's forwarding capability (price), which is then used by an auction-based forwarder list selection to guarantee the QoS of different types of traffic flows in an adversarial network scenario. The metric considers the node's trust level (i.e., direct and indirect), available resources (i.e., energy and bandwidth), and service potential (i.e., ETX) for each type of flow. It is defined as the summation of two prices (or equivalently, costs), resources ($c_{rsc}$) and service ($c_{srv}$) as follows:

\begin{equation}
TDSM(i,j,k)=c_{rsc }(j)+c_{srv} (i,j,k)~~~~,
\end{equation}
with:
\begin{equation}
\begin{aligned}
& c_{rsc} (j,k)=\\  &\frac{ \frac{1}{2} (\frac{E_{tot_j} - E_{res_j}}{E_{tot_j}} + \frac{E_{cons_j}}{E_{res_j}}) + \frac{k_{max}}{k_{max} -k+1} .(\log_2 (1+ \frac{B_{new_j}}{B_{tot_j} - B_{cur_j}})) }{1+ \frac{k_{max}}{k_{max} -k+1}}
\end{aligned}
\end{equation}
and

\begin{equation}
c_{srv }(i,j,k) = \frac{(1-Trust(i,j)) + \frac{k}{k_{max}} . \frac{ETX(j,dst)}{ETX(i,dst)}}{1+\frac{k}{k_{max}}} ~~~,
\end{equation}

wherein $E_{tot_j}$, $E_{res_j}$, and $E_{cons_j}$  are node $j$'s initial energy, residual energy, and required forwarding energy, respectively. $B_{new_j}$, $B_{tot_j}$, and $B_{cur_j}$ denote node $j$'s new task bandwidth requirement, total bandwidth available, and current bandwidth consumed, respectively. Finally, $k_{max}$ and $k$ represent the highest priority in the traffic flows and the priority of the $k$-type traffic flow. 

\end{itemize}

\par
 We conclude this section with the following remarks:
\begin{itemize}
\item We have attempted to make the treatment of the OR metrics as exhaustive as possible. While there may be a slight chance that a few metrics are missing, we intentionally chose not to mention a couple of metrics whose credibility could not be verified.
\item Our focus, herein, is on wireless networks only, and we opted not to mention OR metrics in other types of networks, for instance, the one in Multi-Layer Social network based Opportunistic Routing \cite{P_60_16}.
\item Due to many differences in the OR metrics, in terms of the employed parameters, the computation method/scope, the features, the target platforms, and the application scenarios, it is impossible to compare their resulting network performances. However, a multitude of simulations comparing the main representatives of OR metric classes is presented in section~\ref{sec4}.   
\item There are several OR protocols in the literature which do not explicitly mention their embedded OR metrics, for instance: MOR\cite{P_2019_3_081656}, Max-SNR \cite{P_2019_3_081719}\cite{P_2019_3_051456}, BAOR \cite{P_2019_3_081847}, EnOR \cite{P_2019_3_081853}, OSTD \cite{P_2019_3_081913}, ELECTION \cite{P_2019_3_081923}, EECOR \cite{P_2019_3_082109}, I-AREOR \cite{P_2020_07_031859}, GEDAR \cite{P_2020_07_031851}, ODYSSE \cite{P_2019_3_082115}, ENS-OR \cite{P_2019_3_082120}, EasyGo \cite{P8_1_1}, CBRT \cite{P_2019_3_082240}, GOCR \cite{P_2019_3_090033}, ECS-OR \cite{P_2019_3_090040}, Parallel-OR \cite{P_2019_3_092101}, CITP \cite{P_2020_07_031845}, SOCGO \cite{P_2019_10_061536}, EEOR \cite{P_2019_3_092109}, Dice \cite{P_2019_3_092121}, AREOR \cite{P_2019_10_041116}, RPSOR \cite{P_2020_07_031830}, POR \cite{P_2019_3_092126},PCR \cite{P_2019_10_301202}, CAOR \cite{P_2019_3_092136}, and PCon \cite{P_2020_03_280106}.
\end{itemize}

\begin{figure*}[t!]
	\centering
	\includegraphics[width=130mm]{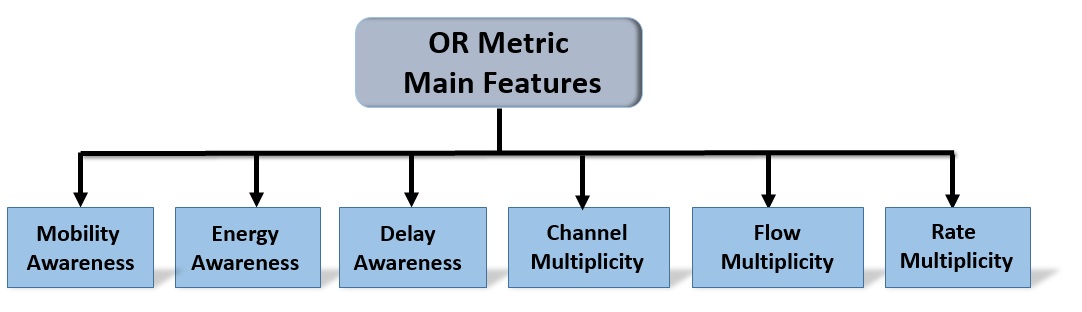}
	\caption{The main features of OR metrics that determine the character of the target networks.}
	\label{fig:fig23}
\end{figure*}

\begin{table*}[!b]
	\vspace{-20pt}
	\renewcommand{\arraystretch}{2}
	\caption {OR Metrics: Quick Reference } \label{table:mostafa}
	\centering
	\resizebox{1.0 \textwidth}{!}{
		\begin{tabular}{|c|c|c|c|c|c|c|c|c|c|c|c|}
			
			\cline{4-12}
			
					\multicolumn{3}{c|}{\multirow{2}{*}{}} & \multicolumn{6}{c|}{\large\bf Features} &
					\multirow{2}{*}{\large\bf Target Platform} &\multirow{2}{*}{\large\bf Related Protocol  }&	\multirow{2}{*}{\large\bf  Other Notable Features }  \\ \cline{1-9}
						
		{\large\bf Computation }&{\large\bf Computation }&{\large\bf Routing Metrics}&   {\large\bf Mobility }  & {\large\bf Energy } & {\large\bf Delay }& {\large\bf Channel }& {\large\bf Flow } & {\large\bf Rate } &     &  & \\ 
		
		{\large\bf Scope}& {\large\bf Method}&	&{\large\bf Awareness} & {\large\bf Awareness}& {\large\bf Awareness} &{\large\bf Multiplicity} & {\large\bf Multiplicity}& {\large\bf Multiplicity}&   &   &   \\ \Xhline{1.8pt}

		\multirow{34}{*}{\key{\LARGE \bf End-to-End}}& \multirow{28}{*}{\large \bf Multi-path} & {\large EAX\cite{P13,P14}}	 & &&& & &   & {\large WN}  & {\large OAPF} &  \\ \cline{3-12}
 
		& &{\large EOTX \cite{P_60_25}}	& & & & & & & {\large  WN} & {\large MORE}  &    \\ \cline{3-12}
		
		& &{\large ExACT \cite{P20}}	& &&& & &\large\tick & {\large WN}  & {\large BitSOR}&     \\ \cline{3-12}

	&&	{\large EMT \cite{P33}}		& & & & & & &  {\large WN} &  {\large LMTOR, MGOR}   &  \\ \cline{3-12}
	
	& &	{\large EATT \cite{P119}\cite{P19}}& &&&  & & \large\tick &{\large WN}  & {\large SMAF, MABF}  &   \\ \cline{3-12}

	&&	{\large OpRENU \cite{P37}\cite{P25}}  & &&& & & & {\large MANET}   & {\large OpRENU}  &    \\ \cline{3-12}	
	
	&&{\large STR \cite{P22}}	& &&& & & &  {\large WN} & {\large  FORLC} &   \\ \cline{3-12} 
	
	&&		{\large OEC\cite{P_50_17,P_50_18}}	& & \large\tick&& & & &  {\large WSN } & {\large EFFORT} &  \\ \cline{3-12}

	&&{\large EDC \cite{P21}\cite{P211}}	& &&& & &   &  {\large WSN}  & {\large ORW }&{\large Duty-cycled}\\ \cline{3-12}

	&&{\large ETC \cite{P36,P36_2_2}}	 & &&& & & &    {\large WMN}   & {\large BOR/AC, BORA} &  \\ \cline{3-12} 
	&&	{\large 	EAB \cite{P36,P36_2_2}}	 & &&& & & &{\large  WMN} & {\large BOR/AC, BORA} & {\large Bandwidth-aware}     \\ \cline{3-12}  	
	
	& &	{\large BCR \cite{P36}}	  & &&& & & &  {\large WMN}     &  {\large BOR/AC, BORA} &{\large  Bandwidth-aware} \\ \cline{3-12}
	&&{\large 	cEAX \cite{P_50_22,P_50_23,P_50_24,P_50_25,P_50_26}}&  &  & & & & & {\large WN}  & {\large ----} &{\large Correlation-aware} \\ \cline{3-12} 	
	&&	{\large MEATT \cite{P38}}	 & &&& \large\tick& &  &  {\large WMN } & {\large CAOR} &{\large Multi-radio}  \\ \cline{3-12}	
	&&{\large 	EAD \cite{P_40}} & &&\large\tick& \large\tick& \large\tick&   &  {\large WN }& {\large ----} & \\ \cline{3-12}	
	
	&&{\large EWATX  \cite{P_60_7}\cite{P_60_24}}	 & &\large\tick&& & & &  {\large WMN }  & {\large MAP } &{\large Multi-Constraint} \\ \cline{3-12}

	&&	{\large M-LLA \cite{P32}}	& \large\tick&&& & & & {\large VANET} & {\large LLA }& {\large Lifetime-aware}     \\ \cline{3-12} 
	
	&&  {\large OECS \cite{P_50_20}}	 & & \large\tick & & & & &{\large WSN}  & {\large ASSORT} &   \\ \cline{3-12}
	
	&&{\large ETC${}^{*}$ \cite{P_50_10} }	& &&\large\tick& & & & {\large WSN} & {\large EoR} & {\large Heterogenous, Duty-cycled}    \\ \cline{3-12}
	
	&&{\large ETCoP \cite{P_50_19}}	 & \large\tick &  & & & & & {\large VANET }&{\large SRPE} &{\large Correlation-aware}    \\ \cline{3-12}

	&&{\large EEP \cite{P_2018_30}}	 &  & \large\tick & & & & & {\large WSN }& {\large EDAD} &{\large Duty-cycled}    \\ \cline{3-12}

	&&{\large EEL \cite{P_50_15}}	&  &  & \large\tick & & & &  {\large UWSN} & {\large UWOR}  &   \\ \cline{3-12}

	&&{\large EC \cite{P_60_11}}	& & \large\tick & & & & &  {\large WN} & {\large REOR} &  {\large Dense Lossy Networks}   \\ \cline{3-12}
	&&{\large ECATX \cite{P_61_20}}	 & &  & & \large\tick& & & {\large CRN} & {\large SAAR}  &  {\large Spectrum-aware}  \\ \cline{3-12}
	
	&& {\large EDTT \cite{P_2020_03_280059}} & &&\large\tick& & & & {\large EH-WSN} &{\large ORDTP} &{\large slotted-time network, dynamic power}     \\ \cline{3-12}
	
	&&	{\large 	SQA \cite{P_2017_4}}	& \large\tick&&& & & & {\large VANET} &{\large ----} &{\large Lifetime-aware}     \\ \cline{3-12}

	&&	{\large EDR \cite{P_2019_3_280407}}	& &&& & & & {\large D2D} & {\large ----}   &  \\ \cline{3-12}
	
	&&	{\large EED \cite{P_2019_3_280407} }	& &&& & & & {\large D2D} & {\large ----}  &   \\ \cline{3-12}

	&&	{\large ETD \cite{P_2018_40}}	& &\large\tick&& & & & {\large IoT} & {\large PoR} & {\large Duty-cycled}   \\ \cline{2-12}

	& \multirow{9}{*}{\large\bf Single-path}&{\large ETX \cite{P6_2,P6}}	& &&& & & & {\large WN} & {\large ----} &  \\ \cline{3-12} 
	
	&&{\large mETX \cite{P_2017_5}}  & & & & & & & {\large WMN} & {\large ----}  & \\  \cline{3-12}
	
	&&{\large ENT \cite{P_2017_5}}  & & & & & & & \large WMN & {\large ---- } &   \\  \cline{3-12}
	
		&&{\large M-Markov \cite{P_2019_2_181359}}  & & & & & & &  {\large WN}& {\large MMSR} &   \\  \cline{3-12}

	&&{\large OLT \cite{P_60_12}}  & & &\large\tick & & & & {\large CRN} & {\large SAOR} & {\large Spectrum-aware} \\ \cline{3-12}
	
	&&{\large ESC \cite{P_61_21}}  & & \large\tick& & & & & {\large WSN} &  {\large EXLIOSE}& {\large Lifetime-aware }  \\  \cline{3-12}
	
	&& {\large M-ORAC \cite{P_40_19}}	& &\large\tick  & & & \large\tick& & {\large WANET} &{\large ORAC}& {\large Backlog-traffic based}  \\ \cline{3-12}

	&&{\large Q-ETX \cite{P_2018_24}}  & & & & & & & {\large CRN} & {\large CANCOR} &  \\  \cline{3-12}

	&&{\large M-SNR \cite{P_2017_3}}  & & & & & & & {\large MANET} & {\large ORGMA} &    
	\\ \Xhline{1.8pt}

		\end{tabular}
		
	}
\end{table*}

\begin{table*}
	\renewcommand{\arraystretch}{2}
	\caption* {TABLE IV: OR Metrics: Quick Reference (continued). } \label{table:mostafa2}
	\centering
	\resizebox{1.0 \textwidth}{!}{
		\begin{tabular}{|c|c|c|c|c|c|c|c|c|c|c|c|}
			
			\cline{4-12}
			
			\multicolumn{3}{c|}{\multirow{2}{*}{}} & \multicolumn{6}{c|}{\large\bf Features} &
			\multirow{2}{*}{\large\bf Target Platform} &\multirow{2}{*}{\large\bf Related Protocol  }&	\multirow{2}{*}{\large\bf  Other Notable Features }  \\ \cline{1-9}
			
			{\large\bf Computation }&{\large\bf Computation }&{\large\bf Routing Metrics}&   {\large\bf Mobility }  & {\large\bf Energy } & {\large\bf Delay }& {\large\bf Channel }& {\large\bf Flow } & {\large\bf Rate } &     &  & \\ 
			
			{\large\bf Scope}& {\large\bf Method}&	&{\large\bf Awareness} & {\large\bf Awareness}& {\large\bf Awareness} &{\large\bf Multiplicity} & {\large\bf Multiplicity}& {\large\bf Multiplicity}&   &   &   \\ \Xhline{1.8pt}

			\multirow{40}{*}{\key{\LARGE\bf  Per-Hop/Local}} &\multirow{29}{*}{\large\bf Multi-path}&{\large EOT \cite{P27} }	& && \large\tick& & & & {\large WN} & {\large GOR} & \\ \cline{3-12}
			
			&&{\large EAR \cite{P33}}	  & & & & & &\large \tick& {\large WN} & {\large LMTOR} &   \\ \cline{3-12}
			
				&&	{\large 	OETT \cite{P33}}	  & & & &  &  & \large\tick &{\large WN}   & {\large LMTOR, MGOR} &   \\ \cline{3-12}

			&&{\large EDRb \cite{P_61_24, P_2019_151}}	 & & \large\tick&& & & & {\large WN} & {\large ----}   & \\ \cline{3-12} 
			&&{\large DDR \cite{P_61_24, P_2019_151}}	 & & & \large\tick& & & & {\large WN} & {\large ----}  &  \\ \cline{3-12} 
			&&{\large MEAR \cite{P35}}	& & & & & &\large\tick & {\large WMN} &  {\large ST-B} &{\large Multicast}   \\ \cline{3-12}   
			&&{\large I-MEAR \cite{P35}}	& & & & & &\large\tick & {\large WMN} & {\large ST-B} & {\large Multicast}   \\ \cline{3-12}

			& &	{\large 	EPA \cite{P23}}  & & & & & & & {\large WN} &   {\large EGOR} &  \\ \cline{3-12}
			
			& &	{\large	OEE \cite{P23} }& &\large\tick & & & & &  {\large WN}& {\large EGOR} &  \\ \cline{3-12}
			
			&&	{\large 	EDP \cite{P24}} & & & &  & & & {\large WMN} &  {\large DPOR} & \\ \cline{3-12}
			
			&&{\large CTT \cite{P_60_14}}	& &&&\large\tick & & & {\large CRN} & {\large OCR} &   \\ \cline{3-12} 
			
			&&{\large FS \cite{P_61_22}}	 & & \large\tick & & & & &{\large WSN}  & {\large ORR} &{\large Duty-cycled}     \\ \cline{3-12}

			&&{\large espeed \cite{P_61_14}}	 & &&\large\tick& & & & {\large WSN} & {\large EQGOR }  &  \\ \cline{3-12} 	
			
			&&{\large 	PSTM \cite{P_61_23}} & &&& & & & {\large WSN} & {\large Cross-Layer OR}&  {\large  Duty-cycled}    \\ \cline{3-12} 	
			&&{\large DUER \cite{P_50_14}}		& & \large\tick  & & & &  & {\large WSN} & {\large U-OR}&    \\ \cline{3-12}

			&&{\large FE \cite{P_50_16}}	 & &  & & &  \large\tick&\large\tick & {\large WN} & {\large MRORNC} & {\large Network Coding } \\ \cline{3-12}

			&&{\large EEPA${}^{*}$ \cite{P_2018_12}}	 & &  & & &  & & {\large UWSN} & {\large PICS, PRCS} & {\large  Correlation-aware} \\ \cline{3-12}
			
			&&{\large FS${}^{*}$ \cite{P_2019_3_231559}}	 & & \large\tick & & &  & &  {\large WSN} & {\large ORR} & {\large Duty-cycled}   \\	\cline{3-12}

			&&{\large EEPA \cite{P_2019_3_240848}}	 & & \large\tick  & & &  & &  {\large UWSN} & {\large OVAR} &   \\	\cline{3-12} 
			
					&&{\large SPDR \cite{P_2018_41}}	 & &  & & &  & &  {\large CRAHN} & {\large GSOR} &   \\	\cline{3-12} 
					&&{\large PDA	\cite{P_2018_41} }& &  & & &  & & {\large CRAHN}  & {\large GSOR} &  \\	\cline{3-12} 
					
						&&{\large ORDP	 \cite{P_2019_3_280633}} & &  & & &  & &  {\large UWSN} & {\large VHGOR} &    \\	\cline{3-12} 
						
					&&{\large M-RSSI \cite{P_2018_43}}	 & &  & & &  & &  {\large WSN} & {\large ODYSSE} & {\large Duty-cycled, Network Coding}   \\	\cline{3-12} 
					&&{\large DSTT \cite{P_2019_1}}	 & &\large\tick  & & &  & &  {\large UAWSN} & {\large EBOR} & {\large Trust-based}  \\	\cline{3-12} 
					&&{\large EFR \cite{P_2019_2} }	 & &  & & &  & \large\tick & {\large CRAHN}  &  {\large LBOR} &  \\	\cline{3-12}
					&&{\large DP \cite{P_2019_2_192147}}	 & &  & & &  & & {\large UOWN}& {\large SectOR }  &  \\	\cline{3-12}
					&&{\large EDP${}^{*}$ \cite{P_2019_2_192147}}	 & &  & & &  &  &{\large UOWN} &{\large SectOR }  &  \\
					\cline{2-12}

					& \multirow{20}{*}{\large\bf Single-path}&{\large M-MPP \cite{P_60_17}}	 & &&& \large\tick& & & {\large CRN} & {\large MPP} &    \\ \cline{3-12} 	
	
		&&	{\large 	E2TX\cite{P30}} & &&& & & & {\large WN} & {\large TOR} & {\large Trust-based}  \\ \cline{3-12}
		
			&&{\large PoS \cite{P_60_13,P_70_2}}	& & & &\large\tick & & &{\large CRN}  & {\large MaxPoS} &  \\ \cline{3-12}

		&&{\large DFD${}_1$ \cite{P_2019_4_010034}\cite{P_2019_4_010059}}	 & & \large\tick & & &  & &  {\large WANET, IoT} & {\large TLG, LinGO} &    \\	\cline{3-12} 
		
		&&{\large DFD${}_2$ \cite{P_2019_4_010041}}	 & \large\tick& \large\tick & & &  & &  {\large MANET} & {\large COR} &   \\	\cline{3-12} 
		
		&&{\large DFD${}_3$ \cite{P_2019_4_010053}}	 & &  & & &\large\tick  & &  {\large FANET} & {\large XLinGO} &    \\	\cline{3-12} 
		&&{\large DFD${}_4$ \cite{P_2019_4_011450}}	 & & \large\tick & & &  & &  {\large WSN} & {\large SCAD} & {\large Duty-cycled }  \\	\cline{3-12} 
		
		&&{\large RTOR \cite{P_50_21}\cite{P_60_10}}	 & &  & & & & & {\large WN} & {\large Trust-based OR}   & {\large Trust-based}\\ \cline{3-12}		
		
		&&{\large TORDP \cite{P_50_21}\cite{P_60_10}}	 & &  & & & & & {\large WN} & {\large Trust-based OR}   & {\large Trust-based}\\ \cline{3-12}
		
		&&{\large GEOTOR \cite{P_50_21}\cite{P_60_10}}	 & &  & & & & & {\large WN} & {\large Trust-based OR}   & {\large Trust-based}\\ \cline{3-12}	
			
		&&{\large ETOR \cite{P_2018_13}}	 & &\large\tick  & & &\large\tick  & & {\large CR-SIoT} & {\large ETOR}& {\large Trust-based}  \\ \cline{3-12}

			&&{\large EPP	\cite{P_2019_3_280633}} & &\large\tick  & & &  & &  {\large UWSN} & {\large VHGOR} &   \\	\cline{3-12} 
			
			&&{\large 4-D \cite{P_2019_3_290326}}	 & &\large\tick  & & &  & &  {\large WSN} & {\large LORA} & {\large Duty-cycled}   \\	\cline{3-12}

	&&{\large ECOR \cite{P_2019_12_051127}} & &\large\tick  & & &  \large\tick& &  {\large CR-SIoT} & {\large ECOR} &  {\large Social-tie-aware} \\ \cline{3-12}  
							
			&& {\large OODT\cite{P_2020_06_230000}} & &\large\tick  & & &  \large\tick& &  {\large CR-SIoT} & {\large OODT} &  {\large Social-tie-aware} \\ \cline{3-12}  
			
			&&{\large M-SGOR \cite{P_2019_3_260013}}	 & &  & & &  & &  {\large WSN} & {\large SGOR} & {\large Trust-based}   \\	\cline{3-12} 
				
			&&{\large RM \cite{P_2019_3_280030}}	 & & \large\tick & & &  & &  {\large WSN-based IoT} & {\large SelGOR} & {\large Trust-based}   \\ 	\cline{3-12} 
			
			&&{\large DSOR \cite{P_2020_03_271947}}	 & & \large\tick & & & \large\tick  & &  {\large MANET} & {\large DSOR} & {\large Trust-based, traffic-differentiated}   \\ 	\cline{3-12}
				
				&&{\large QAD \cite{P_2019_5_181545}} &\large\tick &  & & &  & &  {\large VANET} & {\large SCAOR} &   \\  \Xhline{1.8pt}
			
		\end{tabular}
		
	}
\end{table*}

\subsection{\bf OR Metrics: A Quick Reference} \label{sec33}
\begin{figure*}[t!]
	\centering
	\includegraphics[width=1.0\linewidth]{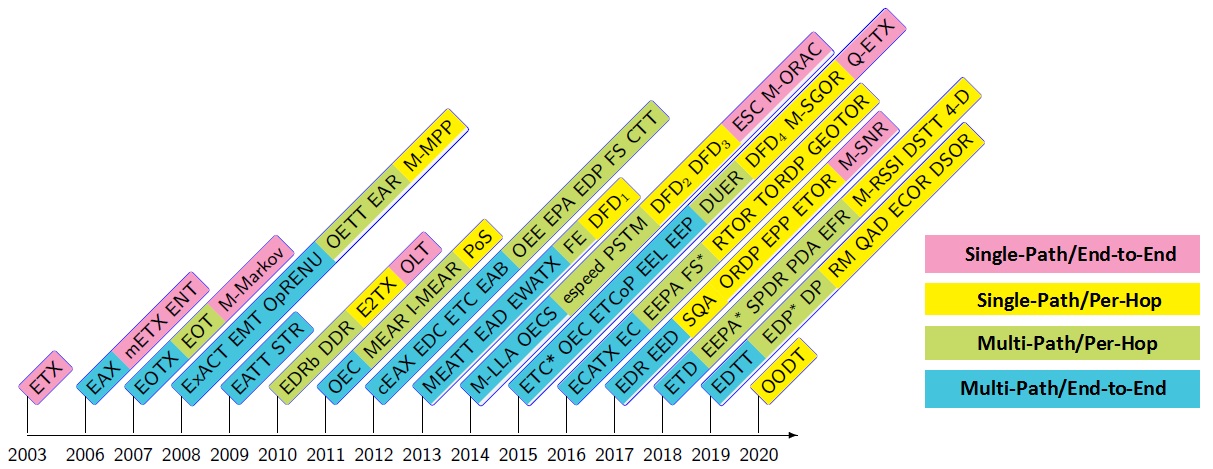}
	\caption{Chronological introduction of OR metrics.}\label{fig:fig11}
\end{figure*}

In this section, OR researchers are provided with a couple of handy and easy-to-grasp indexes of OR metrics. We start by listing and explaining the most important OR metric features that, to some extent, determine their corresponding target wireless networks (Fig.~\ref{fig:fig23}).
\begin{itemize}
\item {\bf Mobility Awareness:} Mobility (movement pattern, speed, etc.) affects the performance of the routing protocol in a network. If the OR metric employed by the routing protocol considers mobility effects, it is called a mobility-aware metric \cite{P32}. Mobility awareness narrows the target networks into mainly MANETs and VANETs.
\item {\bf Energy Awareness:} OR metrics that include energy considerations of the relaying nodes, such as the remaining energy (residual energy), the energy consumption rate, and the capability of energy renewal, are called energy-aware \cite{P_61_21, P_40_19, P_50_17, P_50_18, P_50_20}. Due to this core energy consideration, the scope of the target networks includes mainly WSNs.
\item {\bf Delay Awareness:} While the delay is an important factor in almost all the WNs, delay-sensitive applications require strict attendance to the temporal behavior of the network. The OR optimization methods applied to networks carrying this type of traffic should employ metrics that directly consider the time information of the traffic. Some potential target networks that employ delay-aware metrics are the
CR networks \cite{P_60_12}, WMNs\cite{P_40}\cite{P27}\cite{P_61_24}, and WSNs \cite{P_50_15}.
\item {\bf Channel Multiplicity:} In WNs, where the PHY and MAC layers support different frequency channels (i.e., channel selection), OR can benefit from increased degrees of freedom. Graphs can model Multi-channel networks with multi-link edges. In these networks, an appropriate channel selection is added to the problem of traditional OR routing. Some OR metrics reflect this capability \cite{P38,P_40,P_61_20}. Due to the complexity of the transceivers involved, the target nodes include WMNs and CRNs.

\item {\bf Rate Multiplicity:} Similar to channel multiplicity, provided that PHY and MAC layers support different transmission rates, OR can increase performance by adapting the transmission rate to the network status.
Wireless networks employing this feature include WMNs \cite{P38, P33, P35,P19,P119}.

\item {\bf Flow Multiplicity:} The OR mechanisms that work well when a single flow is present in the network might not do so in scenarios where multiple flows are present. In fact, increasing the number of flows results in some mutual impacts that are not traditionally reflected in single-flow metrics. Moreover, increasing the number of flows beyond one leads to inter-flow priority and queuing issues in common forwarding nodes. While the multi-flow consideration can be a concern of the OR mechanism itself, in some cases, this consideration is embedded inside the metric. Multi-flow metrics exist in two forms: metrics with different values for each flow \cite{P_40} and metrics with a single value that considers the combined effects of all flows crossing a single common node. 
In the latter case, inter-flow NC is usually also present \cite{P_50_16}. WMNs are potential candidates for employing multiple flow metrics in their OR mechanisms.
\end{itemize}
For each metric, Table \ref{table:mostafa} summarizes all the classification information from section \ref{sec3}, the features information reported earlier in this section, the target platform, and the employing OR protocol (if applicable) in a visual-friendly manner for quick referencing.  
One can easily select an OR metric from Table \ref{table:mostafa} for the routing application in hand, by merely looking for the appropriate features and target platform.  
\\The timeline of the introduction of OR metrics, illustrated in Fig.~\ref{fig:fig11}, illustrates the constant-since-birth importance of OR metrics and their evolutionary path.\\
Since the OR protocols use the OR metrics, to avoid unnecessary literature searching, Table \ref{tab:table1} lists the OR metrics with several OR protocol deployments alongside their paper references.

\begin{table*}
	\centering
	\caption{List of OR metrics employed by several OR protocols. }
	\label{tab:table1}
	\resizebox{1.0 \textwidth}{!}{
	\begin{tabular}{cc}
		\Xhline{1.5pt}
		{\bf Routing Metric} & {\bf Protocol} \\
		
		\Xhline{1.5pt}
		& EXOR \cite{P6_2}, Economy\cite{P7}, SOAR\cite{P8}, MORE\cite{P9}, CodeOR\cite{p_9_5}, XCOR\cite{p_9_6}\\
		ETX	& O3\cite{P_5_1}, SlideOR\cite{P_60_1}, CCACK\cite{p_9_4}, OMNC\cite{P_60_3}, Consort\cite{P_50_6}, MaxOPP\cite{P_5_2}, \\
		& ORPL \cite{P12}, CAOR\cite{P_60_18}, TORP\cite{P_60_19}, ORPSN\cite{P_60_20}, CAR \cite{P_2019_3_032343}, CORMAN \cite{p_9_1}\\
		& ExOR compact\cite{P_60_21}, NCOR\cite{P_60_22}, WBFL\cite{P_2019_3_032334}, E-MORE \cite{P_2020_07_031912}, ViMOR \cite{P_2019_3_081939} \\
		
		\midrule
		
		EAX &  LCAR\cite{P17}, OAPF \cite{P13} , ORDF \cite{P_2018_321},  ABF,SAF\cite{P119}\cite{P19}\\
		\midrule
		EATT & MABF, SMAF\cite{P119}\cite{P19}, CoRoute\cite{P_61_18}, CRAHN\cite{P_61_19}, PLASMA \cite{P15}, ROMP\cite{P18}, DSMA\cite{P_70_1}, COMO\cite{P_2019_3_081634} \\
		\midrule
		EDC & ORW\cite{P21}\cite{P211}, ORPL\cite{P_61_15}, ORPL-DT \cite{P_2017_7} DOF\cite{P_61_16}, Staffetta\cite{P_61_17}, COF\cite{P_2019_10_151334}\\
		\midrule
		M-MPP &  MPP\cite{P_60_17}, MaxPoS \cite{P_60_13}\\ 	
		\midrule
		PoS & MaxASA, MinTT\cite{P_60_13}\cite{P_70_2}, MPOS  \cite{P_70_3} \\ 	
		\midrule
		FS${}^*$ & ORR \cite{P_2019_3_231559}, MORR \cite{P_2019_3_240152}  \\
		\midrule
		EPA & EGOR\cite{P23}, GEDAR\cite{P_61_10}\cite{P_61_11}, HydroCast\cite{P_61_12}, GOC\cite{P_61_13}, EQGOR\cite{P_61_14}, U-OR\cite{P_50_14}  \\
		\Xhline{1.5pt}
	\end{tabular}}
\end{table*}

	\begin{figure*}[htp]
		\begin{center}
			
			\subfigure{\includegraphics[scale=0.27]{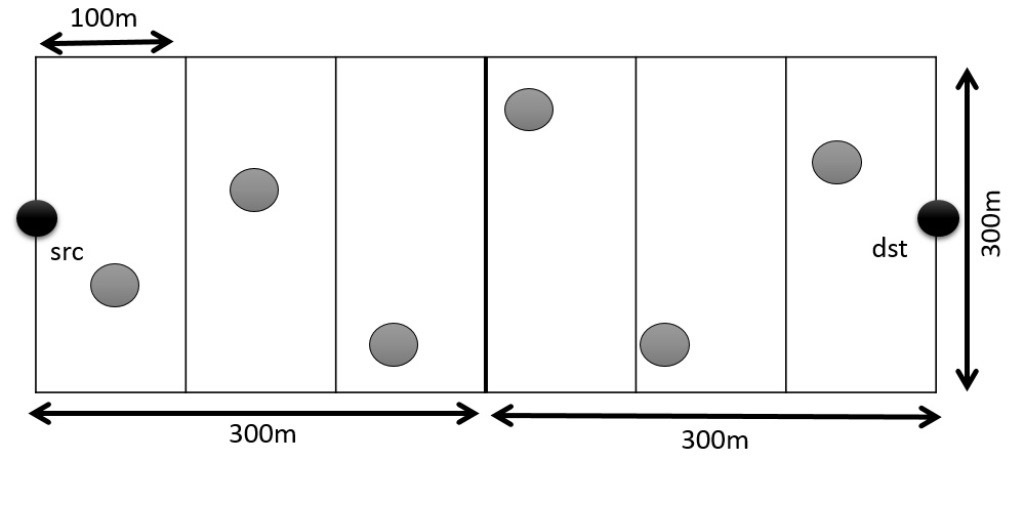}\includegraphics[scale=0.20]{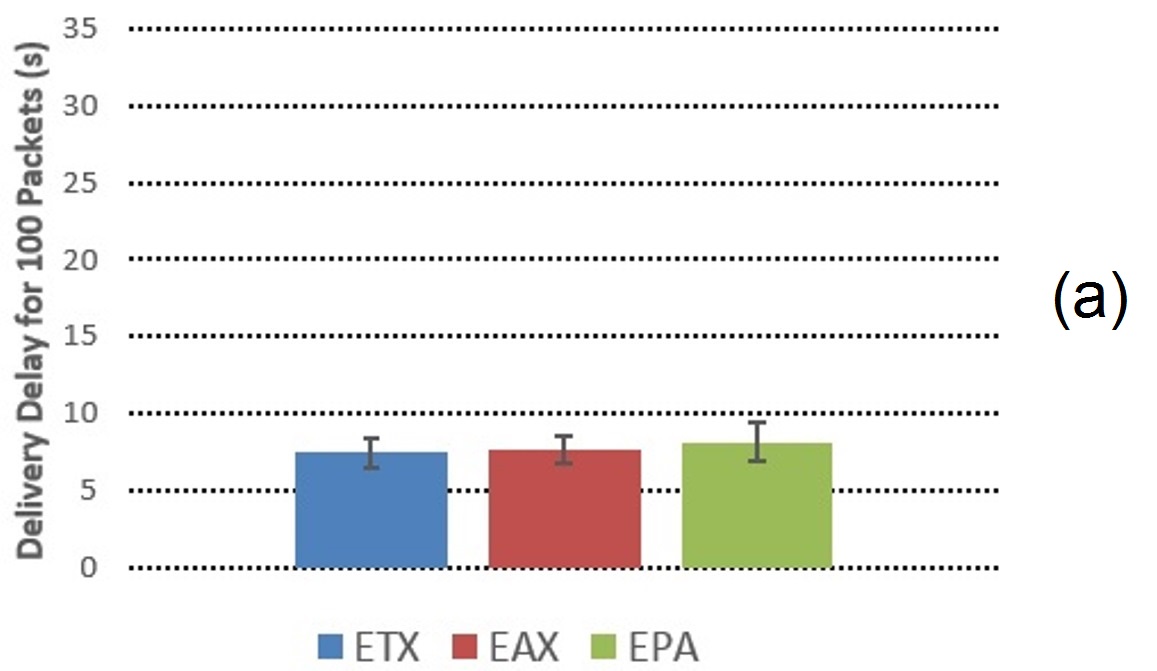} }
			\subfigure{\includegraphics[scale=0.27]{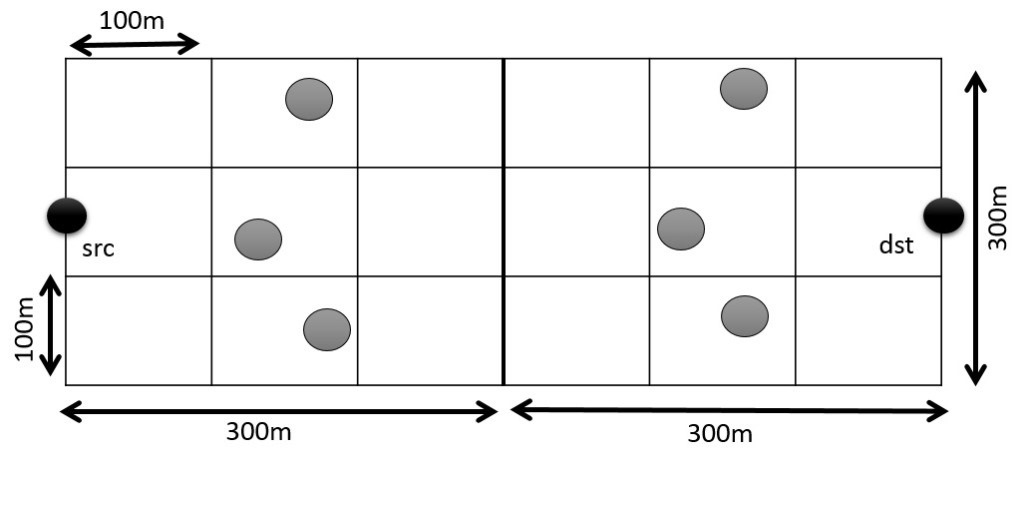}\includegraphics[scale=0.20]{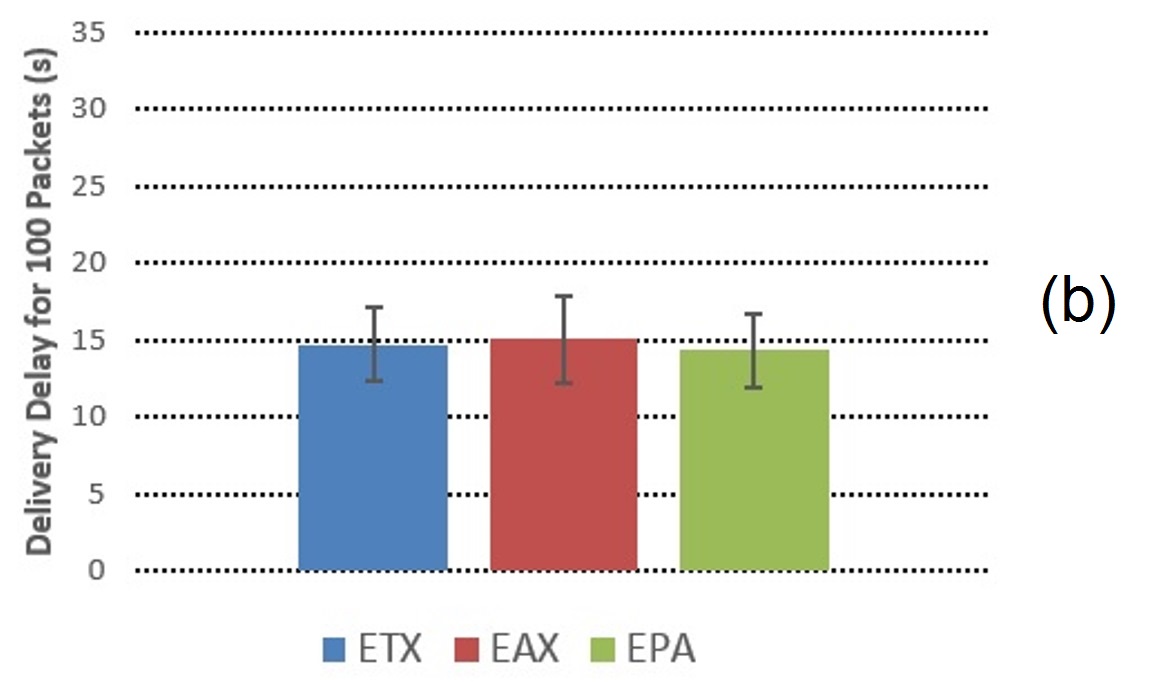}}
			\subfigure{\includegraphics[scale=0.27]{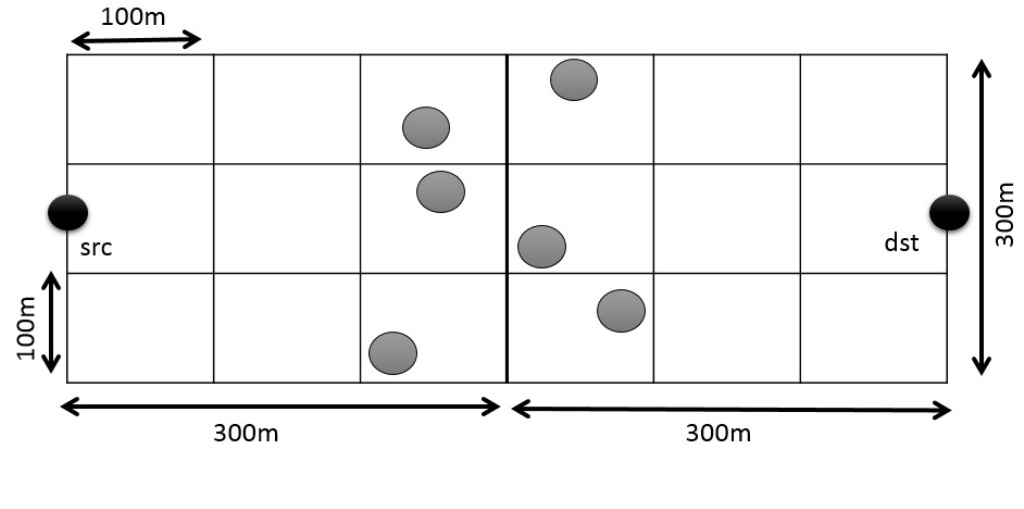}\includegraphics[scale=0.20]{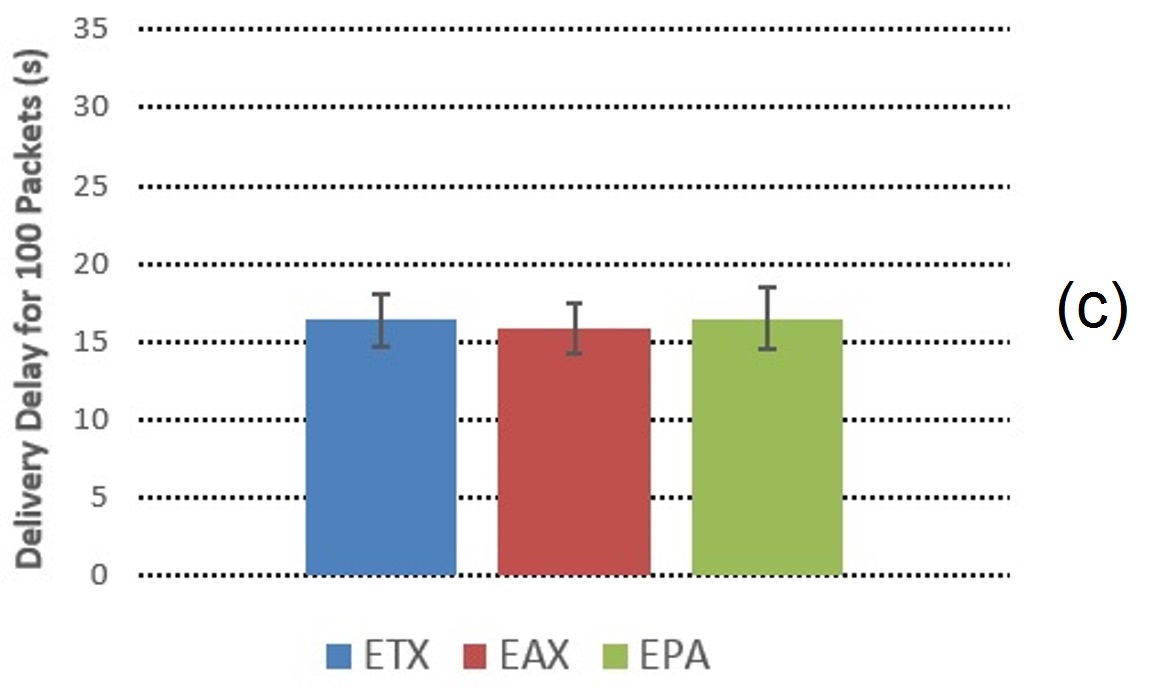}}
			\subfigure{\includegraphics[scale=0.27]{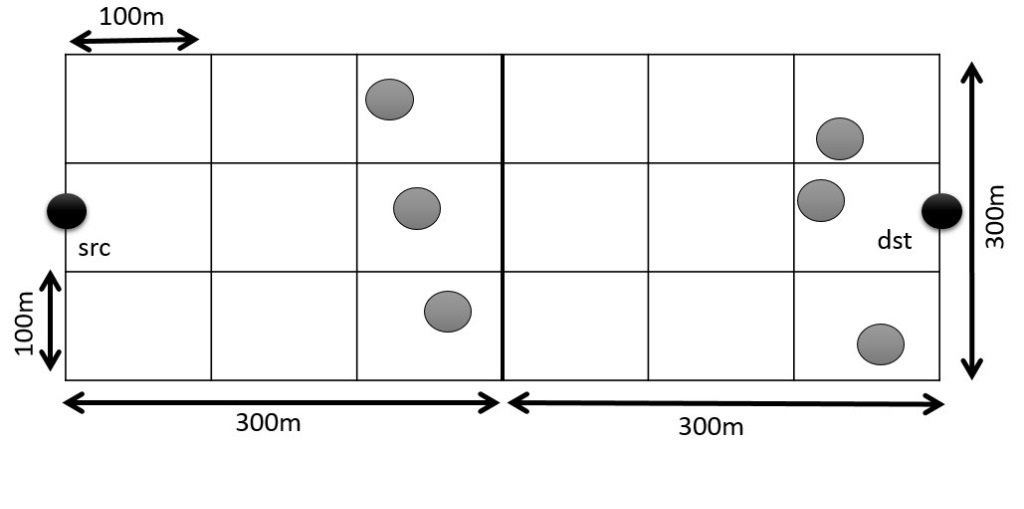}\includegraphics[scale=0.20]{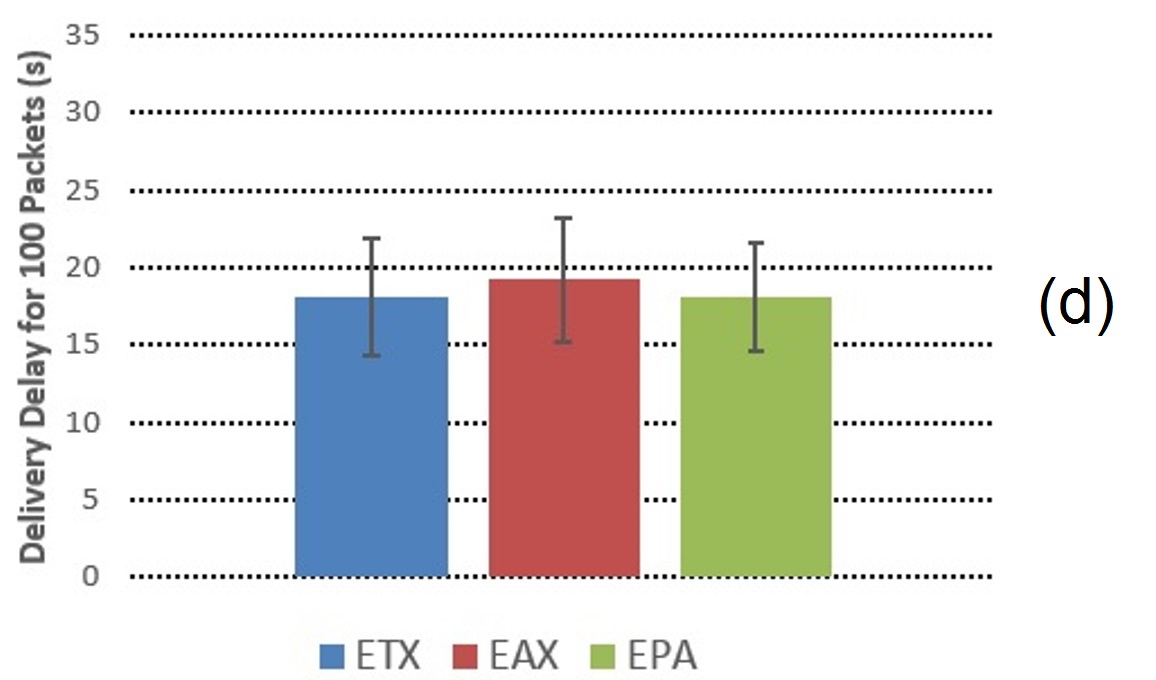}}
			\subfigure{\includegraphics[scale=0.27]{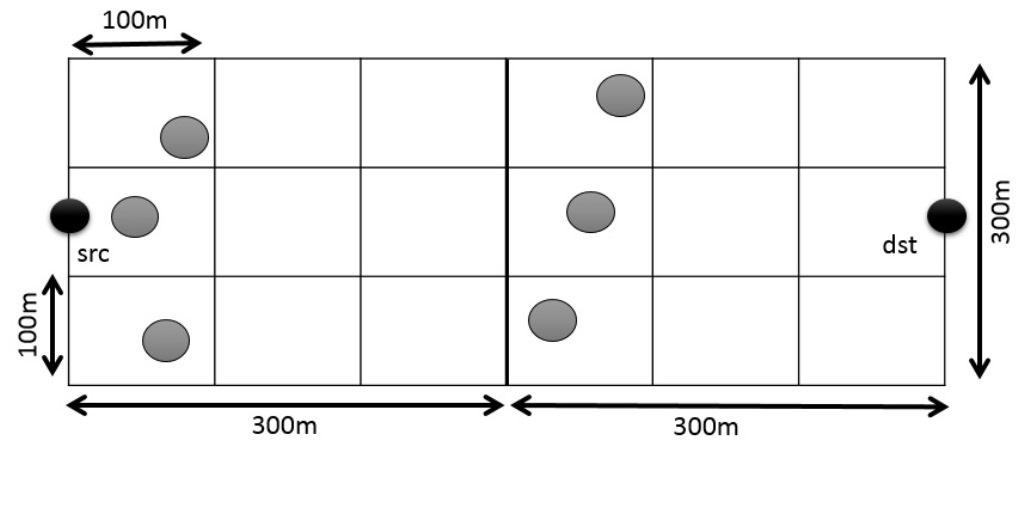}\includegraphics[scale=0.20]{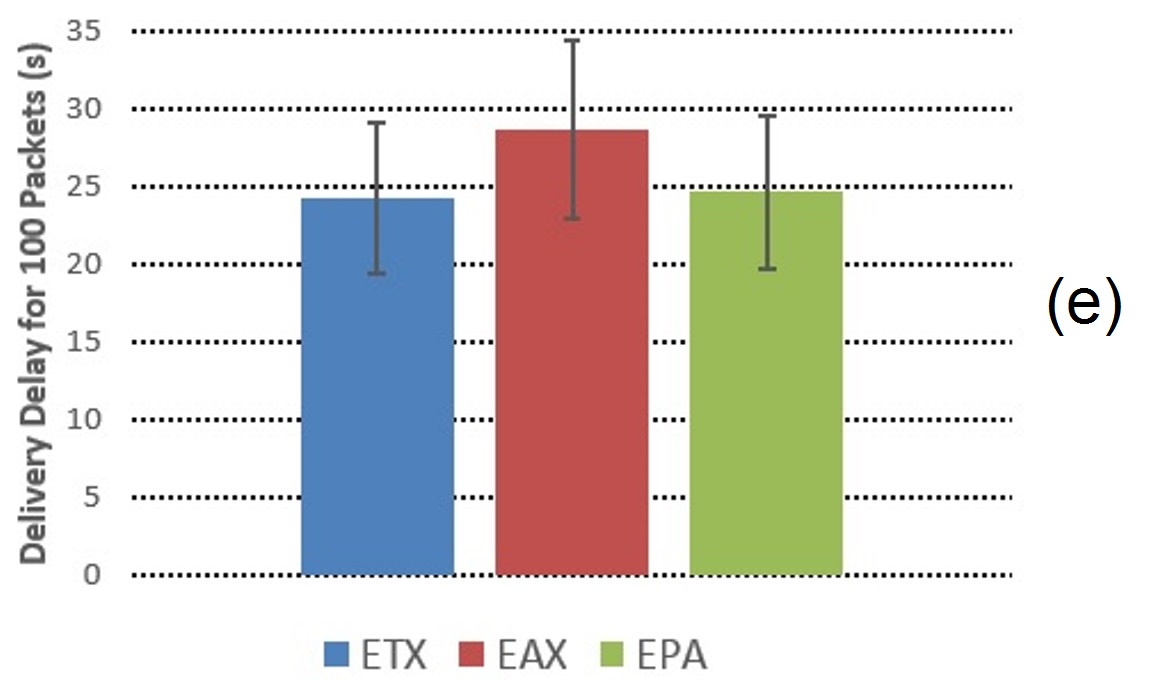}}

		\end{center}
		
		\captionsetup[subtable]{position=top}
		\caption{Investigation of how the representative OR metrics, ETX, EAX, and EPA, perform in a variety of topology scenarios with different progress patterns. Left: Topology scenarios. Right: Performance results.}
		\label{fig:24s}
	\end{figure*}
\subsection{ \bf OR Metrics: Comparative Simulations} \label{sec4}
We would like to, herein, investigate the very much critical dependency of the network performance on the choice of the OR metric, OR protocol, and the network topology. In some cases, the dependence might not be so obvious. Regarding the latter, it is interesting to see that {\it an OR protocol applied to a particular network topology might favor a specific OR metric in terms of performance.}\\          
In the following, we consider only OR metric representatives of the single-path/end-to-end, multi-path/end-to-end, and the multi-path/per-hop classes and compare their performances, through simulation, under the same OR protocol for different topology scenarios. The exclusion of the single-path/per-hop class of OR metrics is because of its obvious underperformance due to both not benefiting from the broadcast nature of the wireless medium and limited vision (i.e., not foreseeing communication holes/dead-ends and unreliable links ahead). The scenarios are intentionally selected to highlight different-in-nature topologies.\newline 
ETX, EAX, and EPA are selected as the representatives of the three OR metric classes. Although not originally introduced as an OR metric, ETX \cite{P5}, the metric used by the first and many subsequent OR protocols, is selected as the representative of the single-hop/end-to-end class. We select EAX, the first OR-dedicated metric introduced, as the representative of the multi-hop/end-to-end OR metric class. EAX defines the opportunistic anypath cost as the expected number of transmissions the source must make for the successful delivery of a packet to the destination. Most of the metrics in this class are derived from EAX  (e.g., by considering extra parameters, as in M-LLA, which considers mobility).    
EPA represents the multi-path/per-hop class of OR metrics. EPA, which defines the geographical advancement of a packet in each hop towards the destination, has inspired many other OR metrics in this class. Regarding the fair comparison noted previously, the above-selected metrics use only the PDP parameter in their definitions.  

To investigate the impact of different choices of OR metrics on network performance, we choose EXOR \cite{P6_2} as the base OR protocol. EXOR, introduced as the first OR protocol, benefits from the broadcast nature of the wireless transmission and considers multiple forwarders in a network with low-quality links. EXOR incorporates the functionality of the MAC layer into the routing protocol by scheduling the transmission times of multiple forwarders. The scheduling is achieved by prioritizing forwarders based on their ETX metrics. In EXOR, the concept of the packet is replaced by a batch, and the batch-map transmission obviates the use of ACKs.\newline
In terms of the topology, we test five different representative topology scenarios with distinguishing natures (left column of Fig.~\ref{fig:24s}). The topology scenarios have been designed to represent different progress patterns while avoiding communication holes. \newline  The EXOR protocol was implemented in the OMNET++ \cite{P_60_23} simulator using the MiXiM wireless framework. OMNET++ was chosen because of its good performance, modularity, and the ability to separate node
behavior from node parameters, where the latter feature facilitates running large parameter studies \cite{P_2017_6}. In line with our investigation purpose, the prioritizing metric in the original EXOR (i.e., ETX) is replaced by EAX and EPA, one at a time. We model the channel fading effect as proposed in \cite{P_70_5}, with R=200 m representing the wireless half-reach radius. The network is a 600 m*300 m rectangle. $src$ and $dst$ are located at the two ends of the rectangle, 600 m apart. Therefore, $src$ cannot directly communicate with $dst$, but it requires at least one forwarding node for communication. The simulation for each topology-metric pair is repeated 50 times to obtain 95\% confidence intervals. The repetitions differ concerning the random locations of the nodes inside their confining blocks. In each run (repetition), $src$ sends one batch of data packets (which includes one hundred 1500B-long data packets). We opted not to implement the EWMA transient-delay smoothing filter in the original EXOR due to its counterproductive impact in multi-hop scenarios. The total data delivery delay is measured as the difference between the transmission start time of the first data packet at $src$ and the reception end time of the last data packet at $dst$.\par
\begin{figure}
		\centering
		\includegraphics[width=70mm]{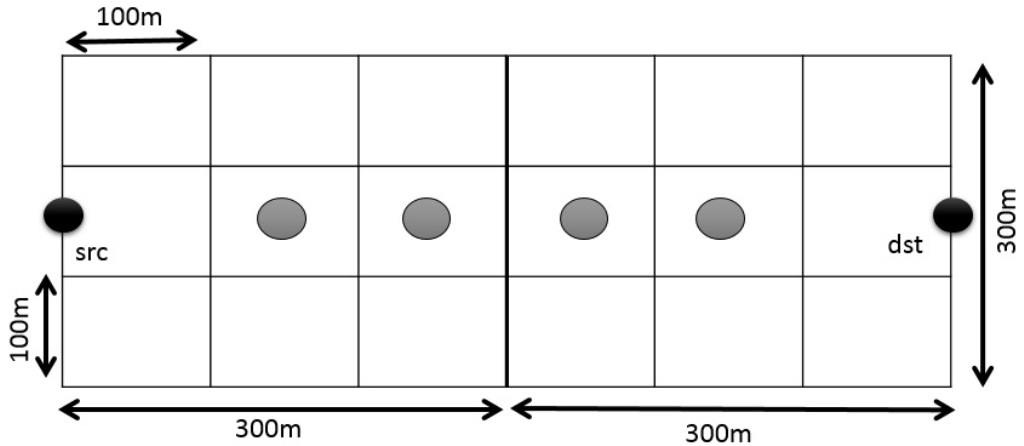}
		\caption{ A pure-longitudinal topology scenario. }
		\label{fig:245s}
	\end{figure}

Some important facts observed based on the simulation results are as follows:
\begin{itemize}
\item{For some topologies (Figs.~\ref{fig:24s}-a,b,c), the delay results are almost indifferent to the choice of metric. As an extreme case, the progress-based (EPA) and path-quality-based (ETX and EAX) prioritization (scheduling) result in the same performance in the pure longitudinal topology scenario of Fig.~\ref{fig:245s}.}
\item {The employed metrics perform differently in each topology. The illustrated topologies are arranged in order of decreasing performance. The best results are observed for the case where the forwarding nodes are arranged in increasing progress preference though being allowed to adopt random locations latitudinally within their confining blocks (i.e., the forwarding nodes are spread longitudinally across the network, Fig.~\ref{fig:24s}-a). 
This result is expected in this case because the forwarding nodes have a wide range of link qualities to the source, which allows them to contribute to the forwarding task incrementally. In fact, in this case, the scheduling concept is in the best possible way. \newline The underperformance of the topology scenario of Fig.~\ref{fig:24s}-b compared to Fig.~\ref{fig:24s}-a is partly due to the existence of one low-quality long hop and partly due to the redundant forwarding contribution role of each of the two forwarding groups.\newline The redundancy problem above is intensified in the topology scenario of Fig.~\ref{fig:24s}-c. The topology scenario of Fig.~\ref{fig:24s}-d shows slightly worse performance than that of Fig.~\ref{fig:24s}-c because of the existence of more low-quality hops. Finally, the worst performance is demonstrated in the topology scenario of Fig.~\ref{fig:24s}-e.  In this scenario, in addition to the existence of low-quality long hops, the cumulation of redundant forwarding nodes around the source (which happens to be the first-hop relay node) gives the source an  overly optimistic view of packets forwarded through the transmission of the batch-map. In fact, this causes the source  to prematurely resume the transmission of the rest of the batch, leading to a more significant number of collisions.}

\item {An interesting observation in Figs.~\ref{fig:24s}-d and ~\ref{fig:24s}-e is that the results of ETX are unexpectedly better than those of EAX. The reason for the difference is that the proximity of the forwarding nodes (i.e., forwarding contribution redundancy) makes the potential superiority of the multi-path consideration of EAX irrelevant. }
\end{itemize}
To investigate how the performance ratings of different metrics vary with differing OR protocols, we compare the delay results of the topology scenario of Fig.~\ref{fig:24s}-e for the EXOR and MORE protocols.   

\begin{figure*}[htp]
	\begin{center}
		\centering
		\subfigure{\includegraphics[scale=0.27]{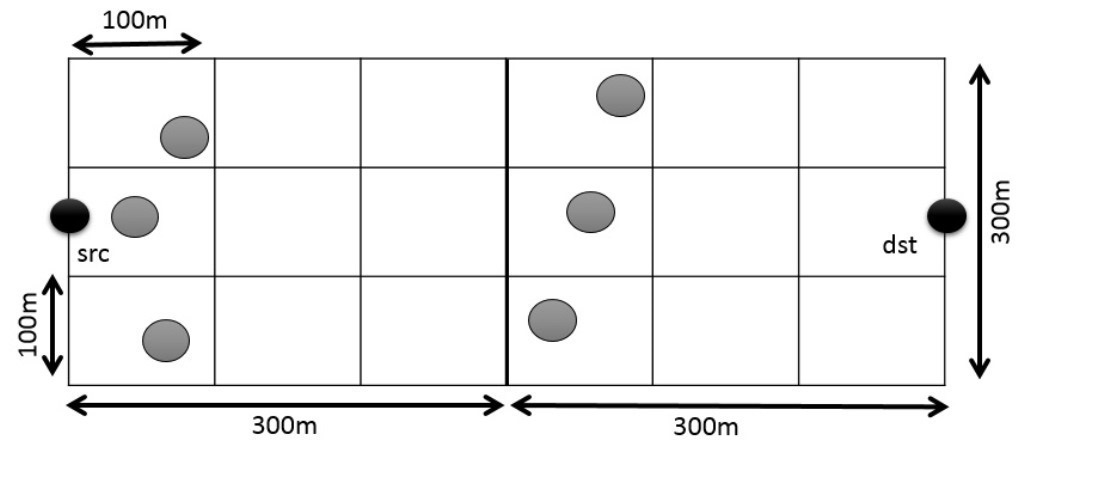}\includegraphics[scale=0.20]{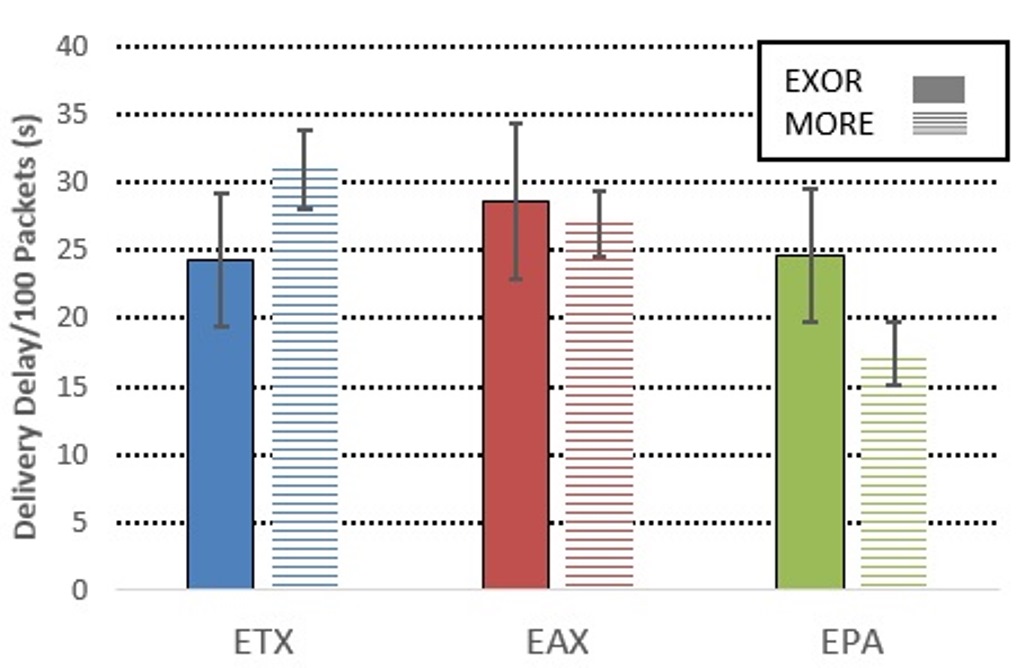}}
		
	\end{center}
	
	\captionsetup[subtable]{position=top}
	\caption{Performance comparison of two flagship OR protocols, EXOR and MORE, using three representative OR metrics, ETX, EAX, and EPA, in the challenging topology scenario of Fig~\ref{fig:24s}-e.  }
	\label{fig:266s}
\end{figure*}

MORE, a MAC-independent opportunistic
routing protocol, is one of the most-cited OR protocols. The excellent performance of MORE is due to the fact that MORE solves the coordination (scheduling) problem between forwarder nodes with the introduction of the concept of transmission credit, and addresses the problem of duplicate transmissions by forwarder nodes through NC employment and makes the use of intermediate ACKs unnecessary (the latter is taken care of by transmitting batch-map in EXOR). MORE inherently uses the ETX metric for its credit calculation.  We need to discuss MORE's credit accumulation and consumption mechanisms more closely to incorporate EAX and EPA into MORE. 

In MORE, $z(i,dst)$ is defined as the expected number of transmission attempts node $i$ should make to deliver one packet to $dst$:

\begin{equation}
	\label{eq39}
	z(i,dst) = \frac{\displaystyle\sum_{j \in \boldsymbol C_i} P(i,j)z(j,dst)}{P_{\boldsymbol C_i}}~~~.
\end{equation}

For each forwarding node $i$, $TXcredit_i$ is defined as the number of transmissions that $i$ should make to the final destination upon successful reception of an innovative (new linearly independent) coded packet from a forwarding node with a higher ETX. 
Each successful reception increases the transmission credit of $i$ by $TXcredit_i$:

\begin{equation}
	\label{eq40}
	TXcredit_i = \frac{z(i,d)}{\displaystyle\sum_{j>i}(z(j,d) p_{ij })} ~~~.
\end{equation}

On the other hand, every transmission attempt by $i$ decreases its transmission credit by $1$. MORE prunes the low-quality forwarding nodes with $z(i,dst) < \frac{1}{10}\sum_{i}z(i,dst)$.\newline
To serve our comparison purpose, we replace $z(i,dst)$ in the credit calculation of eq.~(\ref{eq40}) with ETX, EAX, and EPA one at a time. For instance, $TXcredit_i$ for the EPA case would be $ \frac{EPA(i,d)}{\sum_{j>i}EPA(j,dst) p_{ij}}$, where the numerator denotes the expected progress of a forwarded packet by $i$, and the denominator is the total progress contribution of $i$ in terms of the forwarded packet.
\par	
The simulation environment, setup, and parameters used for MORE are the same as those used for EXOR. The side-by-side simulation results of both protocols employing the three OR metrics (ETX, EAX, and EPA) in the topology scenario of Fig.~\ref{fig:24s}-e are illustrated in Fig.~\ref{fig:266s}. MORE clearly achieves its best result using EPA, whereas EXOR favors ETX for this specific topology scenario.\par   

In this section, we demonstrated how the network performance (total delivery delay) is impacted by choice of OR protocol, OR metric, and the topology scenario.

\subsection{\bf OR Metrics: A Scrutiny}\label{Sec:Scrutiny}

\begin{figure}
\centering
\includegraphics[width=80mm]{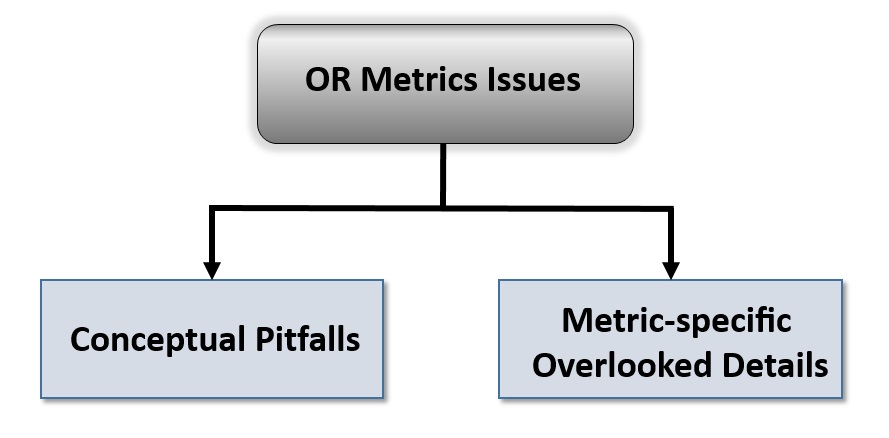}
\caption{Categorization of OR metric issues.}
\label{fig:fig26}
\end{figure}

A closer and critical look at the OR metrics reveals some serious issues that are worth discussing in detail. We have divided these issues into two groups: conceptual pitfalls related to fundamental missed ideas in the general approach to OR metrics and overlooked details related to the definition of some specific OR metrics (Fig.~\ref{fig:fig26}).

\subsubsection{Conceptual Pitfalls}
In this subsection, we raise some general concerns about how OR metrics are computed.
\begin{itemize}

\item {\it OR Metric-Computation Directionality:} Most of the metrics are calculated from the destination to the source. In other words, the metric rates a node based on the cost of delivering a packet from that node to the destination. This value is then used to prioritize this particular node amongst all other members of the same hyperarc \cite{P_71_2} receiving the packet. The general problem with this approach is that the quality of the incoming links is not accounted for in the process of prioritization, which might degrade the OR performance to worse than that provided by traditional routing. To illustrate this issue, consider the network of Fig.~\ref{fig:fig15}. All the intermediate nodes have small differences in their outgoing link qualities to the destination. A typical OR protocol prefers $A$, $B$, and $C$ over $D$ according to any link-quality-based OR metric (e.g., ETX), whereas $D$ is clearly the best forwarding node. All the packets received by nodes A, B, and C combined would most likely be received by node D as well. In other words, an accumulation of packets in D occurs due to its incoming high-quality link. Furthermore, according to the link probabilities in Fig.~\ref{fig:fig15}, these packets have almost the same chance of reception by $dst$ if D rather than the others transmit them. Therefore, unnecessary transmission attempts by A, B, and C can be prevented by letting D transmit first. To investigate the above, we simulated the network in Fig.~\ref{fig:fig15} under the EXOR protocol for two cases: ETX-based node prioritization (A-B-C-D) and manipulated node prioritization (D-A-B-C). The total delivery delay results are shown in Fig.~\ref{fig:fig333}. The results confirm that {\it by looking only forward, the OR metrics miss the reality.}
\newline The conclusion we are attempting to make is that by introducing new metrics that also consider the quality of how error-free incoming packets feed a forwarding node, the OR performance can be improved beyond the levels achieved by the classic metrics. However, the measurement of a node's feed quality may not always be straightforward. For instance, in multi-hop scenarios, the traffic descends on an intermediate node from different paths (Fig.~\ref{fig:fig36}).

\begin{figure}[h!]
	\centering
	\includegraphics[width=60mm]{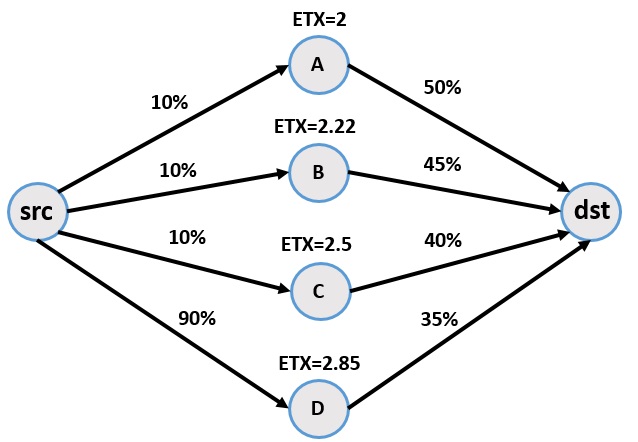}
	\caption{Backward-only metric evaluation problem: the metric prefers $A$ over $B$, $B$ over $C$, and $C$ over $D$. However, the forwarding order should be exactly reversed to achieve better performance.}
	
	\label{fig:fig15}
\end{figure}

\begin{figure}[h!]
	\centering
	\includegraphics[scale=0.25]{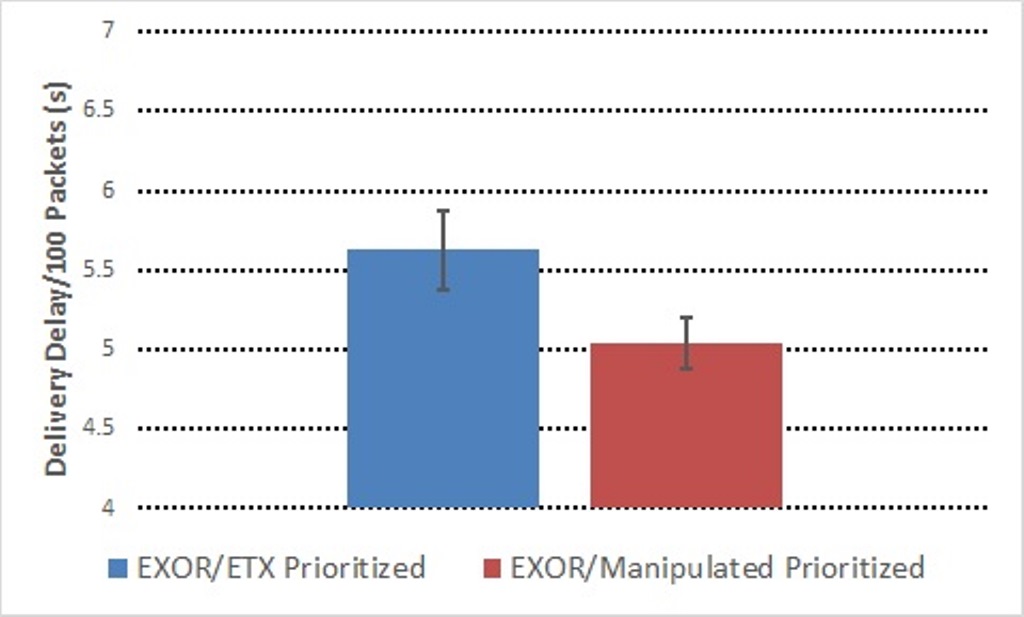}
	\caption{The performance results of applying EXOR to the topology scenario in Fig.~\ref{fig:fig15} using ETX and manipulated prioritization.}
	\label{fig:fig333}
\end{figure}

\begin{figure}[h!]
	\centering
	\includegraphics[width=80mm]{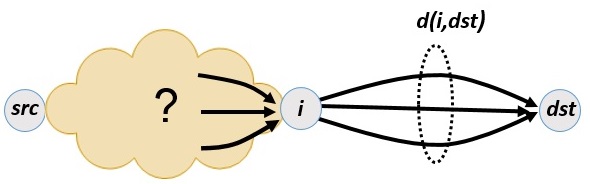}
	\caption{While classic metrics calculate the quality of delivery at the destination from an intermediate node $i$, the importance of the quality of packet reception at $i$ is ignored.}
	
	\label{fig:fig36}
\end{figure}
\item {\it ACK Scope Impact:} In the OR metrics that consider the calculation of the expected number of transmissions required, an important issue is the way in which a subject packet is acknowledged in the corresponding parent OR protocol. In Fig.~\ref{fig:fig772}, where  $p_1$ and $p_2$ denote the delivery probabilities to and from an intermediate node $i$, if the packets from the source to the destination are verified on a per-link manner (i.e., store-process-forward), the required expected number of transmissions is equal to $\frac{1}{p_1} + \frac{1}{p_2}$. However, in an end-to-end verified case, since the intermediate node has to wait until the completion of the transit packet (i.e., store-forward), the expected number of transmissions required is equal to $\frac{2}{p_1 . p_2}$, which is greater or at least equal to $ \frac{1}{p_1} + \frac{1}{p_2}$. The expected number of transmissions required can be equal to $\frac{1}{p_1 . p_2} $ only in cut-through wired networks. Interested researchers are cautioned about any confusion regarding this issue. 

\begin{figure}[h!]
	\centering
	\includegraphics[width=80mm]{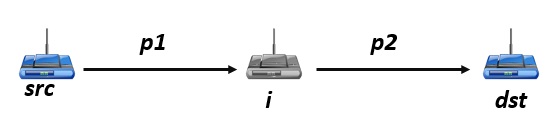}
	\caption{A simple two-hop $src-dst$ network transmission.}
	
	\label{fig:fig772}
\end{figure}

\item {\it Links' Order Impact:}
Another critical issue, which was initially raised by \cite{P_71_1}, is the impact of the links' order.\cite{P_71_1} states that if the MAC layer's transmission-attempt limit is considered, then the location of the low-quality link on the path is an important characteristic. A first glance at Fig.~\ref{fig:fig141} might lead to the conclusion that, in terms of link quality, the transmission costs of delivering one packet through both paths are equal. However, if the MAC layer's transmission-attempt limit is 2, then packet delivery over the upper path costs 20 transmissions while it costs 18 transmissions over the lower path. Additionally, the above is valid if per-link verification (link-layer acknowledgment) is used.

\begin{figure}[h!]
	\centering
	\includegraphics[width=80mm]{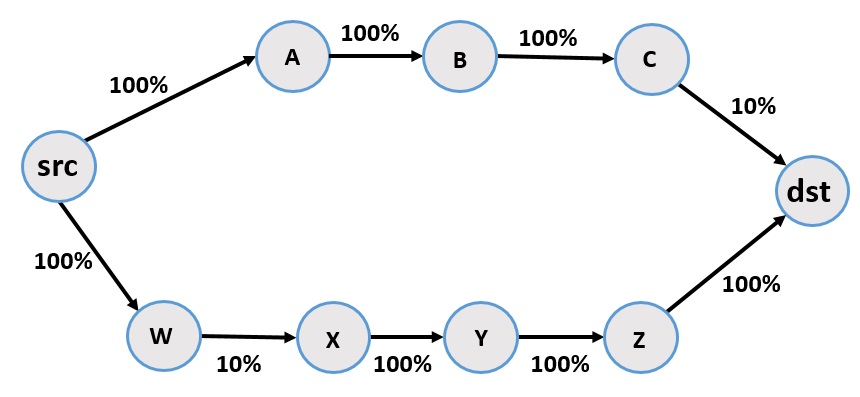}
	\caption{Two similar paths with different link orders. }
	
	\label{fig:fig141}
\end{figure}
	

\end{itemize}

\subsubsection{Overlooked Metric-specific Details }
Throughout this subsection, we identify details that have been overlooked in the computation of specific OR metrics.

\begin{itemize}

\item {\bf ETX:} According to~\cite{P5}, if the forward and reverse delivery ratios of a link are $d_f$ and $d_r$, respectively (Fig.~\ref{fig:fig13}), then:

\begin{equation}
ETX=\frac{1}{d_f . d_r}~~~,
\end{equation}

\begin{figure}[h!]
	\centering
	\includegraphics[width=50mm]{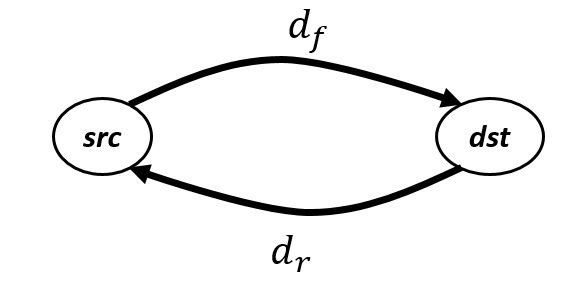}
	\caption{Forward (packet) and reverse (acknowledgment) delivery probabilities in a simple $src-dst$ transmission scenario.}
	
	\label{fig:fig13}
\end{figure} 

which is understood to be valid for per-link acknowledgment scenarios. However, since packets should be received correctly before being acknowledged (store-process-acknowledge) and since the packet and its ACK substantially differ in length, the correct average number of transmissions, in terms of the forwarding packet transmissions, should be: \begin{equation}
\frac{1}{d_f} +\frac{\frac{L_{ACK}}{L}}{ d_r}~~.
\end{equation}

 How this error could affect the final results depends on the numerical values of $d_f$, $d_r$, and $\frac{L_{ACK}}{L}$.
\item {\bf ETC:} In an attempt to determine ETC, \cite{P_50_10}  calculates $FAR$, the wake-up ratio of all the forwarding candidates, by finding the union of all the wake-up periods (eq.~(2) in \cite{P_50_10}). There appears to be an error in this derivation since the union of all the wake-up periods requires additional terms according to eq.~(\ref{eq50}) (beyond the second term): 

\begin{equation}
\begin{aligned}
\label{eq50}
|\cup_{i=1}^{n} A_i| = &\sum_{i=1}^{n}|A_i| - \sum_{1\leq i<j\leq n}(|A_i\cap A_j|)+\\ &...+\sum_{1\leq i<j<k\leq n} |A_i\cap A_j\cap A_k|-\\ 
&...+(-1)^{n-1}|A_1\cap...\cap A_n|  
\end{aligned}
\end{equation}

which might be significant.
\item {\bf EPA/EOT/EAR/EDRb/DDR/MEAR/OEE/DUER/ FE/espeed:} Distance proximity might not always be an appropriate packet advancement criterion since a geographically closer-to-destination node might have no forwarding path. For instance, in the network of Fig.~\ref{fig:fig77}, all the nodes on the upper path are geographically closer to the destination than node 2 on the lower path, but the upper path is a communication dead-end.
 \begin{figure}
 	\centering
 	\includegraphics[width=90mm]{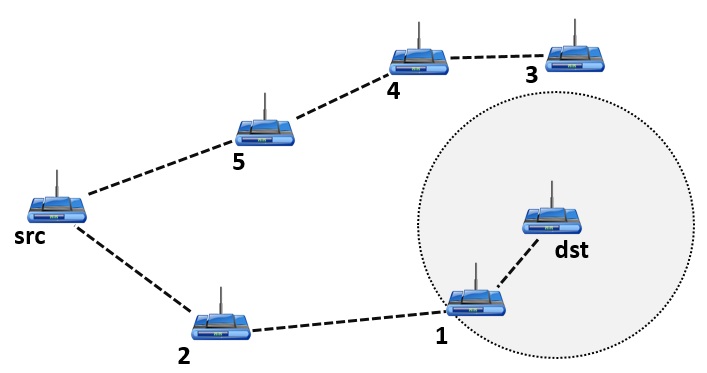}
 	\caption{An example topology scenario illustrating the incompetence of distance-proximity-based metrics in avoiding dead-end communication paths. }
 	
 	\label{fig:fig77}
 \end{figure}
\item {\bf ETCoP:} \cite{P_50_19} raises a link order issue similar to that mentioned previously in this section, though without relying on the MAC layer's transmission attempt limit (see Fig.~\ref{fig:fig772}). Furthermore, \cite{P_50_19} states that a packet originating from $src$ and destined for $dst$ in Fig.~\ref{fig:fig772} can experience only the following (in a per-link verification scenario):

\begin{itemize}
	\item the data packet may be dropped on the first link (with probability $1-p_1$),
	\item the data packet may be successfully delivered on the first link and be dropped on the second link (with probability $p_1 (1-p_2)$),
	 
	\item the data packet is successfully delivered on both links (with probability $p_1 p_2$).
\end{itemize}  
The total expected number of transmissions is $(1-p_1) + 2  p_1  (1-p_2) +  2  p_1  p_2 $. With a similar extension,  \cite{P_50_19} estimates the expected number of transmissions of an n-hop path under the failed receptions of the end node, $Fc(n)$, from which ETCoP is calculated as in eq.~(16) in \cite{P_50_19}.\newline We have to dispute the two-link derivation of the total expected number of transmissions above if no additional constraint is imposed since the packet could theoretically experience an unlimited number of failures on the first link before getting through. In other words, we believe that the sample space of possible outcomes of packet experience in \cite{P_50_19} is not complete. The exact expectation, accounting for all the possible packet experiences, should be:
\begin{equation}
	\sum_{m=0}^{\infty}\sum_{n=0}^{\infty} {(1-p_1)}^m p_1 {(1-p_2)}^n p_2(m+n+2)~~,
\end{equation}
which simplifies to the familiar expression of $\frac{1}{p_1} + \frac{1}{p_2}$. The latter shows that without any additional constraint (e.g., as in \cite{P_71_1}) and in a per-link verification scenario, the link order does not matter, as it was concluded in our earlier discussion.

\item {\bf EATT/OETT/MEATT/EAD/EOT/OEE/FE:}
 In the definition of this group of metrics, we observe the co-presence of the packet length and the probability delivery ratio parameters. This co-presence may undermine the generally agreed-upon concept of inter-dependency between these two parameters presented by, for instance, the classic independent-bit-errors model of: 
 
\begin{equation}
	P_{PKT} = \displaystyle(1-p_{_b})^{L}~~~,
\end{equation}
where $p_{_b}$ and  $P_{PKT}$ denote the bit error and packet error probabilities or some other empirically obtained relationship, as in \cite{P_71_3}.   

\item {\bf STR:} Regarding the network of Fig.~1 in \cite{P22}, we were not able to generate the same results as in Table I.

\end{itemize}

\section{\large\bf OR Metric: Future Works}\label{sec5}

\begin{figure}[h]
	\centering
	\includegraphics[width=95mm]{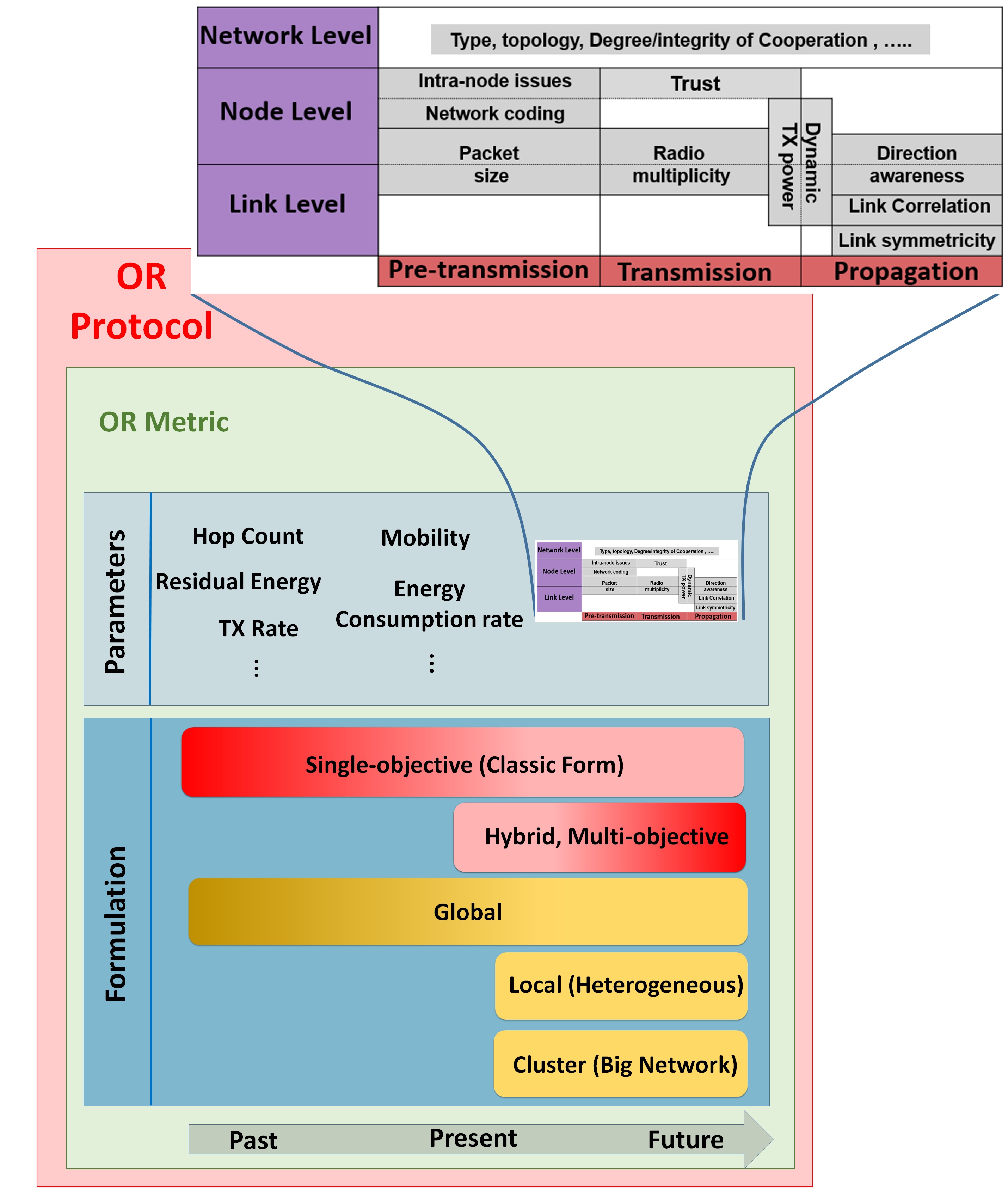}
	\caption{Future research suggestions framework.}
	\label{fig:futureworks}
\end{figure}
Simply speaking, the OR is transmission scheduling among a CFS selected based on some OR metrics, in multi-hop networks. Thus, the metric design significantly affects the network's throughput. Fast-paced emergence of either new network types or new applications on existing networks, following the growth in 5G, Smart City, and IoT technologies, necessitates the development of new OR metrics which reflect the new environments and generate adequate performance.\\
 The multi-faceted nature of the OR metric requires a structured approach to its development process. Figure~\ref{fig:futureworks} proposes a development framework, consistent with the OR metric design process of Fig.~\ref{fig:process}, in which:
\begin{itemize}
\item the development is divided into two main fronts, parameters, and formulations,
\item the time-evolution of the research focus has been demonstrated,
\item and for each constituent, the level of research activity is illustrated by different color shades wherein darker shades represent higher activities.  	
\end{itemize}

\subsection {Parameters Development:}
Devising OR metrics with new parameters, in a systematically manner, can be considered in three different levels, the network, node, and link levels (the magnified part of Fig.~(\ref{fig:futureworks})).\\
\begin{itemize}
\item Network-level parameters:\\
 Some example network-level parameters for future exploration are as follows:
\begin{itemize}
\item 
	{\it Network Type (platform): }Emergence of new types of network or new missions (applications) in existing networks, for instance VENs \cite{P_2019_2_101347}, Internet of Vehicles (IoV) \cite{P_2019_4_192011}\cite{P_2019_4_192019}, 5G \cite{P_2019_4_192038}, UAVNET \cite{P_2019_1_279,P_2019_1_278,P_2019_4_010053}, molecular communications \cite{P_2019_4_192054}, and D2D \cite{P_2018_300}\cite{P_2019_3_280407}\cite{P_2019_4_192124}, have introduced new parameters describing the specific aspects of their corresponding networks or missions. Devising OR metrics, which are aware of these parameters, helps to achieve better routing performance. For example, while OR was originally designed for WMNs featuring static and under-single-administration topology, nowadays, it has found its way in highly dynamic networks such as VANETs and UAVNETs. In these environments, the metric should adequately reflect the impact of high node mobility on the wireless link quality.

\item{\it Degree/Integrity of Cooperation: }With very few exceptions, almost all the OR proposals assume that the transmission-overhearing nodes participate in the forwarding process willingly, trustfully, and indiscriminately. The variety of new emerging wireless applications demands wireless networks with different missions and organizations. While OR was originally designed for WMNs featuring a fixed and predetermined organization, usually under a single administration, OR can now be deployed in structurally relaxed (ad hoc) networks.  In such circumstances, exposure to security lapses and selfishness might impact the integrity and the degree of cooperation in OR. A metric aware of the degree/integrity of cooperation will result in better CFS selections in less-friendly scenarios.
\item {\it Trust:} The concept of trust has prevalently been used to differentiate between the goodwill of nodes. Trust is defined as a node's degree of subjective belief about the future behavior of other entities in the network in a given context \cite{P_60_8}, \cite{P_60_9},\cite{P_2017_8}. In the context of OR, trust can be considered to be the level of reliance on (cooperation of) a node to forward a packet if it is required to do so as a member of a CFS. Different trust computational approaches have been introduced as different trust models. Our study shows that E2TX \cite{P30}, RTOR \cite{P_50_21} and its close relatives, TORDP and GEOTOR \cite{P_60_10}, are metrics that accommodate trust via a very simple direct trust model. ETOR \cite{P_2018_13} extends the trust model to include indirect trust. However, all the aforementioned trust-based metrics act on a link level. We believe that there is ample room for research exploring more involved trust models acting over a larger scope, such as a path rather than just a link.
\par On another front, nodes might not contribute equally to the OR in networks that are not under a single administration. For example, a node might prioritize its self-generated traffic or a specific flow over other transit ones.
\item{\it Topology:}
Concerning the effect of the topology-OR metric pair on the network's performance, extensive simulations in section \ref{sec4} revealed implicit dependencies. In the same manner, it is evident that the node density impacts the size and quality of CFS. Also, end-to-end metrics perform better in dense networks, while per-hop metrics are generally better suited to sparse networks — all of the above points to the importance of topology in OR problems. As a consequence, further topology-aware metric studies are encouraged.
\end{itemize}
\item {Node-/Link-level parameters:}
\\Regarding the node- and link-level considerations, one can divide the packet delivery process into pre-transmission (preparation), transmission, and propagation phases. \\
The pre-transmission phase constitutes all the processes that occur to get a packet to the head of a sender node's transmission queue. These processes are exclusively internal to the sender node (node-level-only considerations). The transmission phase includes the selection of the transmission parameters (e.g., transmission power, transmission rate/channel), which, on the one hand, requires the sender node's corresponding capabilities, and on the other hand, impacts the packet delivery quality on the forwarding link(s) (joint node-/link-level considerations). Also, in the propagation phase, the parameters of the physical channel (path loss, shadowing, multi-path fading, etc.), as link-level-only considerations, contribute to the packet delivery quality. \\
In the following, we present some suggestions for future research on node-/link-level parameters.
\begin{itemize}
 \item{\it Packet size: }As mentioned in Section \ref{Sec:Scrutiny}, experimental results (e.g., \cite{P_2019_5_021616}) and mathematical models (e.g., equation packet length) relate the PDP, the core parameter in OR metrics, to the packet size and the condition of the wireless channel \cite{P_71_3}. While shorter packets result in higher PDPs, the network goodput deteriorates due to the additional overhead of the packets. \par
 According to the best of our knowledge, almost all previous OR metric research assumed a fixed packet size. We believe that answering questions similar to the following might open promising research avenues: {\it What is the best packet size for achieving an optimum OR throughput in a given network topology? Is it beneficial to use multiple simultaneous packet sizes in a single OR flow? How does, if at all, variable packet sizes help in dealing with long-term network dynamics?}
 
 \item{\it Intra-node considerations: } OR metrics are traditionally more concerned with link-quality issues (exterior to node) than forwarding capabilities and the internal processes of the involved nodes. We would like to address the intra-node phenomena that impact the forwarding capabilities of the nodes in the OR context. \\
 With the main focus being on the packet retransmission limit in the context of WSNs, \cite{P_2017_9} discusses the internal processes involved from the reception of a packet until the transmission of the packet. Similarly, we model a forwarding node as the succession of a receiver, an input queue, a processing entity, an output queue, and a transmitter, as illustrated in Fig.~\ref{fig:fig311}. In single-radio scenarios, reception and transmission share common circuitry.\\
 
 \begin{figure}[h!]
 	\centering
 	\includegraphics[width=85mm]{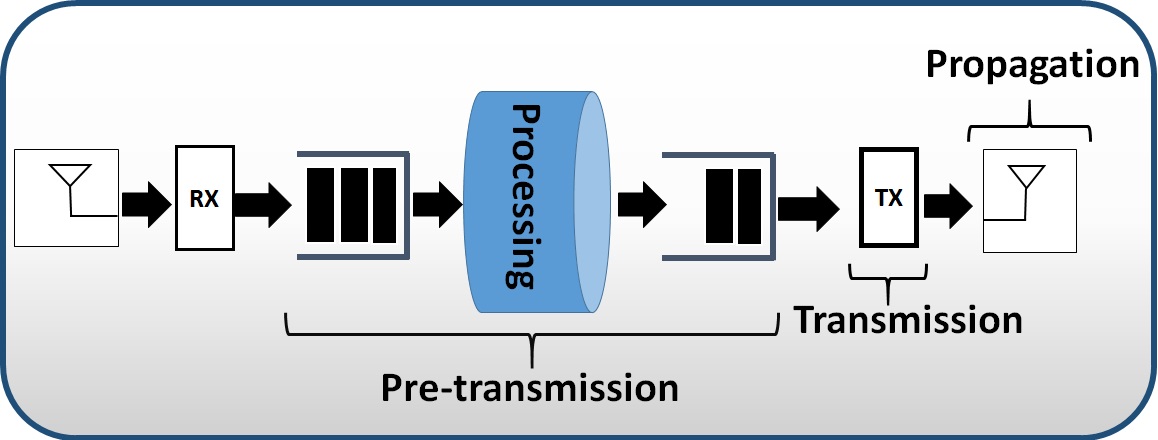}
 	\caption{A simplified model for packet flow inside a node.}
 	
 	\label{fig:fig311}
 \end{figure}
 The forwarding capability of a node in OR is strongly related to the dwelling time of the packet in the succession of the input queue, processing entity, and output queue (pre-transmission phase in Fig.~\ref{fig:fig311}). While the processing power (hardware/software) is inherent to the node, the length of the queues is dependent not only on the node's processing power but also on the node's involvement in the underlying network transactions. Low processing power (either because of the hardware limitations, or the complexity at the overlaying OR/higher-layer protocols) can delay the transportation of packets from the input to the output. Moreover, a node located along a succession (or at the intersection) of good-quality links is more likely to be repeatedly selected as a forwarding relay, which will result in crowded queues (i.e., congestion) or packet drops due to limited queue lengths. Figure \ref{fig:fig322} illustrates a circumstance where the classic OR prioritization of the nodes on the top path over those on the lower path (higher link qualities) does not necessarily result in better performance. Therefore, the development of new OR metrics that also reflect the intra-node capabilities/limitations (e.g., \cite{P_2018_24} considers the output queue length), is worth additional research effort. 
 
 	\begin{figure}[h!]
 		\centering
 		\includegraphics[width=90mm]{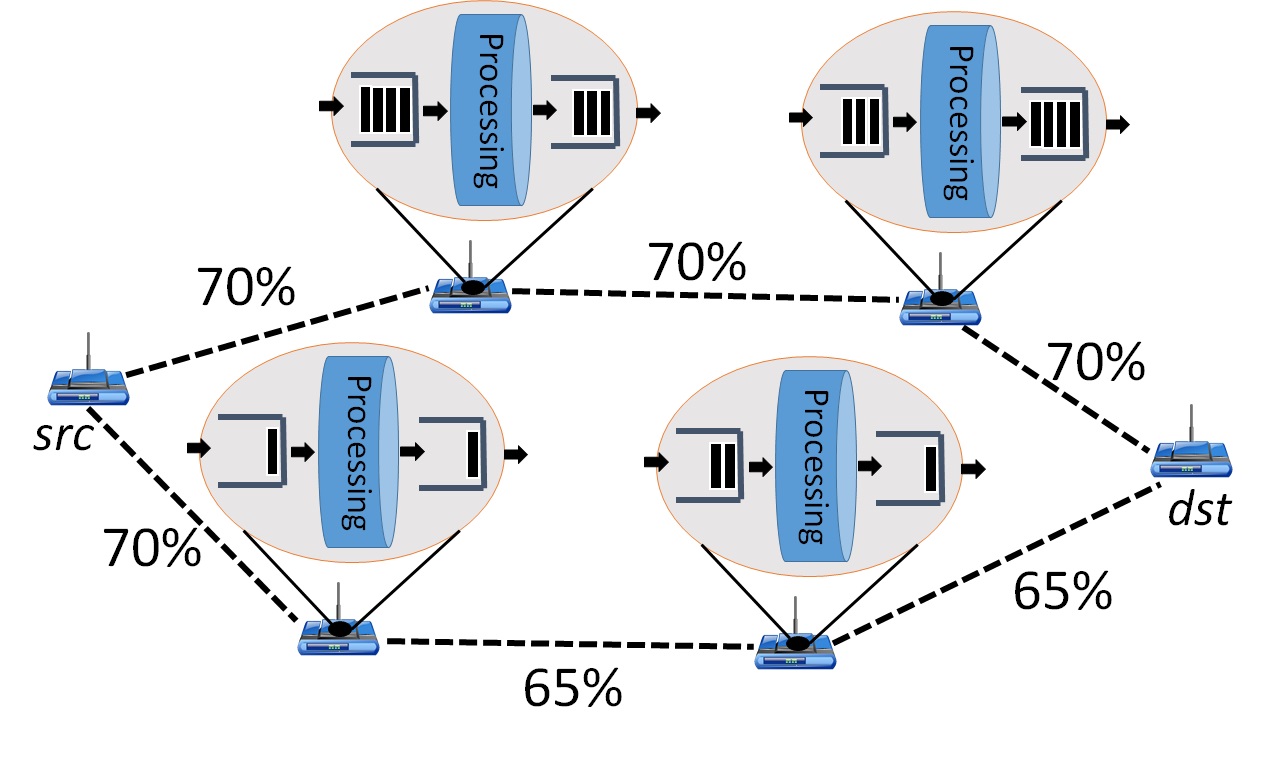}
 		\caption{A probable circumstance illustrating congestion on a high-quality path.}
 		
 		\label{fig:fig322}
 	\end{figure}
 	\item{\it Network Coding: }OR and NC both attempt to reduce the number of required  transmissions, the former through diversity over links and the latter through diversity over transmissions. An interesting potential research topic is to investigate how the introduction of coding-aware metrics, as opposed to already studied joint coding-OR techniques using non-coding aware metrics \cite{P_2019_1_24}, affects the OR's total performance. Thus far, the only related work is the FE metric \cite{P_50_16}.
 	\item{\it Link Correlation: }Link correlation, defined as the channel similarities between multiple receivers and one single data source, has been addressed in several works \cite{P_2017_10}\cite{P_2017_11}. Link correlation results in the receptions at multiple receivers listening to the same transmission being dependent. 
 	On the other hand, OR, by its very nature, deploys the diversity over multiple links for more efficient delivery. Therefore, OR is more promising when links are independent and capable of incremental cooperation. The impact of link correlation on the OR performance has been studied in \cite{P_50_22,P_50_23,P_50_24,P_50_25,P_50_26}. Furthermore, a simple method  for estimating link correlations is presented in \cite{P_2017_10}. cEAX \cite{P_50_24} has introduced the impact of link correlation on the $\boldsymbol{F}_i$ selection mechanism in EAX \cite{P13,P14}, enforcing an additional constraint on the size of $\boldsymbol{F}_i$. The authors in \cite{P_50_23} have suggested using the packet loss joint probability among all the forwarders in the same level when listening to a sender from the previous level. Link-correlation-aware OR metrics increase network throughput by avoiding reception redundancy among highly correlated links.  
 	We believe that there remains room for further study on this topic mainly because:
 		\begin{itemize}
 			\item the link correlation is the most counter-effective factor in  the OR performance,   
 			\item the traditional link correlation computations rely on statistical operations on long streams of data rendering it impractical for heavily fluid wireless networks such as UAVs and VANETs,
 			\item to the best of our knowledge, the link correlation in recently-popular directional antenna scenarios is not addressed yet.
 		\end{itemize}
 		\item{\it 	Link symmetricity: }In almost all the previous OR-related studies, links (the PDP) are considered to be symmetrical. First, physical transmission paths in wireless channels are not necessarily reversible in terms of PDP. Second, since packets of different sizes (e.g., data packet vs. acknowledgment packet) flow in different directions over a link, considering earlier discussions, the assumption of symmetrical links (equal PDPs) may not be accurate.\\ 
 		{\it Direction awareness:}
 		OR and directional wireless transmission are intended to improve efficiency by, to some extent, contradicting strategies. OR attempts to benefit from the inherent broadcast nature of the wireless transmission by getting all the overhearing nodes to cooperate in furthering a flow. However, directional transmission attempts to let multiple independent flows exist concurrently by limiting the mutual interference between them (i.e., fewer nodes involved in furthering a specific flow). From another perspective, the optimization scope of OR includes a single flow, whereas the optimization scope of directional transmission is the entire network. 
 		
 		\begin{figure}[h!]
 			\centering
 			\includegraphics[width=70mm]{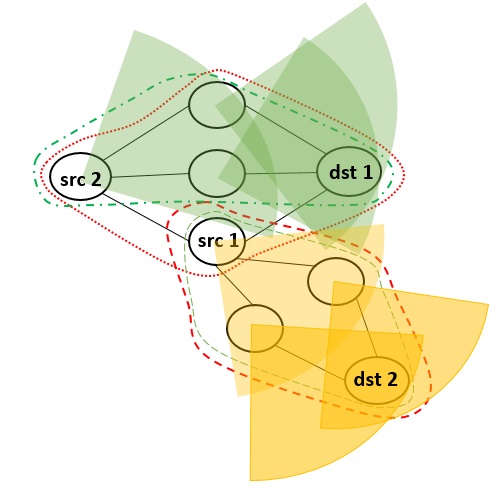}
 			\caption{Concurrent multi-flow transmissions using directional antennas.}
 			
 			\label{fig:fig344}
 		\end{figure}
 		To clarify the potential compromise between the two approaches, we show two source-destination pairs with overlapping wireless space in Fig.~\ref{fig:fig344}. When employing the conventional omni-directional OR, the two transmissions should occur in tandem since $src2$ overhears $src1$'s transmission, and, to the benefit of the first flow, $src2$ can potentially serve as a member of $src1$'s CFS. On the other hand, when using the directional-OR configured as illustrated, $src2$ cannot hear $src1$'s transmission and, to the benefit of the total throughput, it can initiate its own flow concurrently. \\
 		In somewhat related efforts, \cite{P_40_2} applies beamforming to enhance performance in a single-flow scenario by concentrating the transmission power on those members of the CFS with a higher packet progress contribution. \cite{P_40_3} tries to limit the number of involved forwarding nodes without any reference to how this limitation will affect other possible concurrent flows.\\
 		With the increasing popularity of directional wireless transmission, particularly in new generations of wireless networks (e.g., 5G \cite{P_2017_12}), incorporating these two otherwise-contradicting approaches through introducing direction-aware OR metrics is promising.

 		\item{\it Dynamic transmission power: }
 		The transmission power of a wireless node directly influences its radio range. Manipulating the transmission power impacts the quality of the surrounding links, thereby resulting in a change in the connectivity of the network through eliminating existing links or creating new links (logical topology change). Therefore, the radio range directly impacts the routing paradigm.\\
 		In traditional routing, short and high-quality links are preferred over long and erroneous wireless links \cite{P5} due to the nonlinear relationship between power and distance in wireless transmission. However, due to the multi-path nature of OR (where even low-quality links are contributing), this preference cannot be extended to OR scenarios without reservations. From another perspective, any change in the connectivity paradigm and/or link qualities will likely impact the values of the OR metrics and, consequently, the forwarder candidate selection results.\\
 		Therefore, incorporating the transmission power into the OR metrics (provided that the variable transmission power is supported at the PHY layer) and investigating the interaction between the transmission power and the underlying OR performance may reveal new interesting research problems.
 		
 		\item{\it Radio multiplicity:}
 		While imposing higher costs on hardware, OR in networks with multiple-radio transceivers can benefit from an extra degree of concurrency, provided that the multi-radio support of the embedded metric is in place.
\end{itemize}		
\end{itemize}

\subsection{Formulation Development}
  The formulation of an OR metric shows how its parameters interact and is a direct result of the adopted design optimization process as per Fig.~(\ref{fig:process}). Traditionally, the majority of the optimization processes have been mathematical-, heuristics-, and empirical-based. As time goes by, wireless communications, particularly under 5G and IoT umbrellas, like other fields of engineering, face the challenge of data hugeness. Because of this, and also because of the advances in computing power, Machine Learning (ML) (and more recently Deep Learning (DL)) methods have found their ways into different parts of wireless communication systems \cite{P_2019_4_192157}\cite{P_2019_4_192214}. While there are several  ML-/DL-based wireless routing works in the literature \cite{P_2019_4_192220,P_2019_5_262231,P_2019_5_262236}, to the best of our knowledge, there are not so many (if any at all) such research efforts regarding OR metrics.\par
  About OR metric objectives, moving towards more complex networks requires the merits of the relaying nodes to be a compromise between more than one not-necessarily-congruent basic objectives. A new research direction can involve devising hybrid metrics as either a function of multiple single-objective metrics \cite{P_2019_1}\cite{P_2019_4_010034} \cite{P_2019_4_010059} \cite{P_2019_4_010053} or a single multi-objective metric.\\
  Conventional wireless networks are homogeneous in type and limited in size. Thus, it is practical to define a single OR metric (i.e., global OR metric) for the whole network and disseminate the metric information during a reasonable time duration. Modern wireless networks are characterized by the high number of short-range nodes (i.e., large size in terms of the number of hops), e.g., in the context of IoT, as well as by the coexistence and cooperation of different-in-nature networks (i.e., heterogeneous networks), e.g., in the context of 5G. In such networks, a single OR metric cannot adequately reflect differing coexisting natures in the network. An attractive research avenue can include an investigation into designing a set of local OR metrics each representing a particular homogeneous neighborhood, in the case of heterogeneous networks, or a specific limited-in-size cluster, in the case of large-scale homogeneous networks.\par
  We conclude this section by expressing the serious need for rethinking the computation directionality of OR metrics, as discussed in section \ref{Sec:Scrutiny}.

\section{\bf Conclusion}\label{sec6}
 The ubiquity of multi-hop ad hoc wireless networks, both new and emerging or classic ones, necessitates greater efficacy in wireless routing algorithms. In this regard, OR, which has proven its merits for more than a decade, remains promising. At the heart of each OR protocol is an embedded OR metric that influences the overall performance. In this work, we start with a tutorial on OR wherein it is studied from a new layered-diversity point of view, and a new OR metric design scheme is introduced. Then, an exhaustive treatment of OR metrics, classified based on their computation methods and scopes and reformulated according to a unified notation, is presented. The latter enables us to take a structured, investigative, and comparative approach to OR metrics, which provides valuable insight for interested researchers. Extensive simulations compare the main representatives of OR metric classes and reveal not-so-obvious dependencies regarding the network performance. Self-explanatory, easy-to-grasp, and visual-friendly quick references are provided, which can be used independently from the rest of the paper. Finally, a new insightful framework for future research directions, consistent with the introduced OR metric design scheme, is developed. We can confidently claim that this tutorial survey is the first complete and exclusive investigation on OR metrics, beneficial to both generalists and OR specialists.
\begin{figure}[h!]
	\includegraphics[width=0.5 \textwidth]{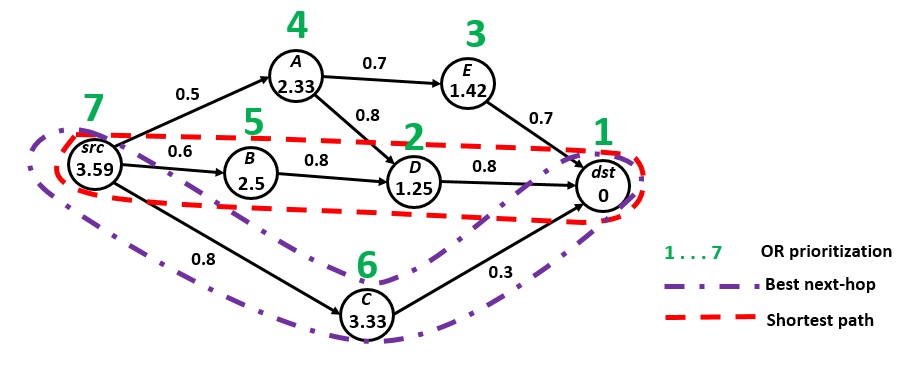}
	\caption{The path(s) selected when using three different routing protocols: the best next-hop
		, the shortest single-path, and the best any-path OR based on EAX.}
	\label{fig:fig111}
\end{figure} 

\section{Appendix \label{Appendix}}
To obtain a better understanding of how OR differs from traditional routing protocols, Fig.~\ref{fig:fig111} gives an example differentiating between the paths traversed from a source ($src$) to a destination ($dst$) using the best next-hop, the shortest single-path, and the best any-path OR protocols. The numbers on the connecting links denote the corresponding delivery probabilities. The best next-hop protocol suggests the path {\it src-C-dst}. The shortest single-path protocol (i.e., Dijkstra) selects the path {\it src-B-D-dst}. Based on the EAX metric \cite{P13,P14} values shown below each node's name (to be detailed shortly) in Fig.~\ref{fig:fig111}, the OR protocol suggests the ordered list of forwarding nodes ($dst,D,E,A,B,C,src$), represented by green numbers above each node.

\bibliographystyle{IEEEtran}
\bibliography{References}{}

\begin{thebibliography}{100}
\providecommand{\url}[1]{#1}
\csname url@samestyle\endcsname
\providecommand{\newblock}{\relax}
\providecommand{\bibinfo}[2]{#2}
\providecommand{\BIBentrySTDinterwordspacing}{\spaceskip=0pt\relax}
\providecommand{\BIBentryALTinterwordstretchfactor}{4}
\providecommand{\BIBentryALTinterwordspacing}{\spaceskip=\fontdimen2\font plus
\BIBentryALTinterwordstretchfactor\fontdimen3\font minus
  \fontdimen4\font\relax}
\providecommand{\BIBforeignlanguage}[2]{{%
\expandafter\ifx\csname l@#1\endcsname\relax
\typeout{** WARNING: IEEEtran.bst: No hyphenation pattern has been}%
\typeout{** loaded for the language `#1'. Using the pattern for}%
\typeout{** the default language instead.}%
\else
\language=\csname l@#1\endcsname
\fi
#2}}
\providecommand{\BIBdecl}{\relax}
\BIBdecl

\bibitem{P_2018_27}
{A. Zanella, N. Bui, A. Castellani, L. Vangelista, and M. Zorzi}, ``Internet of
  things for smart cities,'' \emph{IEEE Internet of Things journal}, vol.~1,
  no.~1, pp. 22--32, 2014.

\bibitem{P_2018_28}
{F. Li, and Y. Wang}, ``Routing in vehicular ad hoc networks: A survey,''
  \emph{IEEE Vehicular technology magazine}, vol.~2, no.~2, 2007.

\bibitem{P_2019_3_032301}
{N. Li, J. Mart{\'\i}nez-Ortega, H. Vicente, S. Antonio}, ``{Probability
  prediction-based reliable and efficient opportunistic routing algorithm for
  VANETs},'' \emph{IEEE/ACM Transactions on Networking (TON)}, vol.~26, no.~4,
  pp. 1933--1947, 2018.

\bibitem{P_2019_2_101347}
A.~Lam and V.~Li, ``Opportunistic routing for vehicular energy network,''
  \emph{IEEE Internet of Things Journal}, vol.~5, no.~2, pp. 533--545, 2018.

\bibitem{P_2019_1_279}
R.~J. L.~Gupta and G.~Vaszkun, ``Survey of important issues in uav
  communication networks,'' \emph{IEEE Communications Surveys \& Tutorials},
  vol.~18, no.~2, pp. 1123--1152, 2016.

\bibitem{P_2019_1_278}
J.~Jiang and G.~Han, ``Routing protocols for unmanned aerial vehicles,''
  \emph{IEEE Communications Magazine}, vol.~56, no.~1, pp. 58--63, 2018.

\bibitem{P_2019_4_010053}
{D. Ros{\'a}rio, Z. Zhao, T. Braun, E. Cerqueira, A. Santos, and I. Alyafawi},
  ``Opportunistic routing for multi-flow video dissemination over flying ad-hoc
  networks,'' \emph{Proceeding of IEEE International Symposium on a World of
  Wireless, Mobile and Multimedia Networks}, pp. 1--6, 2014.

\bibitem{P_2018_29}
{A. Al-Fuqaha, M. Guizani, M. Mohammadi, M. Aledhari, and M. Ayyash},
  ``Internet of things: A survey on enabling technologies, protocols, and
  applications,'' \emph{IEEE Communications Surveys \& Tutorials}, vol.~17,
  no.~4, pp. 2347--2376, 2015.

\bibitem{P_2019_2_279}
{A. AlZubi, M. Al-Ma'aitah, and A. Abdulaziz}, ``A best-fit routing algorithm
  for non-redundant communication in large-scale iot based network,''
  \emph{Computer Networks}, 2019.

\bibitem{P_2018_300}
{A. Asadi, Q. Wang, and V. Mancuso}, ``A survey on device-to-device
  communication in cellular networks,'' \emph{IEEE Communications Surveys \&
  Tutorials}, vol.~16, no.~4, pp. 1801--1819, 2014.

\bibitem{P_2019_3_280407}
{Y. Li, K. Chi, H. Chen, Z. Wang, and Y. Zhu}, ``{Narrowband Internet of Things
  systems with opportunistic D2D communication},'' \emph{IEEE Internet of
  Things Journal}, vol.~5, no.~3, pp. 1474--1484, 2018.

\bibitem{P_2017_12}
A.~R. M.~Agiwal and N.~Saxena, ``{Next generation 5G wireless networks: A
  comprehensive survey},'' \emph{IEEE Communications Surveys \& Tutorials},
  vol.~18, no.~3, pp. 1617--1655, 2016.

\bibitem{P_2018_32}
{I. Akyildiz, W. Lee, M. C. Vuran, and S. Mohanty}, ``A survey on spectrum
  management in cognitive radio networks,'' \emph{IEEE Communications
  magazine}, vol.~46, no.~4, 2008.

\bibitem{P_2019_2_5113}
{Y. Chen, J. Zhang, and I. Marsic}, ``Link-layer-and-above diversity in
  multihop wireless networks,'' \emph{IEEE Communications Magazine}, vol.~47,
  no.~2, pp. 118--124, 2009.

\bibitem{P_2019_1_23}
W.~Stallings, \emph{Data \& Computer Communications--International Edition,
  Sixth}.\hskip 1em plus 0.5em minus 0.4em\relax Pearson Education India, 2007.

\bibitem{P_2019_2_51029}
{F. Adachi, M. Sawahashi, and H. Suda}, ``{Wideband DS-CDMA for next-generation
  mobile communications systems},'' \emph{IEEE communications Magazine},
  vol.~36, no.~9, pp. 56--69, 1998.

\bibitem{P_2019_2_51035}
{R. Nee and R. Prasad}, \emph{OFDM for wireless multimedia
  communications}.\hskip 1em plus 0.5em minus 0.4em\relax Artech House, Inc.,
  2000.

\bibitem{P_2019_2_51037}
{E. Larsson, O. Edfors, F. Tufvesson, and T. Marzetta}, ``{Massive MIMO for
  next generation wireless systems},'' \emph{IEEE communications magazine},
  vol.~52, no.~2, pp. 186--195, 2014.

\bibitem{P_2019_2_51040}
B.~Vucetic and J.~Yuan, \emph{Space-time coding}.\hskip 1em plus 0.5em minus
  0.4em\relax John Wiley \& Sons, 2003.

\bibitem{P_2019_2_51105}
{Z. Yang, and H. Fang, and C. Pan}, ``{ACE with frame interleaving scheme to
  reduce peak-to-average power ratio in OFDM systems},'' \emph{IEEE
  Transactions on Broadcasting}, vol.~51, no.~4, pp. 571--575, 2005.

\bibitem{P_2019_4_191937}
{M. Pearlman, Z. Haas, P. Sholander, and S. Tabrizi}, ``On the impact of
  alternate path routing for load balancing in mobile ad hoc networks,'' in
  \emph{2000 First Annual Workshop on Mobile and Ad Hoc Networking and
  Computing. MobiHOC}.\hskip 1em plus 0.5em minus 0.4em\relax IEEE, 2000, pp.
  3--10.

\bibitem{P_2019_1_26}
H.~A.-R. M.~Hasan and F.~Al-Turjman, ``A survey on multipath routing protocols
  for qos assurances in real-time wireless multimedia sensor networks,''
  \emph{IEEE Communications Surveys \& Tutorials}, vol.~19, no.~3, pp.
  1424--1456, 2017.

\bibitem{P_2019_1_24}
{S. Kafaie, Somayeh, Y. Chen, O. Dobre, and A. Hossam}, ``{Joint Inter-Flow
  Network Coding and Opportunistic Routing in Multi-Hop Wireless Mesh Networks:
  A Comprehensive Survey},'' \emph{IEEE Communications Surveys \& Tutorials},
  vol.~20, no.~2, pp. 1014--1035, 2018.

\bibitem{P_2019_1_25}
B.~H. M.~Iqbal, B.~Dai and S.~Yu, ``Survey of network coding-aware routing
  protocols in wireless networks,'' \emph{Journal of Network and Computer
  Applications}, vol.~34, no.~6, pp. 1956--1970, 2011.

\bibitem{p_9_7}
S.~Katti and H.~Balakrishnan, ``Symbol-level network coding for wireless mesh
  networks,'' in \emph{ACM SIGCOMM Computer Communication Review}, vol.~38,
  no.~4, 2008, pp. 401--412.

\bibitem{P6}
S.~Biswas and R.~Morris, ``Opportunistic routing in multi-hop wireless
  networks,'' \emph{ACM SIGCOMM Computer Communication Review}, vol.~34, no.~1,
  pp. 69--74, 2004.

\bibitem{P9}
{S. Chachulski, D. Katabi, S. Katti, M. Jennings}, ``Trading structure for
  randomness in wireless opportunistic routing,'' \emph{ACM SIGCOMM}, pp.
  169--180, 2007.

\bibitem{P_2018_25}
{A. Gharaibeh, M. Salahuddin, S. Jahed Hussini, A. Khreishah, I. Khalil, M.
  Guizani, and A. Al-Fuqaha}, ``Smart cities: A survey on data management,
  security, and enabling technologies,'' \emph{IEEE Communications Surveys \&
  Tutorials}, vol.~19, no.~4, pp. 2456--2501, 2017.

\bibitem{P_2018_20}
A.~Lam and V.~Li, ``Opportunistic routing for vehicular energy network,''
  \emph{IEEE Internet of Things Journal}, vol.~5, no.~2, pp. 533--545, 2018.

\bibitem{P_2018_33}
{N. Li, J. Martínez-Ortega, V. Hernández Díaz, and J. Fernandez},
  ``Probability prediction-based reliable and efficient opportunistic routing
  algorithm for vanets,'' \emph{IEEE/ACM Transactions on Networking (TON)},
  vol.~26, no.~4, pp. 1933--1947, 2018.

\bibitem{P32}
J.~Rak, ``{LLA: A new anypath routing scheme providing long path lifetime in
  VANETs},'' \emph{IEEE communications letters}, vol.~18, no.~2, pp. 281--284,
  2014.

\bibitem{P_2017_7}
{D. Kang, H. Kim, and S. Bahk}, ``{ORPL-DT: Opportunistic Routing for Diverse
  Traffic in Multihop IoT Networks},'' in \emph{IEEE Global Communications
  Conference, GLOBECOM}, 2017, pp. 1--6.

\bibitem{P_2018_13}
L.~L. X.~Zhong, R.~Lu and S.~Zhang., ``{ETOR: Energy and Trust Aware
  Opportunistic Routing in Cognitive Radio Social Internet of Things},'' in
  \emph{IEEE Global Communications Conference, GLOBECOM}, 2017, pp. 1--6.

\bibitem{P_2018_21}
{F. Singh, J. K. Vijeth, and C. Siva Ram Murthy}, ``Parallel opportunistic
  routing in iot networks,'' in \emph{Wireless Communications and Networking
  Conference (WCNC)}.\hskip 1em plus 0.5em minus 0.4em\relax IEEE, 2016, pp.
  1--6.

\bibitem{P_2018_22}
{I. Nunes, C. Celes, I. Nunes, P. OS Vaz de Melo, and A. AF Loureiro},
  ``Combining spatial and social awareness in d2d opportunistic routing,''
  \emph{IEEE Communications Magazine}, vol.~56, no.~1, pp. 128--135, 2018.

\bibitem{P_2018_26}
{Y. Han, H. Wu, Z. Yang, and D. Li}, ``A new data transmission strategy in
  mobile d2d networks—deterministic, greedy, or planned opportunistic
  routing?'' \emph{IEEE Transactions on Vehicular Technology}, vol.~66, no.~1,
  pp. 594--609, 2017.

\bibitem{P_2018_23}
{J. Oksanen, B. Kaufman, V. Koivunen, and H. Vincent Poor}, ``Robotics inspired
  opportunistic routing for cognitive radio using potential fields,''
  \emph{IEEE Transactions on Cognitive Communications and Networking}, vol.~1,
  no.~1, pp. 45--55, 2015.

\bibitem{P_2018_24}
{Y. Qin, X. Zhong, Y. Yang, L. Li, and Y. Ye}, ``Combined channel assignment
  and network coded opportunistic routing in cognitive radio networks,''
  \emph{Computers \& Electrical Engineering}, vol.~52, pp. 293--306, 2016.

\bibitem{P6_2}
S.~Biswas and R.~Morris, ``Exor: opportunistic multi-hop routing for wireless
  networks,'' \emph{ACM SIGCOMM Computer Communication Review}, vol.~35, no.~4,
  pp. 133--144, 2005.

\bibitem{P_2020_06_251106}
{H. ALshaer and E. Horlait}, ``{An optimized adaptive broadcast scheme for
  inter-vehicle communication},'' vol.~5, pp. 2840--2844, 2005.

\bibitem{P8}
{E. Rozner, J. Seshadri, Y. Mehta, and L. Q. L. Qiu}, ``Soar: Simple
  opportunistic adaptive routing protocol for wireless mesh networks,''
  \emph{IEEE transactions on Mobile computing}, vol.~8, no.~12, p. 1622, 2009.

\bibitem{P_8_1}
M.~K. A.~Zubow and J.-P. Redlich, ``{Multi-channel opportunistic routing in
  Multi-Hop Wireless Networks},'' \emph{the 13th European Wireless Conference},
  2007.

\bibitem{P_60_6}
M.~Lu and J.~Wu, ``Opportunistic routing algebra and its applications,'' in
  \emph{In Proceeding of IEEE INFOCOM}, 2009, pp. 2374--2382.

\bibitem{P_60_4}
T.~Le and Y.~Liu, ``Opportunistic overlay multicast in wireless networks,'' in
  \emph{IEEE Global Telecommunications Conference, GLOBECOM}, 2010, pp. 1--5.

\bibitem{P_41_20}
C.~Westphal, ``{Opportunistic routing in dynamic ad hoc networks: The OPRAH
  protocol},'' in \emph{IEEE International Conference onMobile Adhoc and Sensor
  Systems (MASS)}, 2006, pp. 570--573.

\bibitem{P37}
M.~L. J.~Wu and F.~Li, ``Utility-based opportunistic routing in multi-hop
  wireless networks,'' in \emph{The 28th International Conference on
  Distributed Computing Systems,ICDCS}.\hskip 1em plus 0.5em minus 0.4em\relax
  IEEE, 2008, pp. 470--477.

\bibitem{P17}
M.~G. H.~Dubois-Ferrière and M.~Vetterli, ``Valuable detours: Least-cost
  anypath routing,'' \emph{IEEE/ACM Transactions on Networking (TON)}, vol.~19,
  no.~2, pp. 333--346, 2011.

\bibitem{p_9_1}
{Z. Wang, Y. Chen, and C. Li}, ``{CORMAN: A novel cooperative opportunistic
  routing scheme in mobile ad hoc networks},'' \emph{IEEE Journal on Selected
  Areas in Communications}, vol.~30, no.~2, pp. 289--296, 2012.

\bibitem{p_9_2}
{Y. Li, A. Mohaisen, Z. Zhang}, ``Trading optimality for scalability in
  large-scale opportunistic routing,'' \emph{IEEE Transactions on Vehicular
  Technology}, vol.~62, no.~5, pp. 2253--2263, 2013.

\bibitem{P36}
{P. Zhao, X. Yang, J. Wang, B. Liu, and J. Wang}, ``{BOR/AC: bandwidth-aware
  opportunistic routing with admission control in wireless mesh networks},'' in
  \emph{INFOCOM, 2012 Proceedings IEEE}.\hskip 1em plus 0.5em minus 0.4em\relax
  IEEE, 2012, pp. 2701--2705.

\bibitem{P_9_2_2}
{Y. Liang, W. Chung, H. Zhang, and S. Kuo}, ``Throughput improvement of
  multi-hop wireless mesh networks with cooperative opportunistic routing,'' in
  \emph{Wireless Communications and Networking Conference (WCNC)}.\hskip 1em
  plus 0.5em minus 0.4em\relax IEEE, 2012, pp. 3035--3039.

\bibitem{P30}
{W. WangBo, H. HuangChuanhe, Y. YangWenzhong, and W. WangTong}, ``Trust
  opportunistic routing protocol in multi-hop wireless networks,'' in
  \emph{International Conference on Wireless Communications, Networking and
  Information Security (WCNIS)}, 2010, pp. 563--567.

\bibitem{P_43_41}
N.~Chakchouk, ``A survey on opportunistic routing in wireless communication
  networks,'' \emph{IEEE Communications Surveys \& Tutorials}, vol.~17, no.~4,
  pp. 2214--2241, 2015.

\bibitem{P_42_21}
{M. Abuelela, S. Olariu, and I. Stojmenovic}, ``{OPERA: Opportunistic packet
  relaying in disconnected vehicular ad hoc networks},'' in \emph{IEEE
  International Conference on Mobile Ad Hoc and Sensor Systems, MASS}, 2008,
  pp. 285--294.

\bibitem{P_43_44}
{Y. Yuan, H. Yang, S. Wong, S. Lu, and W. }, ``{ROMER: resilient opportunistic
  mesh routing for wireless mesh networks},'' in \emph{IEEE workshop on
  wireless mesh networks (WiMesh)}, vol.~12, 2005.

\bibitem{P_50_4}
B.~Raffaele and M.~Nurchis, ``{Survey on diversity-based routing in wireless
  mesh networks: Challenges and solutions},'' \emph{Computer Communications},
  vol.~33, no.~3, pp. 269--282, 2010.

\bibitem{P_50_5}
L.~S. P.~Spachos and D.~Hatzinakos, ``Performance comparison of opportunistic
  routing schemes in wireless sensor networks,'' in \emph{Annual Communication
  Networks and Services Research Conference (CNSR)}.\hskip 1em plus 0.5em minus
  0.4em\relax IEEE, 2011, pp. 271--277.

\bibitem{P_50_6}
D.~Y. X.~Fang and G.~Xue, ``Consort: node-constrained opportunistic routing in
  wireless mesh networks,'' in \emph{In Proceeding of IEEE INFOCOM}, 2011, pp.
  1907--1915.

\bibitem{P7}
W.~S. C-J.~Hsu, H-I.~Liu, ``Economy: a duplicate free opportunistic routing,''
  in \emph{Proceedings of the 6th international conference on mobile
  technology, application \& systems}.\hskip 1em plus 0.5em minus 0.4em\relax
  ACM, 2009, p.~17.

\bibitem{p_9_2_1}
{R. Ahlswede, N. Cai, S. R. Li, and R. W. Yeung}, ``Network information flow,''
  \emph{IEEE Transactions on information theory}, vol.~46, no.~4, pp.
  1204--1216, 2000.

\bibitem{p_9_666}
{M. Abdolahi, F. Eshghi, M. Kelarestaghi, and M. Bag-Mohammadi}, ``{QADC-MORE:
  A new QoS-aware dynamic credit MORE protocol for wireless mesh networks},''
  in \emph{IEEE Symposium on Computers and Communication (ISCC)}, 2015, pp.
  883--888.

\bibitem{P4}
{S. Katti, H. Rahul, D. Katabi, M. Medard, and J. Crowcroft}, ``{XORs in the
  air: practical wireless network coding},'' \emph{IEEE/ACM Transactions on
  Networking (ToN)}, vol.~16, no.~3, pp. 497--510, 2008.

\bibitem{P_60_2}
W.~C.-C. D.~Koutsonikolas and Y.~C. Hu, ``Efficient network-coding-based
  opportunistic routing through cumulative coded acknowledgments,''
  \emph{IEEE/ACM Transactions on Networking (TON)}, vol.~19, no.~5, pp.
  1368--1381, 2011.

\bibitem{p_9_3}
{P. Li, S. Guo, S. Yu, and A. V. Vasilakos}, ``{CodePipe: An opportunistic
  feeding and routing protocol for reliable multicast with pipelined network
  coding},'' in \emph{In Proceeding of IEEE INFOCOM}, 2012, pp. 100--108.

\bibitem{p_9_4}
C.-C.~W. D.~Koutsonikolas and Y.-C. Hu, ``{CCACK: Efficient network coding
  based opportunistic routing through cumulative coded acknowledgments},'' in
  \emph{In Proceeding of IEEE INFOCOM}, 2010, pp. 1--9.

\bibitem{p_9_5}
{Y. Lin, Y. Lin, B. Li, B. Liang, and B.Liang}, ``{CodeOR: Opportunistic
  routing in wireless mesh networks with segmented network coding},'' in
  \emph{IEEE International Conference on Network Protocols, ICNP}, 2008, pp.
  13--22.

\bibitem{p_9_6}
Y.-C.~H. D.~Koutsonikolas and C.~Wang, ``{XCOR: Synergistic interflow network
  coding and opportunistic routing},'' \emph{ACM MobiCom SRC}, 2008.

\bibitem{P_43_45}
{Y. Yan, B. Zhang, J. Zheng and J. Ma}, ``Core: a coding-aware opportunistic
  routing mechanism for wireless mesh networks,'' vol.~17, no.~3, 2010.

\bibitem{P_43_46}
{Y. Yan, B. Zhang, H. Mouftah, and J. Ma}, ``Practical coding-aware mechanism
  for opportunistic routing in wireless mesh networks,'' in \emph{IEEE
  International Conference on Communications, 2008. ICC}, 2008, pp. 2871--2876.

\bibitem{p_9_8}
J.~X. W.~Hu and Z.~Zhang, ``pmore: Exploiting partial packets in opportunistic
  routing,'' in \emph{IEEE Global Telecommunications Conference (GLOBECOM )},
  2011, pp. 1--6.

\bibitem{p_9_9}
{D. Koutsonikolas, Y. Hu, and C. Wang}, ``{Pacifier: High-Throughput, Reliable
  Multicast without``Crying Babies''in Wireless Mesh Networks},'' in \emph{In
  Proceeding of IEEE INFOCOM}, 2009, pp. 2473--2481.

\bibitem{P_40_1}
M.~S. R.~Baumann, S.~Heimlicher and A.~Weibel, ``A survey on routing metrics,''
  \emph{TIK report}, vol. 262, pp. 1--53, 2007.

\bibitem{p_9_10}
L.~C.-A. A.~Darehshoorzadeh and V.~Pla, ``Modeling and comparison of candidate
  selection algorithms in opportunistic routing,'' \emph{Computer Networks},
  vol.~55, no.~13, pp. 2886--2898, 2011.

\bibitem{p_9_11}
W.~C. Y.~Li and Z.~Zhang, ``Optimal forwarder list selection in opportunistic
  routing,'' in \emph{IEEE 6th International Conference on Mobile Adhoc and
  Sensor Systems, MASS}, 2009, pp. 670--675.

\bibitem{P_2019_2_71950}
{M. Youssef, M. Ibrahim, M. Abdelatif, L. Chen, and A. Vasilakos}, ``Routing
  metrics of cognitive radio networks: A survey,'' \emph{IEEE Communications
  Surveys \& Tutorials}, vol.~16, no.~1, pp. 92--109, 2014.

\bibitem{P_2020_06_272108}
A.~Trivi{\~n}o-Cabrera and S.~Canadas-Hurtado, ``{Survey on opportunistic
  routing in multihop wireless networks},'' \emph{International Journal of
  Communication Networks and Information Security}, vol.~3, no.~2, pp.
  170--177, 2011.

\bibitem{P25}
A.~Boukerche and A.~Darehshoorzadeh, ``Opportunistic routing in wireless
  networks: Models, algorithms, and classifications,'' \emph{ACM Computing
  Surveys (CSUR)}, vol.~47, no.~2, p.~22, 2015.

\bibitem{P_2019_5_181545}
{V. Sadatpour, F. Zargari, and M. Ghanbari}, ``{A Collision Aware Opportunistic
  Routing Protocol for VANETs in Highways},'' \emph{Wireless Personal
  Communications}, pp. 1--14, 2019.

\bibitem{P_2019_2_192147}
{A. Celik, N. Saeed, B. Shihada, Al-Naffouri, A. Tareq, and M. Alouini},
  ``{SectOR: Sector-Based Opportunistic Routing Protocol for Underwater Optical
  Wireless Networks},'' \emph{IEEE Wireless Communications and Networking
  Conference (WCNC)}, 2019.

\bibitem{P_2019_3_280030}
{C. Lyu, X. Zhang, Z. Liu, and C. Chi}, ``{Selective Authentication Based
  Geographic Opportunistic Routing in Wireless Sensor Networks for Internet of
  Things Against DoS Attacks},'' \emph{IEEE Access}, vol.~7, pp.
  31\,068--31\,082, 2019.

\bibitem{P_2019_2}
j.~Z. J.~W. W.~Duan, X.~Tang and G.~Zhou, ``{Load Balancing Opportunistic
  Routing for Cognitive Radio Ad Hoc Networks},'' \emph{Wireless Communications
  and Mobile Computing}, vol. 2018, 2018.

\bibitem{P_2019_12_051127}
{X. Zhong, L. Li, S. Zhang, and R. Lu}, ``{ECOR: An Energy Aware Coded
  Opportunistic Routing for Cognitive Radio Social Internet of Things},''
  \emph{Wireless Personal Communications}, pp. 1--20, 2019.

\bibitem{P_2019_1}
Z.~Jin, Z.~Ji, and Y.~Su, ``{An Evidence Theory Based Opportunistic Routing
  Protocol for Underwater Acoustic Sensor Networks},'' \emph{IEEE Access},
  vol.~6, pp. 71\,038--71\,047, 2018.

\bibitem{P_2018_43}
{I. Amdouni, C. Adjih, N. AitSaadi, and P. Muhlethaler}, ``{Extensive
  Experimentations on Opportunistic Routing in Wireless Sensor Networks},''
  \emph{Sensors}, vol.~18, no.~9, p. 3031, 2018.

\bibitem{P_2018_41}
S.~X. J.~W. X.~Tang, J.~Zhou and K.~Zhou., ``{Geographic Segmented
  Opportunistic Routing in Cognitive Radio Ad Hoc Networks Using Network
  Coding},'' \emph{IEEE Access}, vol.~6, pp. 62\,766--62\,783, 2018.

\bibitem{P_2020_06_230000}
{X. Zhong, L. Li, Y. Zhang, B. Zhang, W. Zhang, and T. Yang}, ``{OODT: Obstacle
  Aware Opportunistic Data Transmission for Cognitive Radio Ad Hoc Networks},''
  \emph{IEEE Transactions on Communications}, 2020.

\bibitem{P_2019_3_051447}
{X. Zhang, L. Tao, F. Yan, D. Sung}, ``{Shortest-Latency Opportunistic Routing
  in Asynchronous Wireless Sensor Networks with Independent Duty-Cycling},''
  \emph{IEEE Transactions on Mobile Computing}, 2019.

\bibitem{P_2019_3_051453}
{Q. Guan, F. Ji, Y. Liu, H. Yu, and W. Chen}, ``{Distance-Vector based
  Opportunistic Routing for Underwater Acoustic Sensor Networks},'' \emph{IEEE
  Internet of Things Journal}, 2019.

\bibitem{P_2019_3_051456}
{H. Shafieirad, R. Adve, and S. Shahbazpanahi}, ``{Max-SNR opportunistic
  routing for large-scale energy harvesting sensor networks},'' \emph{IEEE
  Transactions on Green Communications and Networking}, vol.~2, no.~2, pp.
  506--516, 2018.

\bibitem{P_2019_3_051506}
{R. Coutinho, A. Boukerche, and A. Loureiro}, ``{PCR: A Power Control-based
  Opportunistic Routing for Underwater Sensor Networks},'' in \emph{Proceedings
  of the 21st ACM International Conference on Modeling, Analysis and Simulation
  of Wireless and Mobile Systems}.\hskip 1em plus 0.5em minus 0.4em\relax ACM,
  2018, pp. 173--180.

\bibitem{P_2019_3_290326}
{A. Hawbani, X. Wang, Y. Al-Sharabi, A. Ghannami, H. Kuhlani, and S. Karmoshi},
  ``{Load-Balanced Opportunistic Routing for Asynchronous Duty-cycled WSN},''
  \emph{IEEE Transactions on Mobile Computing}, 2018.

\bibitem{P_2020_07_031859}
M.~K. P.~Chithaluru, F. Al-Turjman and T.~Stephan, ``{I-AREOR: An
  Energy-balanced Clustering Protocol for implementing Green IoT in smart
  cities},'' \emph{Sustainable Cities and Society}, p. 102254, 2020.

\bibitem{P_2020_07_031845}
O.~D. R.~Zhao, N.~Li and X.~Shen, ``{CITP: Collision and Interruption Tolerant
  Protocol for Underwater Acoustic Sensor Networks},'' \emph{IEEE
  Communications Letters}, 2020.

\bibitem{P_2020_07_031851}
e.~a. M.~Awais, ``{Towards Void Hole Alleviation: Enhanced GEographic and
  Opportunistic Routing Protocols in Harsh Underwater WSNs},'' \emph{IEEE
  Access}, 2020.

\bibitem{P_2019_10_041116}
R.~T. P.~Chithaluru and K.~Kumar, ``{AREOR--Adaptive ranking based energy
  efficient opportunistic routing scheme in Wireless Sensor Network},''
  \emph{Computer Networks}, vol. 162, p. 106863, 2019.

\bibitem{P_2019_10_151334}
{D. Liu, Z. Cao, Y. He, X. Ji, M. Hou, and H. Jiang}, ``{Exploiting Concurrency
  for Opportunistic Forwarding in Duty-Cycled IoT Networks},'' \emph{ACM
  Transactions on Sensor Networks (TOSN)}, vol.~15, no.~3, p.~31, 2019.

\bibitem{P_2020_03_280106}
R.~Coutinho and A.~Boukerche, ``{PCon: A Novel Opportunistic Routing Protocol
  for Duty-Cycled Internet of Underwater Things},'' \emph{IEEE Symposium on
  Computers and Communications (ISCC)}, pp. 1--6, 2019.

\bibitem{P_2020_07_031830}
{M. Ismail, M. Islam, A. Ahmad, F. Khan, A. Qazi, Z. Khan, Z. Wadud, and M.
  Al-Rakhami}, ``{Reliable Path Selection and Opportunistic Routing Protocol
  for Underwater Wireless Sensor Networks},'' \emph{IEEE Access}, 2020.

\bibitem{P_2019_2_71937}
{C. Zhang, C. Li, and Y. Chen}, ``{Joint Opportunistic Routing and Intra-Flow
  Network Coding in Multi-Hop Wireless Networks: A Survey},'' \emph{IEEE
  Network}, vol.~33, no.~1, pp. 113--119, 2019.

\bibitem{P_2019_2_71944}
S.~Batabyal and P.~Bhaumik, ``Mobility models, traces and impact of mobility on
  opportunistic routing algorithms: A survey,'' \emph{IEEE Communications
  Surveys \& Tutorials}, vol.~17, no.~3, pp. 1679--1707, 2015.

\bibitem{P_2020_06_282301}
W.~Moreira and P.~Mendes, ``Survey on opportunistic routing for
  delay/disruption tolerant networks,'' \emph{SITI Technical Report
  SITI-TR-11-02}, 2010.

\bibitem{P_2019_4_091555}
{A. Van Bemten, J. Guck, C. Machuca, and W. Kellerer}, ``{Routing metrics
  depending on previous edges: The Mn taxonomy and its corresponding
  solutions},'' \emph{IEEE International Conference on Communications (ICC)},
  pp. 1--7, 2018.

\bibitem{P_2019_5_261405}
{A. Guirguis, F. Digham, K. Seddik, M. Ibrahim, K. Harras, and M. Youssef},
  ``{Primary User-Aware Optimal Discovery Routing for Cognitive Radio
  Networks},'' \emph{IEEE Transactions on Mobile Computing}, vol.~18, no.~1,
  pp. 193--206, 2019.

\bibitem{P5}
J.~B. D.~Couto, D.~Aguayo and R.~Morris, ``{A High-Throughput Path Metric for
  Multi-Hop Wireless Routing},'' in \emph{in Proceedings of ACM MobiCom}, 2003,
  pp. 134--146.

\bibitem{P_2017_5}
C.~Koksal and H.~Balakrishnan, ``Quality-aware routing metrics for time-varying
  wireless mesh networks,'' \emph{IEEE Journal on selected areas in
  communications}, vol.~24, no.~11, pp. 1984--1994, 2006.

\bibitem{P_2019_2_181359}
{Y. Wu, S. Das, and R. Chandra}, ``Routing with a markovian metric to promote
  local mixing,'' in \emph{IEEE International Conference on Computer
  Communications (INFOCOM)}, 2007, pp. 2381--2385.

\bibitem{P_60_12}
L.~Shih-Chun and K.~Chen, ``Spectrum aware opportunistic routing in cognitive
  radio networks,'' in \emph{IEEE Global Telecommunications Conference,
  GLOBECOM}, 2010, pp. 1--6.

\bibitem{P_61_21}
B.~A. N.~Sghaier and A.~Mellouk, ``On enhancing network-lifetime in
  opportunistic wireless sensor networks,'' in \emph{IEEE Wireless
  Communications and Networking Conference (WCNC)}, 2014, pp. 2617--2622.

\bibitem{P_40_19}
{Y. Qin, L. Li, X. Zhong, Y. Yang and Y. Ye}, ``Opportunistic routing with
  admission control in wireless ad hoc networks,'' \emph{Computer
  Communications}, vol.~55, pp. 32--40, 2015.

\bibitem{P_2017_3}
C.~J. D.~Kang, H.~Kim and S.~Bahk, ``{ORGMA: Reliable opportunistic routing
  with gradient forwarding for MANETs},'' \emph{Computer Networks}, vol. 131,
  pp. 52--64, 2018.

\bibitem{P13}
{Z. Zhong, J. Wang, S. Nelakuditi, and G.-H. Lu}, ``On selection of candidates
  for opportunistic anypath forwarding,'' \emph{ACM SIGMOBILE Mobile Computing
  and Communications Review}, vol.~10, no.~4, pp. 1--2, 2006.

\bibitem{P14}
Z.~Zhang and S.~Nelakuditit, ``On the efficacy of opportunistic routing,'' in
  \emph{IEEE Conference on Sensor, Mesh and Ad Hoc Communications and Networks,
  SECON}, 2007, pp. 441--450.

\bibitem{P15}
{R. Laufer, P. B. Velloso, L. F. M. Vieira, and L. Kleinrock}, ``{PLASMA: A new
  routing paradigm for wireless multihop networks},'' in \emph{In Proceeding of
  IEEE INFOCOM}.\hskip 1em plus 0.5em minus 0.4em\relax IEEE, 2012, pp.
  2706--2710.

\bibitem{P18}
{W. Yang, H. ChuanHe, Wang Bo, Z. ZhenYu, and W. Tong}, ``{A reliable multicast
  for MANETs based on opportunistic routing and network coding},'' in
  \emph{IEEE International Conference on Wireless Communications, Networking
  and Information Security (WCNIS)}, 2010, pp. 540--545.

\bibitem{P_60_25}
S.~Chachulski, ``{Trading Structure for Randomness in Wireless Opportunistic
  Routing},'' in \emph{Master Thesis, MIT University}, 2007.

\bibitem{P20}
{C. Gray, N. Santhapuri, and S. Nelakuditi}, ``On bit-rate selection for
  opportunistic routing,'' in \emph{IEEE Conference Sensor, Mesh and Ad Hoc
  Communications and Networks Workshops, SECON}, 2008, pp. 1--6.

\bibitem{P33}
W.~L. K.~Zeng and H.~Zhai, ``Capacity of opportunistic routing in multi-rate
  and multi-hop wireless networks,'' \emph{IEEE Transactions on Wireless
  Communications}, vol.~7, no.~12, p. 5118, 2008.

\bibitem{P19}
{R. Laufer, H. Dubois-Ferrire, and L. Kleinrock}, ``Multirate anypath routing
  in wireless mesh networks,'' in \emph{In Proceeding of IEEE INFOCOM}, 2009,
  pp. 37--45.

\bibitem{P119}
{R. Laufer, H. Dubois-Ferriere, and L. Kleinrock}, ``Polynomial-time algorithms
  for multirate anypath routing in wireless multihop networks,'' \emph{IEEE/ACM
  Transactions on Networking}, vol.~20, no.~3, pp. 742--755, 2012.

\bibitem{P37_2}
M.~Lu and J.~W. F.~Li, ``Efficient opportunistic routing in utility-based ad
  hoc networks,'' \emph{IEEE Transactions on Reliability}, vol.~58, no.~3, pp.
  485--495, 2009.

\bibitem{P22}
L.~Y.-a. L.~Yanhua and L.~Pengkui, ``Link probability based opportunistic
  routing metric in wireless network,'' in \emph{International Conference on
  Communications and Mobile Computing, CMC}, vol.~2.\hskip 1em plus 0.5em minus
  0.4em\relax IEEE, 2009, pp. 308--312.

\bibitem{P_50_17}
{C. Hung, K. lin, C. Hsu, C. Chou, and C. Tu}, ``On enhancing network-lifetime
  using opportunistic routing in wireless sensor networks,'' in \emph{In
  Proceedings of 19th International Conference on Computer Communications and
  Networks}.\hskip 1em plus 0.5em minus 0.4em\relax IEEE, 2010, pp. 1--6.

\bibitem{P_50_18}
{C. Hung, K. lin, C. Hsu, and C. Chou}, ``{EFFORT: energy-efficient
  opportunistic routing technology in wireless sensor networks},''
  \emph{Wireless communications and mobile computing}, vol.~13, no.~8, pp.
  760--773, 2013.

\bibitem{P21}
E.~Ghadimi and O.~Landsiedel, ``Opportunistic routing in low duty-cycle
  wireless sensor networks,'' \emph{ACM Transactions on Sensor Networks
  (TOSN)}, vol.~10, no.~4, p.~67, 2014.

\bibitem{P211}
{E. Ghadimi, O. Landsiedel, P. Soldati, and M. Johansson}, ``A metric for
  opportunistic routing in duty cycled wireless sensor networks,'' in
  \emph{IEEE Conference on Sensor, Mesh and Ad Hoc Communications and Networks
  (SECON)}.\hskip 1em plus 0.5em minus 0.4em\relax IEEE, 2012, pp. 335--343.

\bibitem{P36_2_2}
P.~Zhao and X.~Yang, ``Opportunistic routing for bandwidth-sensitive traffic in
  wireless networks with lossy links,'' \emph{Journal of Communications and
  Networks}, vol.~18, no.~5, pp. 806--817, 2016.

\bibitem{P_50_22}
{WS. Wang, S. Kim, Y. Liu, G. Tan, and T. He}, ``Corlayer: A transparent link
  correlation layer for energy efficient broadcast,'' in \emph{Proceedings of
  international conference on Mobile computing \& networking}.\hskip 1em plus
  0.5em minus 0.4em\relax ACM, 2013, pp. 51--62.

\bibitem{P_50_23}
{S. Wang, A. Basalamah, S. Kim, S. Guo, Y. Tobe, and T. He},
  ``Link-correlation-aware opportunistic routing in wireless networks.''
  \emph{IEEE Transactions Wireless Communications}, vol.~14, no.~1, pp. 47--56,
  2015.

\bibitem{P_50_24}
{A. Basalamah, S. Kim, S. Guo, T. He, and Y. Tobe}, ``Link correlation aware
  opportunistic routing,'' in \emph{In Proceeding of IEEE INFOCOM}, 2012, pp.
  3036--3040.

\bibitem{P_50_25}
{S. Wang, S. Kim, Y. Liu, G. Tan, and T. He}, ``{A Unified Metric for
  Correlated Diversity in Wireless Networks.}'' \emph{IEEE Transactions
  Wireless Communications}, vol.~15, no.~9, pp. 6215--6227, 2016.

\bibitem{P_50_26}
------, ``Corlayer: A transparent link correlation layer for energy-efficient
  broadcast,'' \emph{IEEE/ACM Transactions on Networking}, no.~6, pp.
  1970--1983, 2015.

\bibitem{P38}
D.~Z. S.~He and K.~Xie, ``Channel aware opportunistic routing in multi-radio
  multi-channel wireless mesh networks,'' \emph{Journal of Computer Science and
  Technology}, vol.~29, no.~3, pp. 487--501, 2014.

\bibitem{P_40}
C.-J. Hsu and H.-I. Liu, ``Route selection for opportunistic routing in
  multi-channel scenario,'' in \emph{IEEE International Conference on
  Communications (ICC)}, 2013, pp. 6294--6299.

\bibitem{P_60_7}
D.~Y. X.~Fang and G.~Xue, ``{MAP: Multiconstrained anypath routing in wireless
  mesh networks},'' \emph{IEEE Transactions on Mobile Computing}, vol.~12,
  no.~10, pp. 1893--1906, 2013.

\bibitem{P_60_24}
P.~G. X~.Fang, D.~Yang and Guoliang, ``{Multi-Constrained Anypath Routing in
  Wireless Mesh Networks},'' in \emph{IEEE Conference on Sensor, Mesh and Ad
  Hoc Communications and Networks, SECON}, 2010, pp. 475--483.

\bibitem{P_50_20}
{C. Hsu, M. Kuo, S. Wang, and C. Chou}, ``Joint design of asynchronous
  sleep-wake scheduling and opportunistic routing in wireless sensor
  networks,'' \emph{IEEE Transactions on Computers}, vol.~63, no.~7, pp.
  1840--1846, 2014.

\bibitem{P_50_10}
T.~Dinh and T.~Gu, ``A novel metric for opportunistic routing in heterogenous
  duty-cycled wireless sensor networks,'' in \emph{IEEE International
  Conference on Network Protocols (ICNP)}, 2015, pp. 224--234.

\bibitem{P_2019_10_041028}
T.~G. N.~Dinh and Y.~Kim, ``{Rendezvous Cost-aware Opportunistic Routing in
  Heterogeneous Duty-cycled Wireless Sensor Network Systems},'' \emph{IEEE
  Access}, 2019.

\bibitem{P_50_19}
{X. Zhang, X. Cao, L. Yan, and D. Sung}, ``A street-centric opportunistic
  routing protocol based on link correlation for urban vanets,'' \emph{IEEE
  Transactions on Mobile Computing}, no.~1, pp. 1--1, 2016.

\bibitem{P_2018_30}
M.~Wymore and D.~Qiao, ``{Opportunistic Many-to-Many Multicasting in
  Duty-Cycled Wireless Sensor Networks},'' in \emph{IEEE International
  Conference on Communications (ICC)}, 2018, pp. 1--7.

\bibitem{P_2018_31}
X.~Z. M.~Wymore, Y.~Peng and D.~Qiao, ``{EDAD}: energy-centric data collection
  with anycast in duty-cycled wireless sensor networks,'' \emph{IEEE Wireless
  Communications and Networking Conference (WCNC)}, pp. 1560--1565, 2015.

\bibitem{P_50_15}
J.~G. C.~Hsu, H.~Liu and C.~Chou, ``Delay-sensitive opportunistic routing for
  underwater sensor networks,'' \emph{IEEE Sensors Journal}, vol.~15, no.~11,
  pp. 6584--6591, 2015.

\bibitem{P_60_11}
{M. Zhao, A. Kumar, P. Han, J. Chong, and Rongxing Lu}, ``A reliable and
  energy-efficient opportunistic routing protocol for dense lossy networks,''
  \emph{IEEE Wireless Communications Letters}, vol.~6, no.~1, pp. 26--29, 2017.

\bibitem{P_61_20}
{W. Jie, H. Yue, L. Hai, and Y. Fang}, ``Spectrum-aware anypath routing in
  multi-hop cognitive radio networks,'' \emph{IEEE Transactions on Mobile
  Computing}, vol.~16, no.~4, pp. 1176--1187, 2017.

\bibitem{P_2017_4}
{V. Sadatpour, F. Zargari, and M. Ghanbari}, ``A new cost function for
  improving anypath routing performance of vanets in highways,'' \emph{Wireless
  Networks}, pp. 1--11, 2017.

\bibitem{P_2018_40}
{D. Xu, W. Jiao, Z. Yin, B. Wu, Y. Peng, X. Chen, F. Chen, F. and D. Fang},
  ``{Enabling robust and reliable transmission in Internet of Things with
  multiple gateways},'' \emph{Computer Networks}, vol. 146, pp. 183--199, 2018.

\bibitem{P_2020_03_280059}
X.~Z. H.~Cheng, C.~Wang, ``{An Opportunistic Routing in Energy-Harvesting
  Wireless Sensor Networks With Dynamic Transmission Power},'' \emph{IEEE
  Access}, vol.~7, pp. 180\,652--180\,660, 2019.

\bibitem{P27}
{K. Zeng, W. Lou, J. Yang, and D. R. Brown}, ``On throughput efficiency of
  geographic opportunistic routing in multihop wireless networks,''
  \emph{Mobile Networks and Applications}, vol.~12, no.~5, pp. 347--357, 2007.

\bibitem{P23}
J.~Y. K.~Zeng and W.~Lou, ``On energy efficiency of geographic opportunistic
  routing in lossy multihop wireless networks,'' \emph{Wireless Networks},
  vol.~18, no.~8, pp. 967--983, 2012.

\bibitem{P26}
Z.~Y. K.~Zeng and W.~Lou, ``Location-aided opportunistic forwarding in
  multirate and multihop wireless networks,'' \emph{IEEE Transactions on
  Vehicular Technology}, vol.~58, no.~6, pp. 3032--3040, 2009.

\bibitem{P28}
W.~L. K.~Zeng and Y.~Zhang, ``Multi-rate geographic opportunistic routing in
  wireless ad hoc networks,'' in \emph{Military Communications Conference,
  MILCOM}, 2007, pp. 1--7.

\bibitem{P29}
W.~L. K.~Zeng and H.~Zhai, ``On end-to-end throughput of opportunistic routing
  in multirate and multihop wireless networks,'' in \emph{In Proceeding of IEEE
  INFOCOM}, 2008, pp. 816--824.

\bibitem{P_40_22}
W.~L. K.~Zeng, Z.~Yang, ``Opportunistic routing in multi-radio multi-channel
  multi-hop wireless networks,'' \emph{IEEE transactions on Wireless
  Communications}, vol.~9, no.~11, pp. 3512--3521, 2010.

\bibitem{P_40_23}
------, ``Opportunistic routing in multi-radio multi-channel multi-hop wireless
  networks,'' \emph{IEEE transactions on Wireless Communications}, vol.~9,
  no.~11, pp. 3512--3521, 2010.

\bibitem{P34}
B.~Z. R.~Draves, J.~Padhye, ``Routing in multi-radio, multi-hop wireless mesh
  networks,'' in \emph{Proceedings of the 10th annual international conference
  on Mobile computing and networking}.\hskip 1em plus 0.5em minus 0.4em\relax
  ACM, 2004, pp. 114--128.

\bibitem{P_61_24}
{Z. Ruifeng, J. Gorce, O. Berder, and O. Sentieys}, ``Lower bound of
  energy-latency tradeoff of opportunistic routing in multihop networks,''
  \emph{EURASIP Journal on Wireless Communications and Networking}, vol. 2011,
  p.~2, 2011.

\bibitem{P_2019_151}
{R. Zhang, Olivier Berder, G. Jean-Marie, and S. Olivier}, ``Energy delay
  tradeoff in wireless multihop networks with unreliable links,'' \emph{Ad Hoc
  Networks}, vol.~10, no.~7, pp. 1306--1321, 2012.

\bibitem{P35}
{G. Zeng, P. Huang, M. Mutka, L. Xiao, and E. Torng}, ``Efficient opportunistic
  multicast via tree backbone for wireless mesh networks,'' in \emph{IEEE
  International Conference on Mobile Adhoc and Sensor Systems (MASS)}.\hskip
  1em plus 0.5em minus 0.4em\relax IEEE, 2011, pp. 600--609.

\bibitem{P24}
A.~Darehshoorzadeh and L.~Cerda-Alabern, ``Distance progress based
  opportunistic routing for wireless mesh networks,'' in \emph{Wireless
  Communications and Mobile Computing Conference (IWCMC)}.\hskip 1em plus 0.5em
  minus 0.4em\relax IEEE, 2012, pp. 179--184.

\bibitem{P_60_14}
L.~X.~C. Y.~Liu and X.~Shen, ``Spectrum-aware opportunistic routing in
  multi-hop cognitive radio networks,'' \emph{IEEE Journal on Selected Areas in
  Communications}, vol.~30, no.~10, pp. 1958--1968, 2012.

\bibitem{P_61_22}
K.~Daehyeok and Y.~Suh, ``Multi-rate combination of opportunistic routing and
  network coding: An optimization perspective,'' in \emph{IEEE Wireless
  Communications and Networking Conference (WCNC)}, 2012, pp. 1947--1952.

\bibitem{P_61_14}
{L. Cheng, J. Niu, J. Cao, S.K. Das, and Y. Gu}, ``Qos aware geographic
  opportunistic routing in wireless sensor networks,'' \emph{IEEE Transactions
  on Parallel and Distributed Systems}, vol.~25, no.~7, pp. 1864--1875, 2014.

\bibitem{P_61_23}
A.~Sayakkara and D.~Kim, ``Cross-layer opportunistic forwarding to reduce
  patterned synchronization effect in highly resource constrained wsns,'' in
  \emph{IEEE Wireless Communications and Networking Conference (WCNC)}, 2014,
  pp. 1962--1967.

\bibitem{P_50_14}
T.~N. R.~Zhang and J.~Morrison, ``Utility energy-based opportunistic routing
  for lifetime enhancement in wireless sensor networks,'' in \emph{IEEE
  International Conference on Communications (ICC)}, 2015, pp. 6324--6330.

\bibitem{P_50_16}
H.~P. M.~Aajami and J.~Suk, ``Combining opportunistic routing and network
  coding: a multi rate approach,'' in \emph{IEEE Wireless Communications and
  Networking Conference (WCNC)}, 2013, pp. 2208--2213.

\bibitem{P_2019_3_240848}
{S. Ghoreyshi, A. Shahrabi, and T. Boutaleb}, ``An opportunistic void avoidance
  routing protocol for underwater sensor networks,'' in \emph{IEEE 30th
  International Conference on Advanced Information Networking and Applications
  (AINA)}.\hskip 1em plus 0.5em minus 0.4em\relax IEEE, 2016, pp. 316--323.

\bibitem{P_2019_3_231559}
J.~So and H.~Heejung, ``Load-balanced opportunistic routing for duty-cycled
  wireless sensor networks,'' \emph{IEEE Transactions on Mobile Computing},
  vol.~16, no.~7, pp. 1940--1955, 2016.

\bibitem{P_2018_12}
K.~H. B.~Weigang, H.~Wang and R.~Zhao, ``{Path Diversity Improved Opportunistic
  Routing for Underwater Sensor Networks},'' \emph{Sensors}, vol.~18, no.~4,
  2018.

\bibitem{P_60_17}
{H. Khalife, S. Ahuja, N. Malouch, and M. Krunz}, ``Probabilistic path
  selection in opportunistic cognitive radio networks,'' in \emph{IEEE Global
  Telecommunications Conference, GLOBECOM}, 2008, pp. 1--5.

\bibitem{P311}
M.~Allman and V.~Paxson, ``{RFC 2988},'' November 2000.

\bibitem{P31}
S.~Latake and G.~R. Shinde, ``{Trust Opportunistic Multicast Routing (TOMR)
  Protocol In Multihop Wireless Network},'' \emph{International Journal of
  Computer Science and Information Technologies}, vol.~4, no.~3, pp. 481--484,
  2013.

\bibitem{P_60_13}
S.~O.~Badarneh and H.~B. Salameh, ``{Opportunistic Routing in Cognitive Radio
  Networks: Exploiting Spectrum Availability and Rich Channel Diversity},'' in
  \emph{IEEE Global Telecommunications Conference, GLOBECOM}, 2011, pp. 1--5.

\bibitem{P_70_2}
O.~Badarneh and H.~Salameh, ``Probabilistic quality-aware routing in cognitive
  radio networks under dynamically varying spectrum opportunities,''
  \emph{Computers \& Electrical Engineering}, vol.~38, no.~6, pp. 1731--1744,
  2012.

\bibitem{P_2019_4_010034}
{Z. Zhao, D. Ros{\'a}rio, T. Braun, E. Cerqueira, H. Xu, and L. Huang},
  ``Topology and link quality-aware geographical opportunistic routing in
  wireless ad-hoc networks,'' \emph{9th International Wireless Communications
  and Mobile Computing Conference (IWCMC)}, pp. 1522--1527, 2013.

\bibitem{P_2019_4_010059}
{D. Ros{\'a}rio, Z. Zhao, A. Santos, T. Braun, and E. Cerqueira}, ``{A
  beaconless opportunistic routing based on a cross-layer approach for
  efficient video dissemination in mobile multimedia IoT applications},''
  \emph{Computer communications}, vol.~45, pp. 21--31, 2014.

\bibitem{P_2019_4_010041}
{Z. Zhao,D. Ros{\'a}rio, T. Braun, and E. Cerqueira}, ``Context-aware
  opportunistic routing in mobile ad-hoc networks incorporating node
  mobility,'' \emph{IEEE Wireless Communications and Networking Conference
  (WCNC)}, pp. 2138--2143, 2014.

\bibitem{P_2019_4_011450}
{Z. Zhao, and T. Braun}, ``{SCAD: Sensor context-aware adaptive duty-cycled
  beaconless opportunistic routing for WSNs},'' \emph{IEEE 26th Annual
  International Symposium on Personal, Indoor, and Mobile Radio Communications
  (PIMRC)}, pp. 2038--2043, 2015.

\bibitem{P_2019_4_021403}
{N. Baccour, A. Koub{\^a}a, L. Mottola, M. Z{\'u}{\~n}iga, H. Youssef, C.
  Boano, and M. Alves}, ``Radio link quality estimation in wireless sensor
  networks: A survey,'' \emph{ACM Transactions on Sensor Networks (TOSN)},
  vol.~8, no.~4, p.~34, 2012.

\bibitem{P_2019_3_260013}
{C. Lyu, D. Gu, X. Zhang, S. Sun, Y. Zhang, and A. Pande}, ``{SGOR: Secure and
  scalable geographic opportunistic routing with received signal strength in
  WSNs},'' \emph{Computer Communications}, vol.~59, pp. 37--51, 2015.

\bibitem{P_50_21}
M.~Salehi and A.~Boukerche, ``Trust-aware opportunistic routing protocol for
  wireless networks,'' in \emph{Proceedings of the ACM symposium on QoS and
  security for wireless and mobile networks}, 2014, pp. 79--86.

\bibitem{P_60_10}
{M. Salehi, A. Boukerche, A. Darehshoorzadeh, and A. Mammeri}, ``Towards a
  novel trust-based opportunistic routing protocol for wireless networks,''
  \emph{Wireless Networks}, vol.~22, no.~3, pp. 927--943, 2016.

\bibitem{P_2019_3_280633}
N.~Kanthimathi, ``{Void handling using Geo-Opportunistic Routing in underwater
  wireless sensor networks},'' \emph{Computers \& Electrical Engineering},
  vol.~64, pp. 365--379, 2017.

\bibitem{P_2019_10_061532}
{P. Liu, X. Wang, A. Hawbani, O. Busaileh, L. Zhao, and A. Al-Dubai}, ``{FRCA:
  A Novel Flexible Routing Computing Approach for Wireless Sensor Networks},''
  \emph{IEEE Transactions on Mobile Computing}, 2019.

\bibitem{P_2020_03_271947}
L.~L. X.~W. X.~Zhong, R.~Lu and Y.~Zheng, ``{DSOR: A Traffic-Differentiated
  Secure opportunistic Routing with Game Theoretic Approach in MANETs},''
  \emph{IEEE Symposium on Computers and Communications (ISCC)}, pp. 1--6, 2019.

\bibitem{P_60_16}
{A. Socievole, E. Yoneki, F. De Rango, and J. Crowcroft}, ``Opportunistic
  message routing using multi-layer social networks,'' in \emph{Proceedings of
  the 2nd ACM workshop on High performance mobile opportunistic systems}, 2013,
  pp. 39--46.

\bibitem{P_2019_3_081656}
{Y. Dai, Ying and J. Wu}, ``Opportunistic routing based scheme with multi-layer
  relay sets in cognitive radio networks,'' in \emph{IEEE Wireless
  Communications and Networking Conference (WCNC)}.\hskip 1em plus 0.5em minus
  0.4em\relax IEEE, 2015, pp. 1159--1164.

\bibitem{P_2019_3_081719}
{H. Shafieirad, R. Adve, and S. ShahbazPanahi}, ``Opportunistic routing in
  large-scale energy harvesting sensor networks,'' in \emph{2016 IEEE Globecom
  Workshops (GC Wkshps)}.\hskip 1em plus 0.5em minus 0.4em\relax IEEE, 2016,
  pp. 1--6.

\bibitem{P_2019_3_081847}
{W. Cui, Y. Yao, and L. Song}, ``Buffer-aware opportunistic routing for
  wireless sensor networks,'' in \emph{14th IEEE Annual Consumer Communications
  \& Networking Conference (CCNC)}.\hskip 1em plus 0.5em minus 0.4em\relax
  IEEE, 2017, pp. 268--271.

\bibitem{P_2019_3_081853}
{R. Coutinho, A. Boukerche, L. Vieira, and A. Loureiro, Antonio AF}, ``{EnOR:
  Energy balancing routing protocol for underwater sensor networks},'' in
  \emph{IEEE International Conference on Communications (ICC)}.\hskip 1em plus
  0.5em minus 0.4em\relax IEEE, 2017, pp. 1--6.

\bibitem{P_2019_3_081913}
N.~Mirjazaee and N.~Moghim, ``An opportunistic routing based on symmetrical
  traffic distribution in vehicular networks,'' \emph{Computers \& Electrical
  Engineering}, vol.~47, pp. 1--12, 2015.

\bibitem{P_2019_3_081923}
{P. Yuan, M. Song, and X. Zhao}, ``Effective and efficient collection of
  control messages for opportunistic routing algorithms,'' \emph{Journal of
  Network and Computer Applications}, vol.~98, pp. 125--130, 2017.

\bibitem{P_2019_3_082109}
{M. Rahman, Y. Lee, and I. Koo}, ``{EECOR: An energy-efficient cooperative
  opportunistic routing protocol for Underwater acoustic sensor networks},''
  \emph{IEEE Access}, vol.~5, pp. 14\,119--14\,132, 2017.

\bibitem{P_2019_3_082115}
{I. Amdouni, C. Adjih, N. Aitsaadi, and P. Muhlethaler}, ``{Experiments with
  ODYSSE: Opportunistic Duty cYcle based routing for wirelesS Sensor
  nEtworks},'' in \emph{IEEE 41st Conference on Local Computer Networks
  (LCN)}.\hskip 1em plus 0.5em minus 0.4em\relax IEEE, 2016, pp. 232--235.

\bibitem{P_2019_3_082120}
{J. Luo, J. Hu, D. Wu, and R. Li}, ``Opportunistic routing algorithm for relay
  node selection in wireless sensor networks,'' \emph{IEEE Transactions on
  Industrial Informatics}, vol.~11, no.~1, pp. 112--121, 2015.

\bibitem{P8_1_1}
{C. Liu, D. Fang, Y. Hu, S. Tang, D. Xu, W. Cui, X. Chen, B. Liu, G. Xu, and H.
  Chen}, ``{EasyGo: Low-cost and robust geographic opportunistic sensing
  routing in a strip topology wireless sensor network},'' \emph{Computer
  Networks}, vol. 143, pp. 191--205, 2018.

\bibitem{P_2019_3_082240}
{N. Li, J. Martinez-Ortega, and H. D{\'\i}az}, ``{Cross-Layer and Reliable
  Opportunistic Routing Algorithm for Mobile Ad Hoc Networks},'' \emph{IEEE
  Sensors Journal}, vol.~18, no.~13, pp. 5595--5609, 2018.

\bibitem{P_2019_3_090033}
{J. Hu, J. Luo, Y. Zheng, and K. Li}, ``Graphene-grid deployment in energy
  harvesting cooperative wireless sensor networks for green iot,'' \emph{IEEE
  Transactions on Industrial Informatics}, 2018.

\bibitem{P_2019_3_090040}
{H. Fradj, R. Anane, and R. Bouallegue}, ``{Energy consumption for
  opportunistic routing algorithms in WSN},'' in \emph{IEEE 32nd International
  Conference on Advanced Information Networking and Applications (AINA)}.\hskip
  1em plus 0.5em minus 0.4em\relax IEEE, 2018, pp. 259--265.

\bibitem{P_2019_3_092101}
{W. Shin, S. Chung, and Y. Lee}, ``Parallel opportunistic routing in wireless
  networks,'' \emph{IEEE Transactions on Information Theory}, vol.~59, no.~10,
  pp. 6290--6300, 2013.

\bibitem{P_2019_10_061536}
S.~Dahiya and P.~Singh, ``{Energy Efficient SOCGO Protocol for Hole Repair Node
  Scheduling in Reliable Sensor System},'' \emph{Wireless Personal
  Communications}, pp. 1--21, 2019.

\bibitem{P_2019_3_092109}
{X. Mao, S. Tang, X. Xu, X. Li, and H. Ma}, ``{Energy-Efficient Opportunistic
  Routing in Wireless Sensor Networks},'' \emph{IEEE Transactions on Parallel
  \& Distributed Systems}, no.~11, pp. 1934--1942, 2011.

\bibitem{P_2019_3_092121}
{Zhang, Xinyu and Li, Baochun}, ``Dice: a game theoretic framework for wireless
  multipath network coding,'' in \emph{Proceedings of the 9th ACM international
  symposium on Mobile ad hoc networking and computing}.\hskip 1em plus 0.5em
  minus 0.4em\relax ACM, 2008, pp. 293--302.

\bibitem{P_2019_3_092126}
{S. Yang, F. Zhong, C. Yeo, B. Lee, and J. Boleng}, ``{Position based
  opportunistic routing for robust data delivery in MANETs},'' in \emph{IEEE
  Global Telecommunications Conference (GLOBECOM)}.\hskip 1em plus 0.5em minus
  0.4em\relax IEEE, 2009, pp. 1--6.

\bibitem{P_2019_10_301202}
A.~B. R.~Coutinho and A.~Loureiro, ``A novel opportunistic power controlled
  routing protocol for internet of underwater things,'' \emph{Computer
  Communications}, 2019.

\bibitem{P_2019_3_092136}
{Y. Yan, B. Zhang, H. Mouftah, and J. Ma}, ``Practical coding-aware mechanism
  for opportunistic routing in wireless mesh networks,'' in \emph{IEEE
  International Conference on Communications (ICC)}.\hskip 1em plus 0.5em minus
  0.4em\relax IEEE, 2008, pp. 2871--2876.

\bibitem{P_5_1}
{M. K. Han, A. Bhartia, L. Qiu, and E. Rozner}, ``{O3: Optimized overlay-based
  opportunistic routing},'' in \emph{Proceedings of the 20th ACM international
  symposium on mobile ad hoc networking and computing}, 2011, p.~2.

\bibitem{P_60_1}
B.~L. Y.~Lin and B.~Li, ``{SlideOR: Online opportunistic network coding in
  wireless mesh networks},'' in \emph{In Proceeding of IEEE INFOCOM}, 2010, pp.
  1--5.

\bibitem{P_60_3}
X.~Zhang and B.~Li, ``Optimized multipath network coding in lossy wireless
  networks,'' \emph{IEEE journal on selected areas in communications}, vol.~27,
  no.~5, pp. 622--634, 2009.

\bibitem{P_5_2}
M.~C. R.~Bruno and M.~Nurchis, ``{MaxOPP: A novel Opportunistic Routing for
  wireless mesh networks},'' in \emph{IEEE Symposium on Computers and
  Communications (ISCC)}, 2010, pp. 255--260.

\bibitem{P12}
F.~T. B.~Pavković and A.~Duda, ``Multipath opportunistic rpl routing over ieee
  802.15. 4,'' in \emph{Proceedings of the 14th ACM international conference on
  Modeling, analysis and simulation of wireless and mobile systems}, 2011, pp.
  179--186.

\bibitem{P_60_18}
{K. Chung, Y. Chou, and W. Liao}, ``{CAOR: Coding-aware opportunistic routing
  in wireless ad hoc networks},'' in \emph{IEEE International Conference on
  Communications (ICC)}, 2012, pp. 136--140.

\bibitem{P_60_19}
B.~S. Carnley, James and S.~K. Makki, ``{TORP: Tinyos opportunistic routing
  protocol for wireless sensor networks},'' in \emph{IEEE Consumer
  Communications and Networking Conference (CCNC)}, 2011, pp. 111--115.

\bibitem{P_60_20}
S.~Thorat and P.~J. Kulkarni, ``Opportunistic routing in presence of selfish
  nodes for manet,'' \emph{Wireless Personal Communications}, vol.~82, no.~2,
  pp. 689--708, 2015.

\bibitem{P_2019_3_032343}
{G. Huang, B. Zhang, and Z. Yao}, ``Data correlation aware opportunistic
  routing protocol for wireless sensor networks,'' in \emph{IEEE International
  Conference on Communications (ICC)}.\hskip 1em plus 0.5em minus 0.4em\relax
  IEEE, 2017, pp. 1--6.

\bibitem{P_60_21}
{C. Zhang, Y. Chen, and C. Li}, ``{ExOR compact: Reliable opportunistic data
  forwarding for wireless mesh networks},'' in \emph{IEEE International
  Conference on Communications (ICC)}, 2016, pp. 1--5.

\bibitem{P_60_22}
X.~Lan and S.~Zhang, ``{MT-NCOR: A practical optimization method for network
  coded opportunistic routing in WMN},'' in \emph{International Conference on
  Computer Communication and Networks (ICCCN)}.\hskip 1em plus 0.5em minus
  0.4em\relax IEEE, 2014, pp. 1--8.

\bibitem{P_2019_3_032334}
{N. Li, M. Jose, H. Hernandez, and L. Santidrian}, ``{Cross-Layer Balanced
  Relay Node Selection Algorithm for Opportunistic Routing in Underwater Ad-Hoc
  Networks},'' in \emph{IEEE 32nd International Conference on Advanced
  Information Networking and Applications (AINA)}.\hskip 1em plus 0.5em minus
  0.4em\relax IEEE, 2018, pp. 556--563.

\bibitem{P_2020_07_031912}
M.~G. S.~Malekyan, M. Bag-Mohammadi and M.~Abdollahi, ``On selection of
  forwarding nodes for long opportunistic routes,'' \emph{Wireless Networks},
  vol.~25, no.~4, pp. 1847--1854, 2019.

\bibitem{P_2019_3_081939}
{K. Choumas, I. Syrigos, T. Korakis, and L. Tassiulas}, ``Video-aware multicast
  opportunistic routing over 802.11 two-hop mesh networks,'' \emph{IEEE
  Transactions on Vehicular Technology}, vol.~66, no.~9, pp. 8372--8384, 2017.

\bibitem{P_2018_321}
{J. Li, X. Jia, X. Lv, Z. Han, J. Liu, and J. Hao}, ``Opportunistic routing
  with data fusion for multi-source wireless sensor networks,'' \emph{Wireless
  Networks}, pp. 1--11.

\bibitem{P_61_18}
{K. Wooseong, M. Gerla, s. Oh, K. Lee, and A. Kassler}, ``{CoRoute: a new
  cognitive anypath vehicular routing protocol},'' \emph{Wireless
  Communications and Mobile Computing}, vol.~11, no.~12, pp. 1588--1602, 2011.

\bibitem{P_61_19}
H.~F. C.~Chao and L.~Zhang, ``An anypath routing protocol for multi-hop
  cognitive radio ad hoc networks,'' in \emph{IEEE Conference on Scalable
  Computing and Communications and Its Associated Workshops (UTC-ATC-ScalCom)},
  2014, pp. 127--133.

\bibitem{P_70_1}
{P. Feng, F. Wu, B. Liu, and C. Dong}, ``{DSMA: Optimal multirate anypath
  routing in wireless networks with directional antennas},'' in \emph{Wireless
  Communications and Mobile Computing Conference (IWCMC)}.\hskip 1em plus 0.5em
  minus 0.4em\relax IEEE, 2014, pp. 381--386.

\bibitem{P_2019_3_081634}
{F. Wu, K. Gong, T. Zhang, G. Chen, and C. Qiao,}, ``{COMO: A game-theoretic
  approach for joint multirate opportunistic routing and forwarding in
  non-cooperative wireless networks},'' \emph{IEEE Transactions on Wireless
  Communications}, vol.~14, no.~2, pp. 948--959, 2015.

\bibitem{P_61_15}
O.~L. S.~Duquennoy and T.~Voigt, ``Let the tree bloom: Scalable opportunistic
  routing with orpl,'' in \emph{Proceedings of the ACM Conference on Embedded
  Networked Sensor Systems}, 2013, p.~2.

\bibitem{P_61_16}
{D. Liu, M. Hou, Z. Cao, J. Wang, Y. He, and Y. Liu}, ``Duplicate detectable
  opportunistic forwarding in duty-cycled wireless sensor networks,''
  \emph{IEEE/ACM Transactions on Networking (TON)}, vol.~24, no.~2, pp.
  662--673, 2016.

\bibitem{P_61_17}
{M. Marco, A. Loukas, M. Zimmerling, M. Zuniga, and K. Langendoen},
  ``Staffetta: Smart duty-cycling for opportunistic data collection,'' in
  \emph{Proceedings of the ACM Conference on Embedded Network Sensor Systems
  CD-ROM}, 2016, pp. 56--69.

\bibitem{P_70_3}
{Y. Mhaidat, M. Alsmirat, O. S. Badarneh, Y. Jararweh, and H. A. Bany Salameh},
  ``A cross-layer video multicasting routing protocol for cognitive radio
  networks,'' in \emph{IEEE International Conference on Wireless and Mobile
  Computing, Networking and Communications (WiMob)}, 2014, pp. 384--389.

\bibitem{P_2019_3_240152}
{A. Khan, N. Javaid, A. Sher, R. Abbasi, Z. Ahmad, and E. Ahmed}, ``{Load
  Balancing and Collision Avoidance Using Opportunistic Routing in Wireless
  Sensor Networks},'' in \emph{IEEE 32nd International Conference on Advanced
  Information Networking and Applications (AINA)}.\hskip 1em plus 0.5em minus
  0.4em\relax IEEE, 2018, pp. 236--243.

\bibitem{P_61_10}
{R. Coutinho, A. Boukerche, L. FM Vieira, and A. AF Loureiro}, ``{GEDAR:
  geographic and opportunistic routing protocol with depth adjustment for
  mobile underwater sensor networks},'' in \emph{IEEE International Conference
  on Communications (ICC)}, 2014, pp. 251--256.

\bibitem{P_61_11}
------, ``Geographic and opportunistic routing for underwater sensor
  networks,'' \emph{IEEE Transactions on Computers}, vol.~65, no.~2, pp.
  548--561, 2016.

\bibitem{P_61_12}
{N. Youngtae, U. Lee, S. Lee, P. Wang, L. FM Vieira, J. Cui, M. Gerla, and K.
  Kim}, ``{HydroCast: Pressure Routing for Underwater Sensor Networks},''
  \emph{IEEE Transactions Vehicular Technology}, vol.~65, no.~1, pp. 333--347,
  2016.

\bibitem{P_61_13}
{S. Zarar, N. Javaid, A. Sher, A.R. Hameed, Z.A. Khan, and U. Qasim},
  ``{Geographic and Opportunistic Clustering for Underwater WSNs},'' in
  \emph{International Conference on Innovative Mobile and Internet Services in
  Ubiquitous Computing (IMIS)}.\hskip 1em plus 0.5em minus 0.4em\relax IEEE,
  2016, pp. 57--62.

\bibitem{P_60_23}
A.~Varga, ``{OMNeT++ – Object-Oriented Discrete Event Simulator}.''\hskip 1em
  plus 0.5em minus 0.4em\relax http://www.omnetpp.org, 2018.

\bibitem{P_2017_6}
{J. Dede, A. Förster, E. Hernández-Orallo, J. Herrera-Tapia, K. Kuladinithi,
  V. Kuppusamy, P. Manzoni, A. Muslim, A. Udugama, and Z. Vatandas},
  ``Simulating opportunistic networks: Survey and future directions,''
  \emph{IEEE Communications Surveys \& Tutorials}, vol.~20, no.~2, pp.
  1547--1573, 2018.

\bibitem{P_70_5}
A.~N. J.~Kuruvila and I.~Stojmenovic, ``Hop count optimal position-based packet
  routing algorithms for ad hoc wireless networks with a realistic physical
  layer,'' \emph{IEEE Journal on selected areas in communications}, vol.~23,
  no.~6, pp. 1267--1275, 2005.

\bibitem{P_71_2}
{S. Deb, M. Effros, T. Ho, D.R. Karger, R. Koetter, D.S. Lun, M. Médard, and
  N. Ratnakar}, ``Network coding for wireless applications: A brief
  tutorial.''\hskip 1em plus 0.5em minus 0.4em\relax IWWAN, 2005.

\bibitem{P_71_1}
{G. Jakllari, S. Eidenbenz, N. Hengartner, S.V. Krishnamurthy, and M.
  Faloutsos}, ``{Link positions matter: A noncommutative routing metric for
  wireless mesh networks},'' \emph{IEEE Transactions on Mobile Computing},
  vol.~11, no.~1, pp. 61--72, 2012.

\bibitem{P_71_3}
M.~Jacobsson and C.~Rohner, ``Estimating packet delivery ratio for arbitrary
  packet sizes over wireless links,'' \emph{IEEE Communications Letters},
  vol.~19, no.~4, pp. 609--612, 2015.

\bibitem{P_2019_4_192011}
{J. Contreras-Castillo, S. Zeadally, and A. Guerrero-Iba{\~n}ez}, ``Internet of
  vehicles: architecture, protocols, and security,'' \emph{IEEE Internet of
  Things Journal}, vol.~5, no.~5, pp. 3701--3709, 2018.

\bibitem{P_2019_4_192019}
{J. Cheng, J. Cheng, M. Zhou, F. Liu, S. Gao, and C. Liu}, ``{Routing in
  internet of vehicles: A review},'' \emph{IEEE Transactions on Intelligent
  Transportation Systems}, vol.~16, no.~5, pp. 2339--2352, 2015.

\bibitem{P_2019_4_192038}
{M. Agiwal, A. Roy, and N. Saxena}, ``{Next generation 5G wireless networks: A
  comprehensive survey},'' \emph{IEEE Communications Surveys \& Tutorials},
  vol.~18, no.~3, pp. 1617--1655, 2016.

\bibitem{P_2019_4_192054}
{T. Suda, Tatsuya and T. Nakano}, ``{Molecular communication: A personal
  perspective},'' \emph{IEEE transactions on nanobioscience}, vol.~17, no.~4,
  pp. 424--432, 2018.

\bibitem{P_2019_4_192124}
{S. Shaikh, and R. Wism{\"u}ller}, ``{Routing in multi-hop cellular
  device-to-device (D2D) networks: A survey},'' \emph{IEEE Communications
  Surveys \& Tutorials}, vol.~20, no.~4, pp. 2622--2657, 2018.

\bibitem{P_60_8}
{C. Qu,L. Ju, Z. Jia, and L. Zheng}, ``Light-weight trust-based on-demand
  multipath routing protocol for mobile ad hoc networks,'' in \emph{IEEE
  International Conference on Trust, Security and Privacy in Computing and
  Communications (TrustCom)}, 2013, pp. 42--49.

\bibitem{P_60_9}
J.~F. M.~Blaze and J.~Lacy, ``Decentralized trust management,'' in \emph{In
  Proceedings of the IEEE Symposium on Security and Privacy}, 1996, pp.
  164--173.

\bibitem{P_2017_8}
{ S. Asaadi, K. Alikhademi, F. Eshghi, and J. Gilbert}, ``{TQOR: Trust-based
  QoS-oriented routing in cognitive MANETs},'' in \emph{IEEE International
  Symposium on Network Computing and Applications (NCA)}, 2017, pp. 1--8.

\bibitem{P_2019_5_021616}
{W. Dong, X. Liu, C. Chen, Y. He, G. Chen, L. Liu, and J. Bu}, ``{DPLC: Dynamic
  packet length control in wireless sensor networks},'' in \emph{2010
  Proceedings IEEE INFOCOM}.\hskip 1em plus 0.5em minus 0.4em\relax IEEE, 2010,
  pp. 1--9.

\bibitem{P_2017_9}
{J. Wang, Y. Liu, Y. He, W. Dong, and M. Li}, ``{QoF: Towards comprehensive
  path quality measurement in wireless sensor networks},'' \emph{IEEE
  Transactions on Parallel and Distributed Systems}, vol.~25, no.~4, pp.
  1003--1013, 2014.

\bibitem{P_2017_10}
{K. Srinivasan, M. Jain, J. Choi, T. Azim, E. Kim, P. Levis, and B.
  Krishnamachari}, ``The $\kappa$ factor: inferring protocol performance using
  inter-link reception correlation,'' in \emph{Proceedings of the international
  conference on Mobile computing and networking}.\hskip 1em plus 0.5em minus
  0.4em\relax ACM, 2010, pp. 317--328.

\bibitem{P_2017_11}
{T. Zhu, Z. Zhong, T. He, and Z. Zhang}, ``{Exploring Link Correlation for
  Efficient Flooding in Wireless Sensor Networks},'' in \emph{NSDI}, vol.~10,
  2010, pp. 1--15.

\bibitem{P_40_2}
W.~L. C.~Luk and O.~Yue, ``Opportunistic routing with directional antennas in
  wireless mesh networks,'' in \emph{In Proceeding of IEEE INFOCOM}, 2009, pp.
  2886--2890.

\bibitem{P_40_3}
{X. Fang, D. Yang, and G. Xue}, ``{DART: Directional anypath routing in
  wireless mesh networks},'' in \emph{IEEE International Conference on Mobile
  Adhoc and Sensor Systems (MASS)}, 2011, pp. 590--599.

\bibitem{P_2019_4_192157}
{J. Jagannath, N. Polosky, A. Jagannath, F. Restuccia, and T. Melodia},
  ``{Machine Learning for Wireless Communications in the Internet of Things: A
  Comprehensive Survey},'' \emph{Ad Hoc Networks}, 2019.

\bibitem{P_2019_4_192214}
{C. Zhang, P. Patras, and H. Haddadi}, ``Deep learning in mobile and wireless
  networking: A survey,'' \emph{IEEE Communications Surveys \& Tutorials},
  2019.

\bibitem{P_2019_4_192220}
{Q. Mao, F. Hu, and Q. Hao}, ``{Deep learning for intelligent wireless
  networks: A comprehensive survey},'' \emph{IEEE Communications Surveys \&
  Tutorials}, vol.~20, no.~4, pp. 2595--2621, 2018.

\bibitem{P_2019_5_262231}
{Y. Tang, N. Cheng, W. Wu, M. Wang, Y. Dai, and X. Shen}, ``{Delay-Minimization
  Routing for Heterogeneous VANETs with Machine Learning based Mobility
  Prediction},'' \emph{IEEE Transactions on Vehicular Technology}, 2019.

\bibitem{P_2019_5_262236}
{A. Koushik, F. Hu, and S. Kumar}, ``{Deep Q-Learning Based Node Positioning
  for Throughput-Optimal Communications in Dynamic UAV Swarm Network},''
  \emph{IEEE Transactions on Cognitive Communications and Networking}, 2019.

\end{thebibliography}

\end{document}